\UseRawInputEncoding

\documentclass[sn-basic,iicol]{sn-jnl}

\usepackage{amsmath,graphicx}
\usepackage{amssymb}
\usepackage{graphicx}
\usepackage{epstopdf}
\usepackage{tabularx}
\usepackage{color}
\usepackage{times}
\usepackage{epsfig}
\usepackage{amsmath,amsthm}
\usepackage{amsmath}
\usepackage{array}
\usepackage{float}
\usepackage{threeparttable}
\usepackage{algorithm}
\usepackage{color}
\usepackage{lineno}
\usepackage{booktabs}
\usepackage{bm}
\usepackage{subfigure}
\usepackage{cases}
\usepackage{epsfig,amssymb,subfigure,amsmath,version}
\usepackage{amssymb,version,fancybox,mathrsfs}
\usepackage{graphicx,mathrsfs,fancyhdr}
\usepackage{subfigure,slashbox,rotating,version,fancybox,pifont,booktabs,epsfig}
\usepackage{multirow}
\usepackage{amssymb}
\usepackage{amsfonts}
\usepackage{mathrsfs}
\usepackage{booktabs}

\begin{document}

\title[Article Title]{CurvPnP: Plug-and-play Blind Image Restoration with Deep Curvature Denoiser}


\author[1]{\fnm{Yutong} \sur{Li}}\email{yutong$\_$li@tju.edu.cn}

\author*[1]{\fnm{Yuping} \sur{Duan}}\email{yuping.duan@tju.edu.cn}

\affil[1]{\orgdiv{Center for Applied Mathematics}, \orgname{Tianjin University}, \orgaddress{\street{No 92, Weijin},  \city{Tianjin}, \postcode{300072}, \country{P.R. China}}}




\abstract{Due to the development of deep learning-based denoisers, the plug-and-play strategy has achieved great success in image restoration problems. However, existing plug-and-play image restoration methods are designed for non-blind Gaussian denoising such as \citet{zhang2021plug}, the performance of which visibly deteriorate for unknown noises. To push the limits of plug-and-play image restoration, we propose a novel framework with blind Gaussian prior, which can deal with more complicated image restoration problems in the real world. More specifically, we build up a new image restoration model  by regarding the noise level as a variable, which is implemented by a two-stage blind Gaussian denoiser consisting of a noise estimation subnetwork and a denoising subnetwork, where the noise estimation subnetwork provides the noise level to the denoising subnetwork for blind noise removal. We also introduce the curvature map into the encoder-decoder architecture and the supervised attention module to achieve a highly flexible and effective convolutional neural network. The experimental results on image denoising, deblurring and single-image super-resolution are provided to demonstrate the advantages of our deep curvature denoiser and the resulting plug-and-play blind image restoration method over the state-of-the-art model-based and learning-based methods. Our model is shown to be able to recover the fine image details and tiny structures even when the noise level is unknown for different image restoration tasks. The source codes are available at \url{https://github.com/Duanlab123/CurvPnP}. }




\maketitle

\section{Introduction}
Due to the degradation of images during acquisition and transmission process, image restoration is a crucial topic in image processing society, which are required to balance between image spatial details and high-level contextualized information. Mathematically, the degraded image $y$ can be expressed as follows
\[y=\mathcal{T}(x)+n,\]
where $\mathcal{T}$ denotes degradation operation, $x$ denotes clean image and $n$ is the additive white Gaussian noises of standard deviation $\sigma_s$. The aim of image restoration is to recover $x$ from $y$. The classical maximum a posterior (MAP) framework can be utilized to estimate $x$ by maximizing the posterior distribution $p(x\vert y)$, which can be formulated as the following optimization problem
\begin{equation}
\hat{x}=\arg\max_x p(x\vert y)=\arg\max_x\frac{p(y\vert x)p(x)}{p(y)}.
\end{equation}
Since maximizing $p(x\vert y)$ amounts to minimizing the log-likelihood, we have
\begin{equation}
\hat{x}=\arg\min_x -\ln p(y\vert x)-\ln p(x),
\label{model0}
\end{equation}
where $p(x)$ delivers the prior of noise-free image $x$ being independent of degraded image $y$. The usual formulation is $p(x)\propto \exp(-\lambda R(x))$, where $R(x)$ denotes the regularization term and $\lambda$ is a positive scalar. More formally, the model \eqref{model0} can be rewritten as below
\begin{equation}
\hat{x} = \arg\min_{x}~~\frac{1}{2\sigma_s^2}\vert\vert y-\mathcal{T}(x)\vert\vert^2+\lambda R(x).
\label{model1}
\end{equation}
Extensive studies have been devoted to image restoration tasks, which can be roughly divided into model-based methods and learning-based methods. The model-based methods are well-known for their highly explanatory and abilities in dealing with different image restoration tasks without long-time training and large dataset. For instance, the multi-scale vector total variation \citep{dong2011multi} was proposed to deal with color image denoising and deblurring tasks. The curvature regularization \citep{ZhongYD21} and Weingarten map regularization \citep{ZhongLD22} were proposed for image restoration, deblurring and inpainting problems.
However, model-based methods often introduce complex regularization terms to obtain satisfactory restoration results leading to high computational costs.
The learning-based methods have fast inference and high image restoration qualities. \citet{Zamir2021MPRNet} proposed a multi-stage image restoration architecture which broke down the challenging image restoration task into sub-tasks to progressively restore a degraded image. \citet{Chen2022Nonblind} proposed a dataset-free deep learning approach for non-blind image deconvolution, which introduces model uncertainty implemented by a specific spatially-adaptive dropout scheme to handle the solution ambiguity and a self-supervised loss to deal with the measurement noise. \citet{zhang2020gated} proposed a gated fusion network consisting of a restoration branch and a base branch, where the features were fed into the image reconstruction module to generate the sharp high-resolution image.
However, different network models have to be trained according to specific image restoration tasks.

Indeed, model-based methods and learning-based methods have complementary advantages. By combining the advantages of two methods, it can provide more effective image restoration methods such as the Plug-and-Play (PnP) strategy \citep{venkatakrishnan2013plug,sreehari2016plug}. Due to the variable splitting method such as half quadratic splitting (HQS) \citep{geman1995nonlinear} and alternating direction method of multipliers (ADMM) \citep{boyd2011distributed}, the subproblems containing prior terms can be treated separately. The
PnP method replaces the denoising subproblem of model-based optimization with denoiser prior, where both the traditional block matching and 3D filtering (BM3D) \citep{Dabov2007Image}, non-local mean (NLM) \citep{buades2005non} and the pre-trained deep learning denoisers  such as IRCNN \citep{zhang2017learning}, FFDNet \citep{zhang2018ffdnet} have been used as the denoiser. Although the pre-trained learning-based denoisers achieve better restoration results, they also have some shortcomings. On one hand, these approaches need to know the noise level of degraded images and interpose the noise level to the denoiser during iteration, which lose the effect on degraded images with unknown noise levels. On the other hand, it is difficult for the existing PnP image restoration to balance the denoising against the recovery of the fine details well.

Curvature can effectively model the plentiful structural information contained in images, which has been widely used to preserve geometric features of the image surface for various image processing tasks \citep{goldluecke2011introducing,schoenemann2012linear,ulen2015shortest,gorelick2016convexity,chen2017global,gong2017curvature,he2020curvature}.  For instance, \citet{brito2016image} minimized the $L^1$-norm of Gaussian curvature for image denoising and theoretically verified its ability in preserving image contrast and sharp edges. \citet{chambolle2019total} proposed the curvature depending energies in the roto-translation space for various problems from shape- and image processing, which was solved by the primal-dual optimization with convergence guarantee. \citet{zhong2020minimizing} introduced the total curvature regularity for image denoising, segmentation and inpainting, which is shown capable to preserve sharp edges and fine details. \citet{liu2022operator} developed an operator-splitting method for solving the Gaussian curvature regularization model, which achieves excellent restoration results for surface smoothing and image denoising applications. \citet{wang2022efficient} discussed the advantages of the curvature regularization methods over deep learning approaches in preserving geometric properties of images, which can be used as complementary to avoid generating unnatural artifacts produced by deep learning method.

\begin{figure*}[t]
\centering
\includegraphics[width=1.0\linewidth]{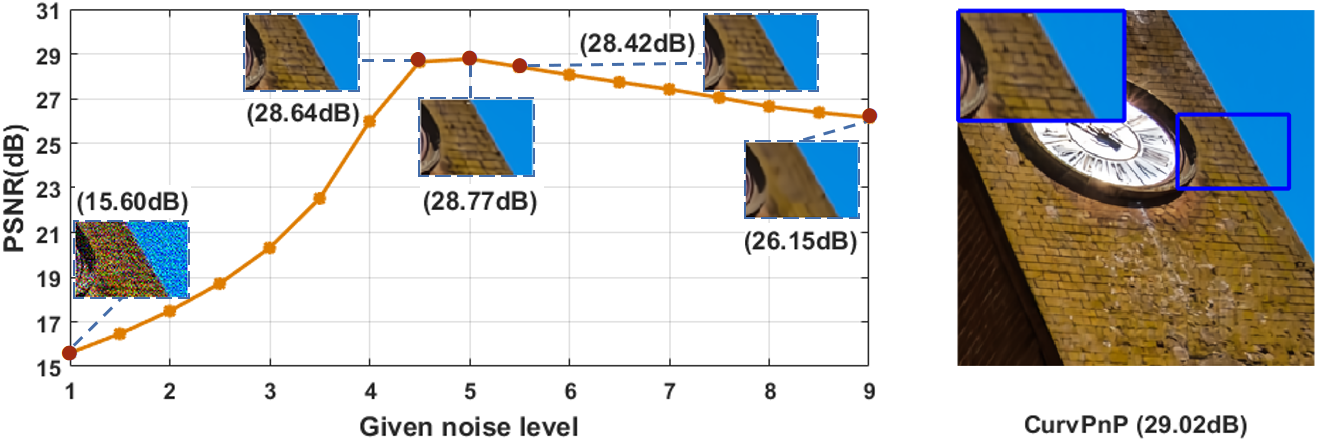}
\caption{Image deblurring results of DPIR with different given noise levels (the left figure) and CurvPnP (the right image) on a blurry image with noise level 5.}
\label{different_n}
\end{figure*}

In this paper, we propose a plug-and-play blind image restoration method with the deep curvature-based denoiser, called CurvPnP, which can ideally solve the blind noise removal problems by regarding the noise level as a variable. To be specific, we set up an effective two-stage blind denoising architecture, consisting of a noise estimation subnetwork and a denoising subnetwork, to deal with noisy images of different noise levels. The noise estimation subnetwork can calculate the noise level in each PnP iteration. Unlike the existing plug-and-play image restoration methods \citep{zhang2017learning,zhang2021plug}, which are designed for non-blind Gaussian denoising, our method does not need to know the noise level in advance. As illustrated in Fig. \ref{different_n}, the restoration performance of the state-of-the-art DPIR \citep{zhang2021plug} degraded obviously once the given noise level is inaccurate. Our CurvPnP can obtain high-quality restoration results without knowing the noise level in advance. The main contributions of this work are as follows

\begin{itemize}
    \item We propose an effective and powerful two-stage blind image restoration method including the noise estimation subnetwork and denoising subnetwork, called C-UNet.
    In particular, we take the advantages of ConvNeXt block and encoder-decoder architecture to build up both subnetworks to maintain efficiency and improve model performance.

    \item We develop a novel denoiser containing rich curvature information, where Gaussian curvature map \citep{ZhongYD21} is not only used as a priori, but also used in the supervised attention module to refine the features. Thanks to the curvature information, our denoiser shows a strong ability to restore the edges and fine structures of images.

    \item The iterative PnP strategy is implemented for solving different image restoration problems such as image deblurring and super-resolution, where our CurvPnP is shown with very competitive results.
\end{itemize}

The rest of the paper is organized as follows.
Sect. \ref{preliminaries} dedicates to review the PnP image restoration methods.
We describe the PnP blind image restoration method in Sect. \ref{curvPnP}. Sect. \ref{denoiser} introduces the deep curvature-based denoiser including the network architecture and implementation details. Experiments on three representative image restoration tasks are provided in Sect. \ref{experiments}. The concluding remarks and possible future works are summarized in Sect. \ref{conlusion}.

\section{Plug-and-play Image Restoration}
\label{preliminaries}
The PnP priors were first proposed as denoisers for image restoration with traditional methods in \citet{venkatakrishnan2013plug}. One of the most widely used denoisers for the plug-and-play framework is BM3D. \citet{dar2016postprocessing} proposed a postprocessing method using the plug-and-play framework with BM3D as the denoiser for compression-artifact reduction. \citet{rond2016poisson} introduced a plug-and-play prior for Poisson inverse problem (P$^4$IP) with BM3D denoiser for image denoising and deblurring problems. \citet{chan2016plug} presented a continuation plug-and-play ADMM scheme, in which BM3D was used as a bounded prior for single image super-resolution and the single photon imaging problem. \citet{ono2017primal} plugged the BM3D as a Gaussian denoiser into the primal update for the primal-dual splitting (PDS) algorithm, whereas the dual update is devoted to handle with both  data-fidelity term and hard constraint. Another widely used denoiser is the nonlocal mean (NLM) denoiser. \citet{unni2018linearized} developed the linearized plug-and-play ADMM, where a fast low-complexity algorithm for doubly stochastic NLM was used as denoiser to deal with super-resolution and single-photon imaging. The weighted nuclear norm minimization (WNNM) \citep{gu2014weighted} and Gaussian mixture model (GMM) \citep{zoran2011from} have also been implemented as denoiser. \citet{kamilov2017plug} proposed the fast iterative shrinkage thresholding algorithm with WNNM denoiser for nonlinear inverse scattering. \citet{yair2018multi} developed a variable splitting method to integrate WNNM denoiser for image inpainting and deblurring. \citet{shi2018plug} plugged the GMM denoiser into ADMM to solve the image restoration inverse problems. \citet{teodoro2019image} built upon the plug-and-play framework by ADMM, and combined it with GMM as denoisers to address image deblurring, compressive sensing reconstruction, and super-resolution problem.

In recent years, the pre-trained deep learning denoisers are widely used for image restoration tasks. \citet{zhangjiawei2017learning} decomposed the non-blind deconvolution problem into image denoising and image deconvolution and trained a fully connected CNN to remove noise. \citet{meinhardt2017learning} replaced the proximal operator of the regularization used in many convex energy minimization algorithms by a denoising neural network in exemplary problems of image deconvolution with different blur kernels and image demosaicking. \citet{zhang2017learning} trained a set of denoisers for image denoising and plugged the learned denoiser prior into the optimization method to solve the image deblurring and super-resolution problems. \citet{gu2018integrating} incorporated a CNN Gaussian denoiser prior and a non-local self-similarity based denoiser prior for image deblurring and super-resolution problems. \citet{tirer2018image} proposed an alternative method for solving inverse problems using off-the-shelf denoisers, which requires less parameter tuning. \citet{he2019optimizing} plugged the residual CNNs into the ADMM algorithm to solve low-dose CT image reconstruction. \citet{li2019learning} plugged a set of CNN denoisers into the split Bregman iteration algorithm for solving the depth image inpainting problem. \citet{zhang2019deep} proposed a principled formulation and framework by extending bicubic degradation based deep single image super-resolution with the PnP framework to recover the low-resolution images with arbitrary blur kernels. \citet{dong2019denoising} implemented both the CNN based denoiser and the back-projection module to solve the super-resolution and deblurring tasks. \citet{tirer2019super} proposed the PnP iterative denoising and backward projection framework to image super-resolution using a set of CNN denoisers. \citet{sun2020block} developed a block coordinate regularization-by denoising algorithm by leveraging the deep denoiser as the explicit regularizer. \citet{bigdeli2020image} used the PnP CNN Maximum a Posteriori (MAP) denoiser to handle image denoising and inpainting tasks. By using the variable splitting technique, \citet{zhao2020multi} separated the fidelity term and regularization term and replaced the image prior model by learned prior implicitly for multi-frame super-resolution. \citet{Zheng2021an} proposed a neural network based method by combining with the deep Gaussian denoisers for image denoising. \citet{zhang2021plug} implicitly served a learning-based non-blind denoiser as the image prior for the PnP image deblurring, super-resolution and demosaicing methods. In \citet{zhang2022deep}, the deep PnP methods with a learning-based denoiser and deep unfolding methods are utilized to solve the image deblurring and super-resolution tasks. The theoretical convergence of the PnP scheme has been established under a certain Lipschitz condition on the denoisers in \citet{pmlr-v97-ryu19a}. \citet{bian2021optimization} solved the MRI reconstruction problem by applying meta-training on the adaptive learned regularization in the variational model. \citet{pmlr-v119-wei20b,Wei2022jmlr} presented the tuning-free PnP proximal algorithm, which can automatically determine the internal parameters such as the penalty parameter, the denoising strength and the terminal time.

\begin{figure*}[t]
\centering
\includegraphics[width=0.9\linewidth]{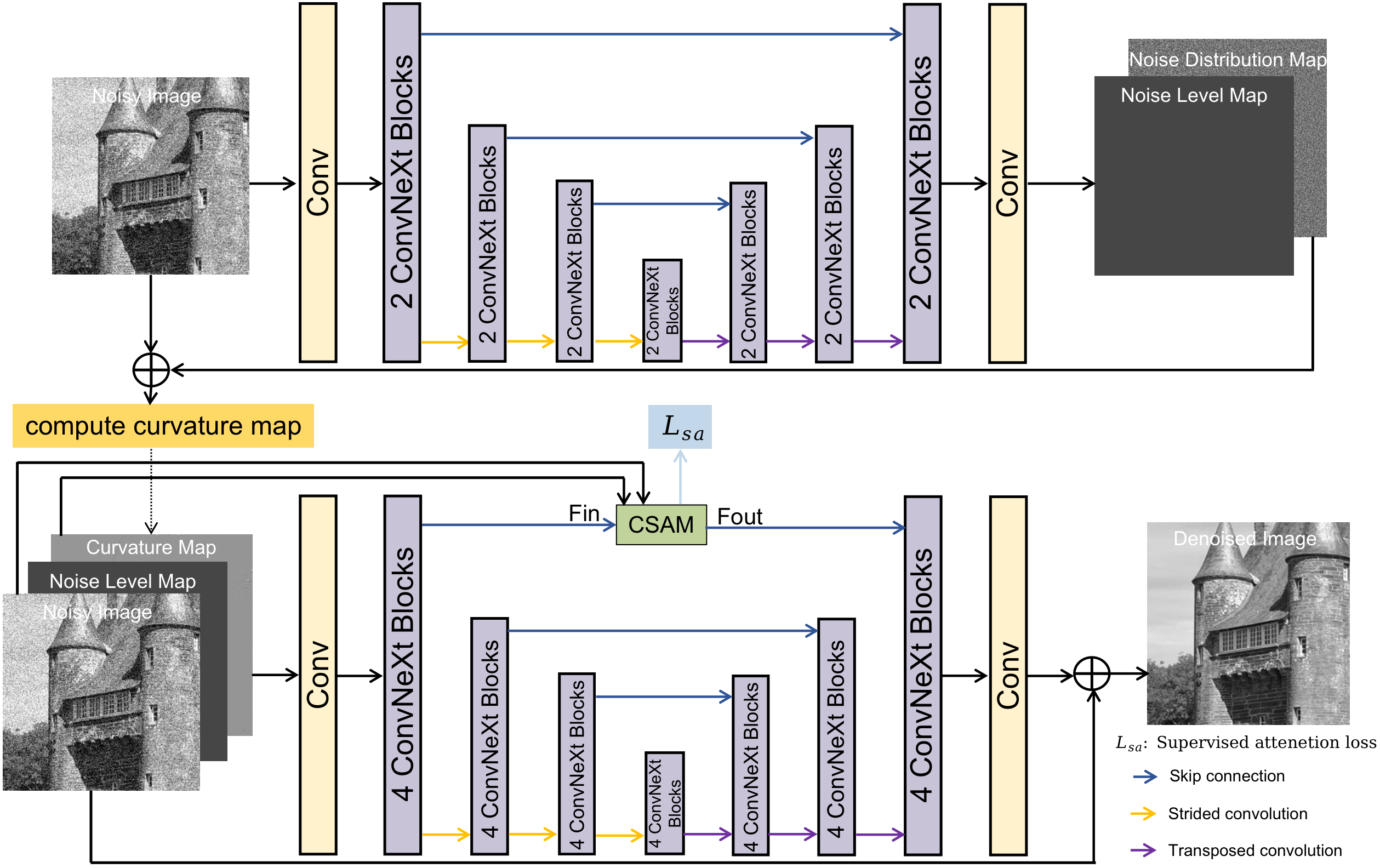}
\caption{The proposed C-UNet for blind image denoising, which consists of a noise estimation subnetwork and a denoising subnetwork.}
\label{netarc}
\end{figure*}

\section{Plug-and-play Blind Image Restoration}
\label{curvPnP}
As aforementioned, the up-to-date PnP method \citep{zhang2021plug} requires the noise level to be known, which is used to estimate the noise level for each iteration. However, once the noise level of the observed degraded image is inaccurate, the restoration result will be degraded obviously. For this, we propose a novel PnP framework by regarding the noise level as a variable. Firstly, to make the data fidelity and regularization separable, we introduce an auxiliary variable $v=x$ and rewrite the restoration model \eqref{model1} into a constrained minimization problem as follows
\begin{equation}\label{con_min}
\min_{x,v}~~\frac{1}{2\lambda\sigma_s^2}\vert\vert y-\mathcal{T}(x)\vert\vert^2+ R(v)\quad s.t. \quad v=x,
\end{equation}
where $\sigma_s$ denotes the noise level of the given degraded image $y$. We then reformulate it into the unconstrained minimization problem by the penalty method as follows
\begin{equation}
\min_{x, v, \sigma} ~~\frac{1}{2\lambda\sigma_s^2}\vert\vert y-\mathcal{T}(x)\vert\vert^2+  R(v)+\frac{1}{2\sigma^2}\vert\vert v-x\vert\vert^2,
\label{Lag}
\end{equation}
where $\sigma>0$ is a variable used to model the noise level of the noisy image $x$. Note that we introduce $\sigma$ to estimate the noise level that varies during iteration. In what follows, we sequentially minimize the two variables $x$ and $v$ iterative and alternatively by
\begin{equation}
x_{k+1} = \arg\min_{x}~~\frac{1}{2\lambda\sigma_s^2}\vert\vert y-\mathcal{T}(x)\vert\vert^2+\frac{1}{2\sigma_{k}^2}\vert\vert v_k-x\vert\vert^2,
\label{subprobx}
\end{equation}
and
\begin{equation}
(\sigma_{k+1}, v_{k+1}) = \arg\min_{\sigma, v}~~\frac{1}{2\sigma^2}\vert\vert v-x_{k+1}\vert\vert^2+ R(v),
\label{subprobv}
\end{equation}
respectively. More specifically, the sub-minimization problem \eqref{subprobx} w.r.t. the data defility is a quadratic problem, which can be solved by the closed-form solution. The specific solution of $x$ depends on the degradation operator $\mathcal{T}$, which will be given in the experiment section.
On the other hand, the solution of \eqref{subprobv} corresponds to the Gaussian denoising problem with the noise level $\sigma$. Obviously, the noise level $\sigma$ varies in the PnP iteration scheme. Unlike the existing Gaussian denoiser, the noise level $\sigma$ should be given in advances. We regard the noise level $\sigma$ as a variable and estimate it by the latest recovered image $x_{k+1}$ as follows
\begin{equation}
\sigma_{k+1}=Noise\_Estimator(x_{k+1}),
\label{solvsigma}
\end{equation}
which is realized by a convolutional neural network.

\begin{figure*}[t]
\centering
\begin{minipage}[htbp]{0.32\linewidth}
\centering
\includegraphics[width=2.5cm]{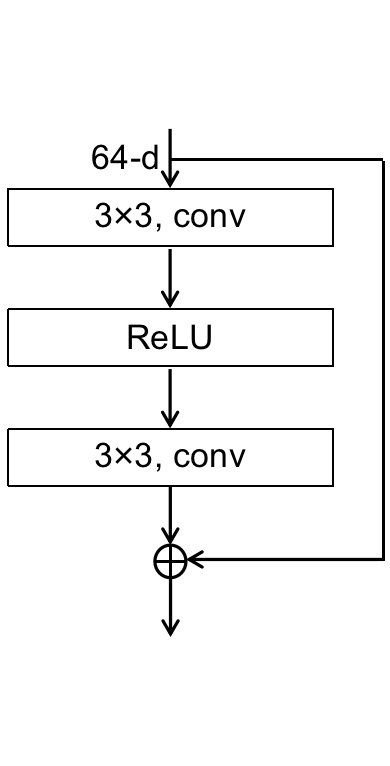}\\
{ \footnotesize{(a) ResNet Block in \citet{zhang2021plug}} }
\end{minipage}
\centering
\begin{minipage}[htbp]{0.32\linewidth}
\centering
\includegraphics[width=2.5cm]{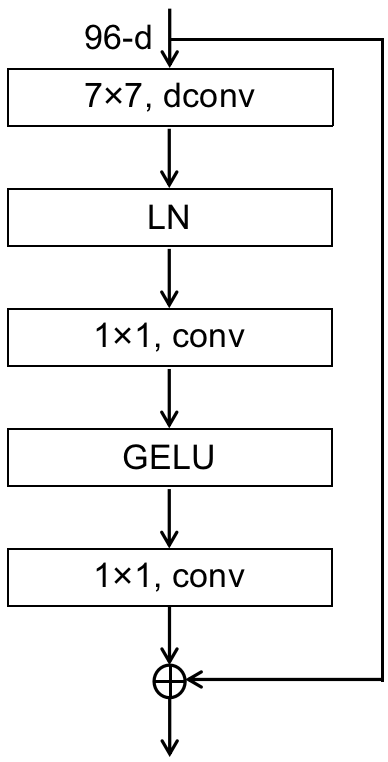}\\
{ \footnotesize{(b) ConvNeXt Block in \citet{liu2022convnet}} }
\end{minipage}
\centering
\begin{minipage}[htbp]{0.32\linewidth}
\centering
\includegraphics[width=2.5cm]{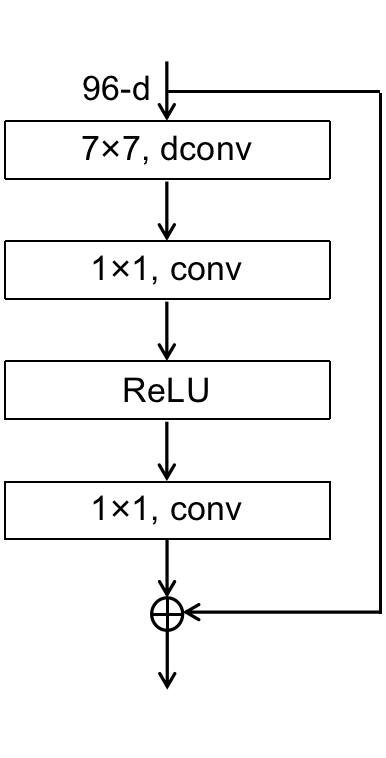}\\
{ \footnotesize{(c) ConvNeXt Block in C-UNet} }
\end{minipage}
\caption{The block design for the ResNet in \citet{zhang2021plug}, the ConvNeXt in \citet{liu2022convnet} and the ConvNeXt used in our network model.}
\label{ConvNeXt}
\end{figure*}

Now, we can solve the variable $v$ by a certain denoiser. Since the noise distribution varies across different image restoration tasks, we introduce an artificial noise control parameter $\rho$ as the coefficient of noise level map to reduce the influence of degenerate operations on noise distribution. Considering that the curvature map contains plentiful image details, we take curvature as a priori, and estimate the solution $v_{k+1}$ from
\begin{equation}
v_{k+1}=Denoiser(x_{k+1},\rho\sigma_{k+1},c_{k+1}),
\label{solvv}
\end{equation}
which is a convolutional neural network called the deep curvature denoiser.

Based on the above discussion, we come up with the PnP blind image restoration method with deep curvature denoiser (shorted by CurvPnP); see Algorithm \ref{alg1}.

\begin{algorithm}
\caption{CurvPnP: Plug-and-play blind image restoration with deep curvature denoisier}\label{alg1}
\begin{algorithmic}
\State \textbf{Input:} Original image $y$, degradation operator $\mathcal{T}$ and positive constants $\lambda$ and $\rho$;
\State \textbf{Initialize:} $\sigma_s=Noise\_ Estimator (y)$, $\sigma_0=50$ and $v_0=y$;
\For{$k= 0, 1, \dots, K-1$}\\
\qquad (i)~~ Compute $x_{k+1}$ from \eqref{subprobx};\\
\qquad (ii)~ Compute $\sigma_{k+1}$ from \eqref{solvsigma};\\
\qquad (iii) Compute $v_{k+1}$ from \eqref{solvv};\\
\qquad (iv)  Check whether the following stopping condition is satisfied:\\
\qquad\quad~ PSNR($v_{k+1}$)$<$PSNR($v_k$);
\EndFor
\State \textbf{Output:} $v$
\end{algorithmic}
\label{CurvPnPalg}
\end{algorithm}

\section{Deep Curvature Denoiser}
\label{denoiser}
\subsection{Network Architecture}
\textbf{C-UNet.} The encoder-decoder architecture \citep{Zamir2021MPRNet} is well-known for its strong ability in  encoding contextual information. However, the classical encoder-decoder UNet lacks the capability to preserve spatial image details and texture structure. Thus, the proposed denoiser, namely C-UNet, integrates the curvature map into UNet for effective denoiser prior modeling. Our C-UNet consists of two subnetworks, i.e., the noise estimation subnetwork and denoising subnetwork, to solve the sub-minimization problem w.r.t. $\sigma$ and $v$, respectively. As illustrated in Fig. \ref{netarc}, both subnetworks are built up using UNet architecture. In particular, the noise estimation subnetwork takes the noisy image $x$ as the input to obtain both the noise level map and noise distribution map. Then we calculate the Gaussian curvature map \citep{ZhongYD21} based on the estimated clean image obtained by removing the noise distribution map from the noisy input. The reason why we choose Gaussian curvature map as a priori is that Gaussian curvature is more suitable for real images in preserving fine structures and details \citep{ZhongYD21}. The denoising subnetwork takes the noisy image, the noise level map and Gaussian curvature map as the input. Similar to \citet{zhang2021plug}, we build up the noise estimation subnetwork and denoising subnetwork based on the encoder-decoder architecture by the 2$\times$2 stride convolution and 2$\times$2 transposed convolution in downsampling and upsampling operations, respectively. Besides, both the ConvNeXt blocks and curvature supervised attention module are introduced for effective denoiser prior modeling.

\begin{figure*}[t]
\centering
\includegraphics[width=1.0\linewidth]{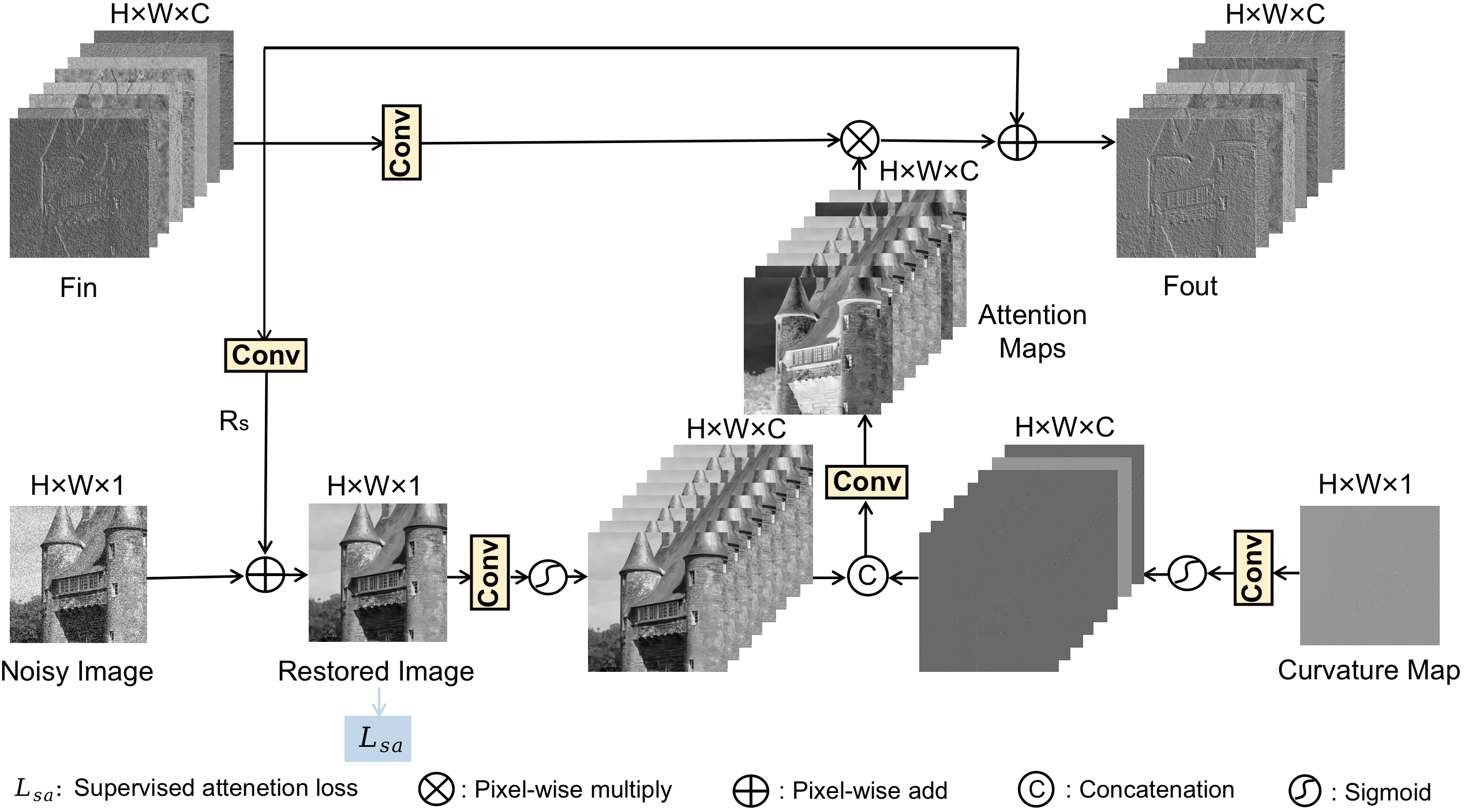}
\caption{The proposed curvature supervised attention module (CSAM).}
\label{SAM}
\end{figure*}

\textbf{ConvNeXt Block.} Inspired by \citet{liu2022convnet}, we integrate the ConvNeXt blocks into UNet for both the noise estimation and denoising subnetworks to promote the ability in feature extraction. Different from the ResNet block (Fig. \ref{ConvNeXt} (a)) used in DRUNet \citep{zhang2021plug}, ConvNeXt block (Fig. \ref{ConvNeXt} (b)) takes the advantage of the depthwise convolution and expand the  network width from 64 to 96 to increase the diversity of features. In addition, the inverted bottleneck is used to avoid loss of information and the large depthwise convolution with kernel size 7$\times$7 is used to improve the performance. In our implementation, we remove the layer normalization and use ReLU to replace GELU to save the memory and raise the effectiveness;  see Fig. \ref{ConvNeXt} (c).

\textbf{Curvature Supervised Attention Module.} The supervised attention module is shown capable to enrich the features in the multi-stage progressive image restoration \citep{Zamir2021MPRNet}. Proceeding from the same purpose, we propose a Curvature Supervised Attention Module (CSAM) in the denoising subnetwork. Fig. \ref{SAM} illustrates the schematic diagram of CSAM, where the curvature map is integrated with the attention maps to enhance the useful features. We use a $1\times1$ convolution to the curvature map followed by the sigmoid activation, which are then integrated with the feature maps obtained by restored images to produce the attention maps through $1\times1$ convolution. By introducing CSAM, the useful features can propagate to the decoder and the less informative features are suppressed by the attention masks.

\subsection{Implementation Details}
For a fair comparison with DRUNet \citep{zhang2021plug}, we use the same training dataset, which consists of 4744 images of Waterloo Exploration Database\footnote{https://ece.uwaterloo.ca/~k29ma/exploration/} \citep{2017Waterloo}, 900 images from DIV2K dataset\footnote{https://data.vision.ee.ethz.ch/cvl/DIV2K/} \citep{2017ntire} and 2650 images from Flick2K dataset\footnote{https://drive.google.com/drive/folders/1AAI2a2BmafbeVExLH-l0aZgvPgJCk5Xm} \citep{lim2017enhanced}. Note that 400 Berkeley segmentation dataset (BSD) is not included because the images of size 180$\times$180 are too small to fit the large training patch size. In order to use a single model to handle noisy image of different noise levels, the noise level of the training data is randomly selected from  the range $[0,50]$.

For the noise estimation subnetwork, we set the initial learning rate as 2e-4 and decrease it by half every 60000 iterations until reaching 1.25e-5. The patch size and batch size are fixed as 256$\times$256 and 64, respectively. The Adam optimizer is used to optimize the noise estimation subnetwork end-to-end with the following loss function
\begin{equation}
L_{N}=\vert\vert X_{NL}-\hat{X}_{NL}\vert\vert_1 + \vert\vert X_{ND}-\hat{X}_{ND}\vert\vert_1,
\end{equation}
where $X_{NL}$ and $\hat{X}_{NL}$ denotes the estimated noise level map and the corresponding ground truth, $X_{ND}$ and $\hat{X}_{ND}$ represents the predicted noise distribution map and its ground truth, respectively.

For the denoising subnetwork, the learning rate starts from 2e-4 and decays by a factor of 0.5 every 100000 iterations and finally ends once it is smaller than 2.5e-5. The denoisier network is trained on 224$\times$224 patches with a batch size of 24. The parameters are also optimized by Adam optimizer by the following loss function
\begin{equation}
L_{I}=\vert\vert X-\hat{X}\vert\vert_1 + L_{sa},
\end{equation}
where $X$ and $\hat{X}$ are the denoised image the corresponding ground-truth image, and $L_{sa}$ is supervised attention loss defined as
\begin{equation}
L_{sa}=\vert\vert X_S-\hat{X}\vert\vert_1,
\end{equation}
with $X_S$ being the restored image in CSAM. The number of channels in each layer from the first scale to the fourth scale are set as 96, 192, 384 and 768 for both noise estimation and denoising subnetworks. It takes about twenty hours and three days twelve hours to train the noise estimation subnetwork and denoising subnetwork on four NVIDIA Geforce RTX 3090 GPUs, respectively.

\section{Experiments}
\label{experiments}
In this section, we evaluate the performance of our CurvPnP on three representative image restoration tasks including gray/color image denoising, deblurring and super-resolution. All traditional restoration methods (including BM3D \citep{Dabov2007Image} and TFOV \citep{yao2020total}) are run on an Intel Core i7 CPU at 3.60GHz, while the learning based methods (including NN+BM3D \citep{Zheng2021an}, IRCNN \citep{zhang2017learning}, CBDNet \citep{guo2019toward}, FDnCNN \citep{zhang2017beyond}, DRUNet \citep{zhang2021plug}, C-UNet, DMPHN \citep{Zhang_2019_CVPR}, MPRNet \citep{Zamir2021MPRNet}, DWDN \citep{dong2020deep}, ZSSR \citep{shocher2018zero}, SRFBN \citep{li2019feedback}, DPIR \citep{zhang2021plug}, DPIR+ and CurvPnP) are implemented on a NVIDIA Geforce RTX 3090 GPU.

\subsection{Datasets}
We perform evaluation on various testing datasets, including Set9\footnote{https://github.com/zhengdharia/Unsupervised$\_$denoising/tree/master/data} \citep{Dmitry2018deep}, Kodak24\footnote{https://github.com/cszn/FFDNet/tree/master/testsets} \citep{Kodak24}, Urban100\footnote{https://github.com/502408764/Urban100} \citep{Huang2015single}, PIPAL\footnote{https://drive.google.com/drive/folders/1G4fLeDcq6uQQmYdkjYUHhzyel\\4Pz81p-} \citep{2020PIPAL}, CC\footnote{https://github.com/csjunxu/MCWNNM-ICCV2017/tree/master/Real$\_$ccno\\ise$\_$denoised$\_$part} \citep{Nam2016holistic} and PolyU\footnote{https://github.com/csjunxu/PolyU-Real-World-Noisy-Images-Dataset/tree/master/CroppedImages} \citep{Xu2018Real} dataset. The Set9 dataset contains 9 classic color images with size of 512$\times$512 or 768$\times$512; the Kodak24 dataset includes 24 color images with size of 500$\times$500; the Urban100 dataset has 100 images of urban scenes of different sizes; the PIPAL dataset contains 200 images of size 288$\times$288; both CC and PolyU datasets are with 15 and 100 images of size 512$\times$ 512.

\begin{figure}[t]
\centering
\hspace{-4.1cm}
\begin{minipage}[htbp]{0.48\linewidth}
\centering
\includegraphics[width=7.6cm]{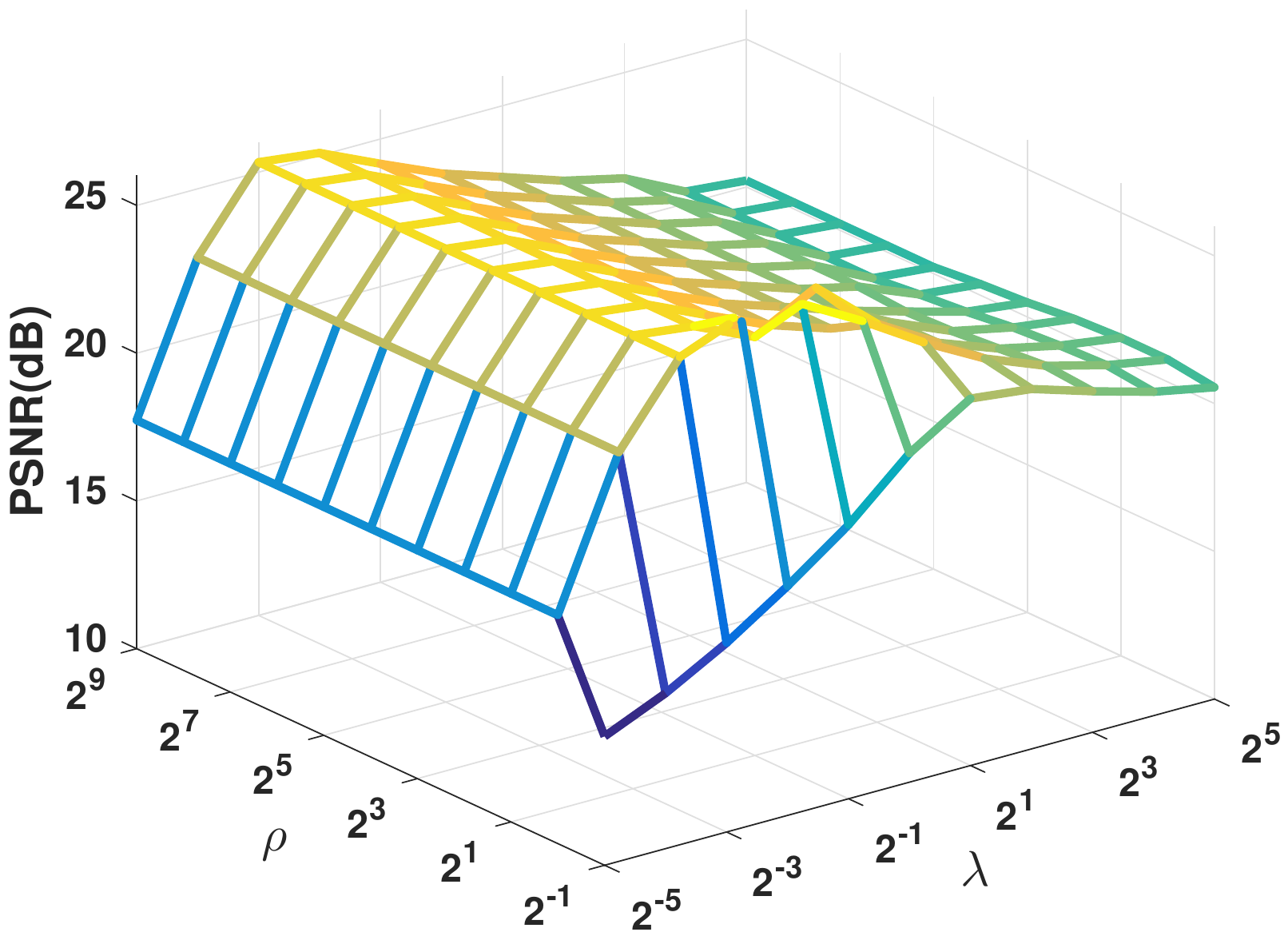}
\end{minipage}
\caption{The PSNR evolutions of $Church$ in Fig. \ref{figdeblur} obtained by different combinations of the parameters $\lambda$ and $\rho$.}
\label{surfdeblur}
\end{figure}

\begin{figure*}[h]
\centering
\begin{minipage}[htbp]{0.137\linewidth}
\centering
\includegraphics[width=2.27cm]{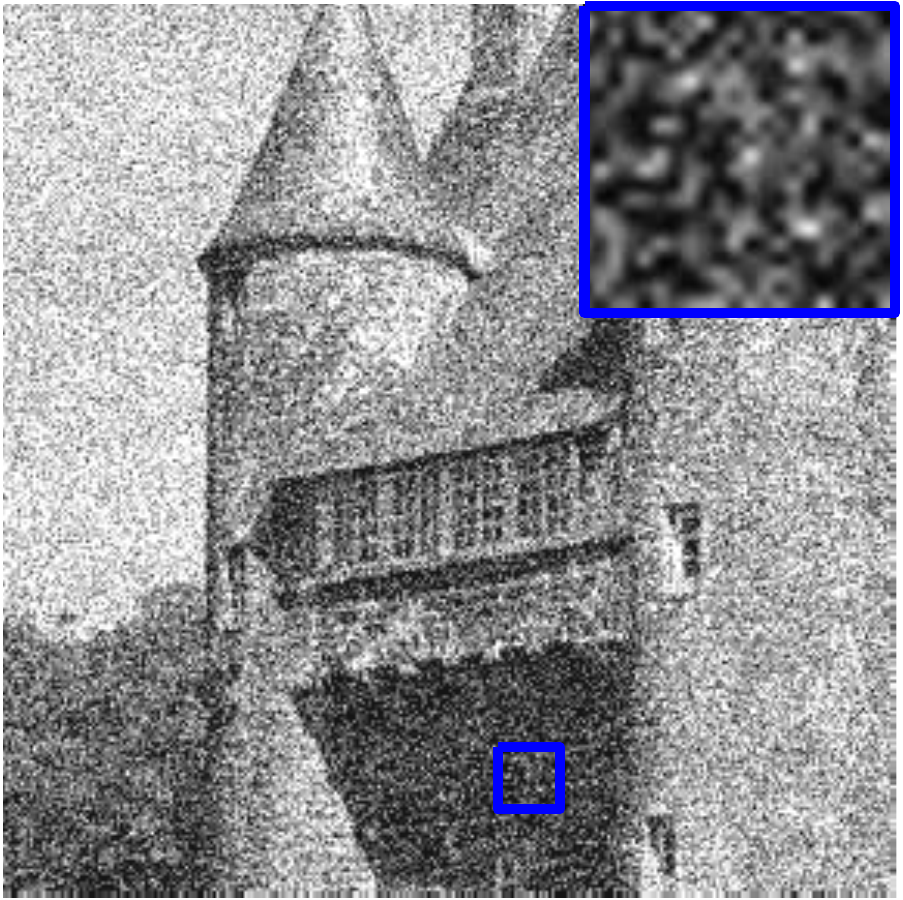}
{ \footnotesize{Noisy Image} }
\end{minipage}
\begin{minipage}[htbp]{0.137\linewidth}
\centering
\includegraphics[width=2.27cm]{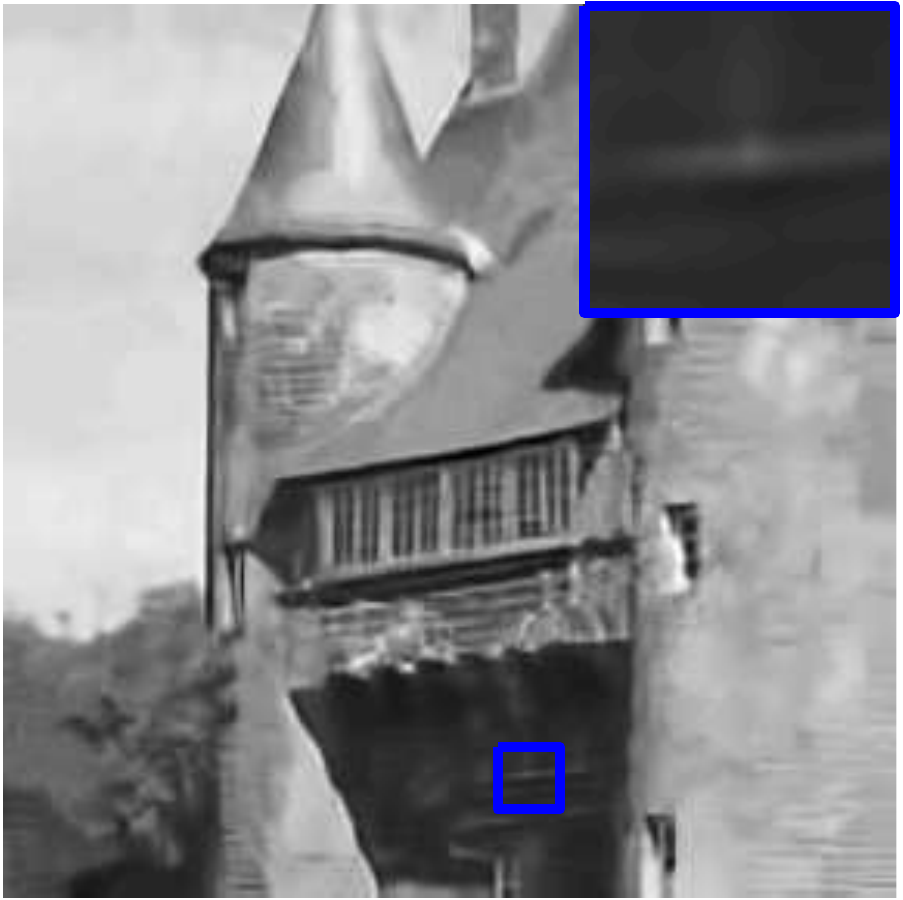}
{ \footnotesize{BM3D (24.28dB)} }
\end{minipage}
\begin{minipage}[htbp]{0.137\linewidth}
\centering
\includegraphics[width=2.27cm]{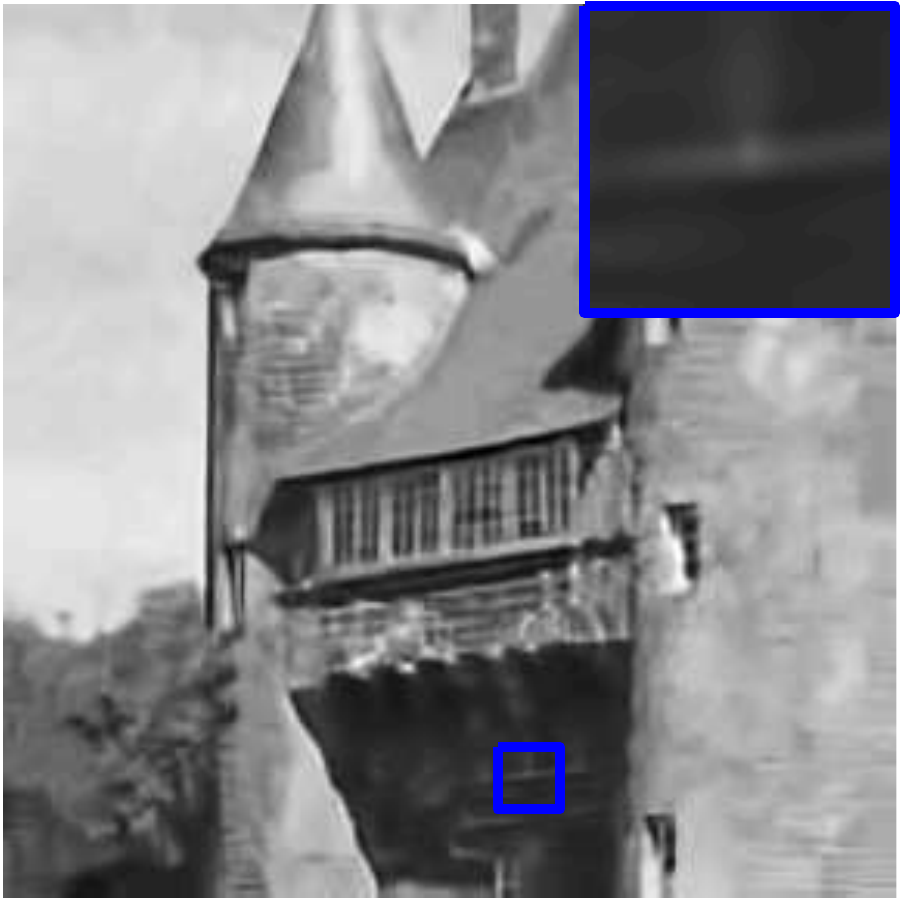}
{ \tiny{NN+BM3D} \footnotesize{(24.33dB)} }
\end{minipage}
\begin{minipage}[htbp]{0.137\linewidth}
\centering
\includegraphics[width=2.27cm]{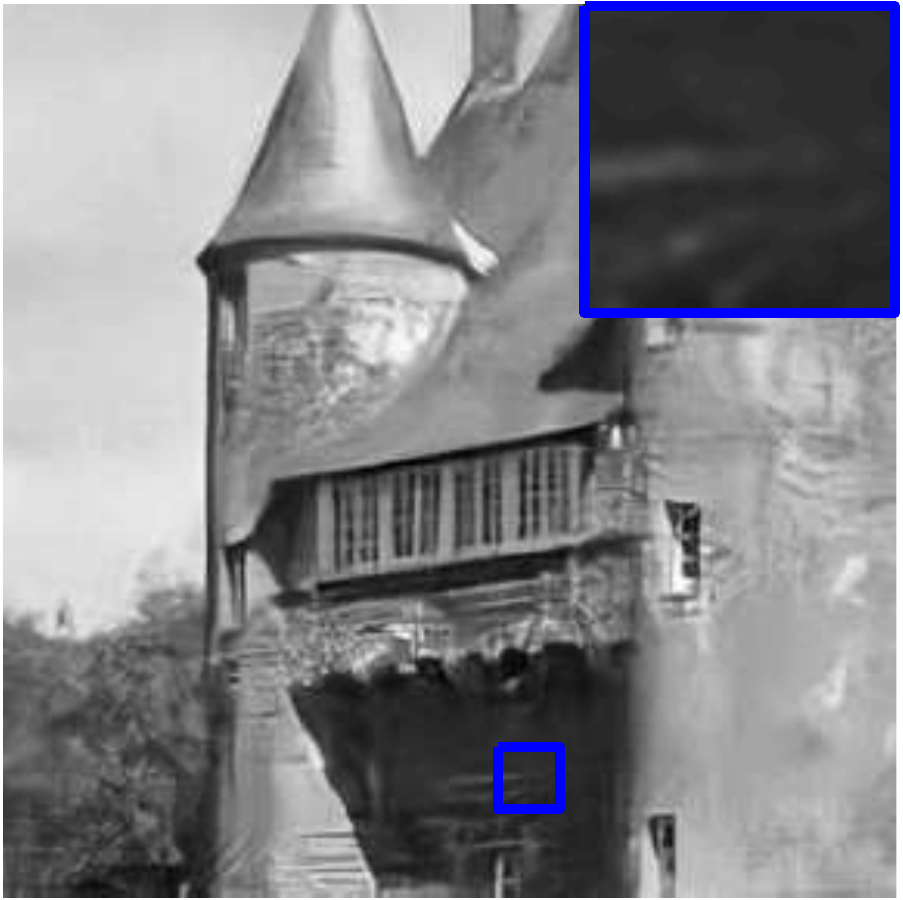}
{ \footnotesize{IRCNN (24.88dB)} }
\end{minipage}
\begin{minipage}[htbp]{0.137\linewidth}
\centering
\includegraphics[width=2.27cm]{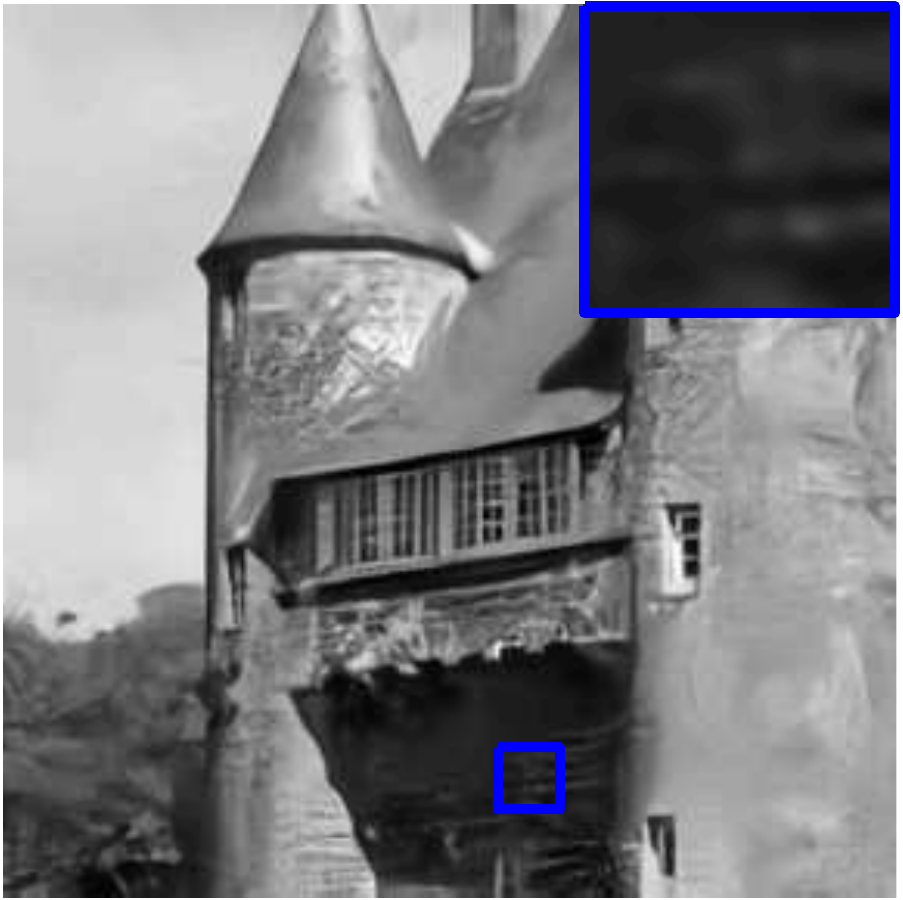}
{ \footnotesize{FDnCNN (24.92dB)} }
\end{minipage}
\begin{minipage}[htbp]{0.137\linewidth}
\centering
\includegraphics[width=2.27cm]{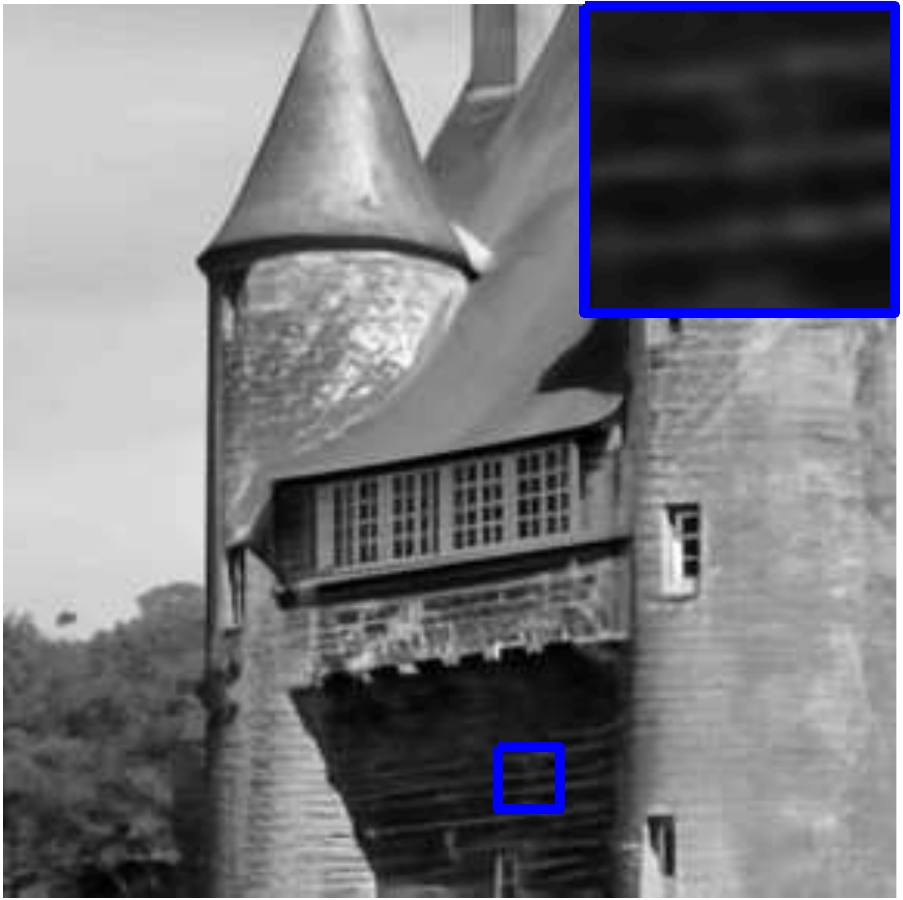}
{ \footnotesize{DRUNet (25.27dB)} }
\end{minipage}
\begin{minipage}[htbp]{0.137\linewidth}
\centering
\includegraphics[width=2.27cm]{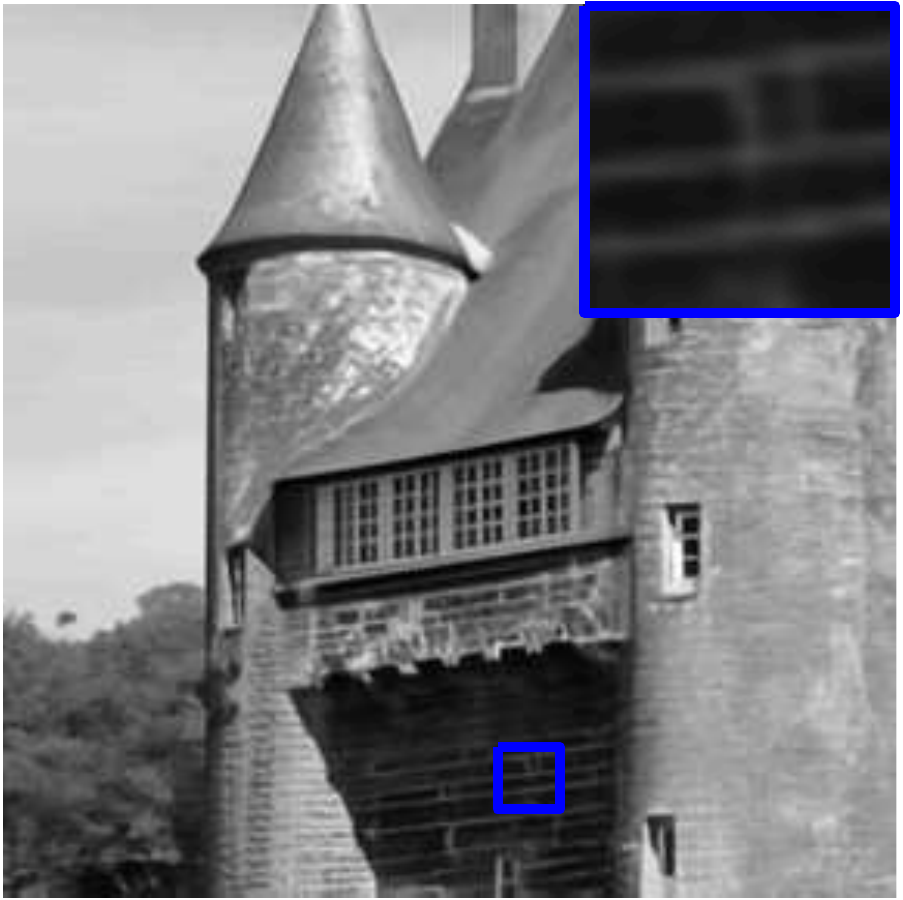}
{ \footnotesize{C-UNet (25.33dB)} }
\end{minipage}\\
\vspace{0.2cm}
\begin{minipage}[htbp]{0.137\linewidth}
\centering
\includegraphics[width=2.27cm]{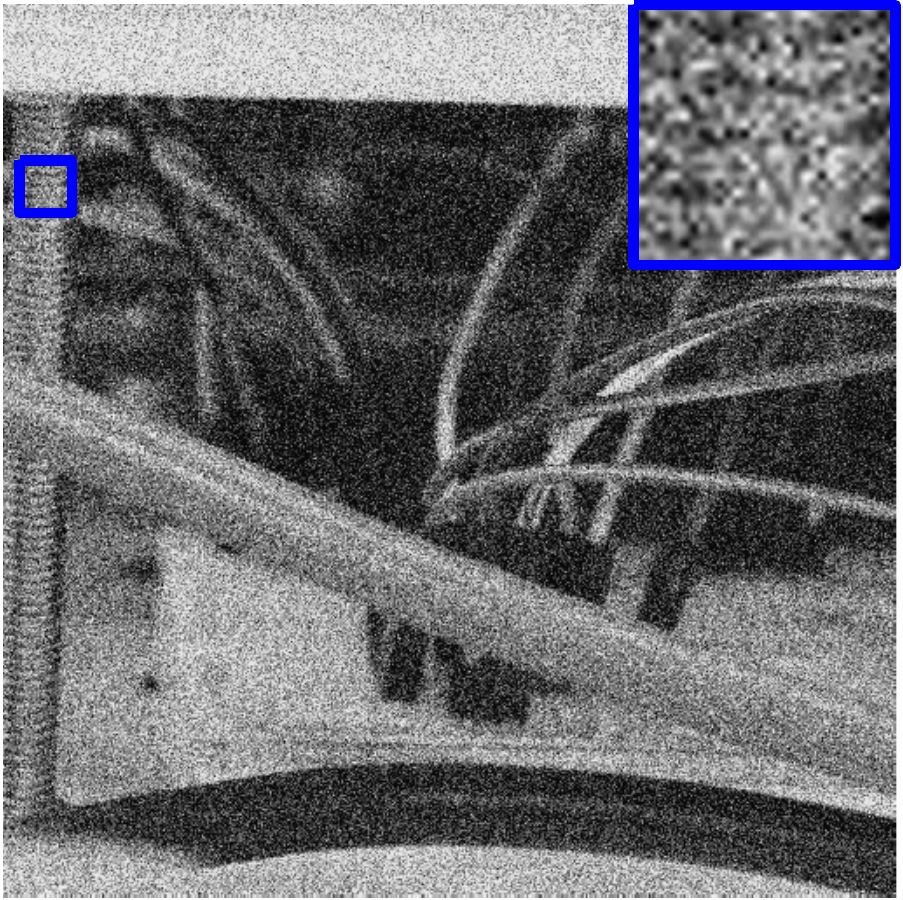}
{ \footnotesize{Noisy Image} }
\end{minipage}
\begin{minipage}[htbp]{0.137\linewidth}
\centering
\includegraphics[width=2.27cm]{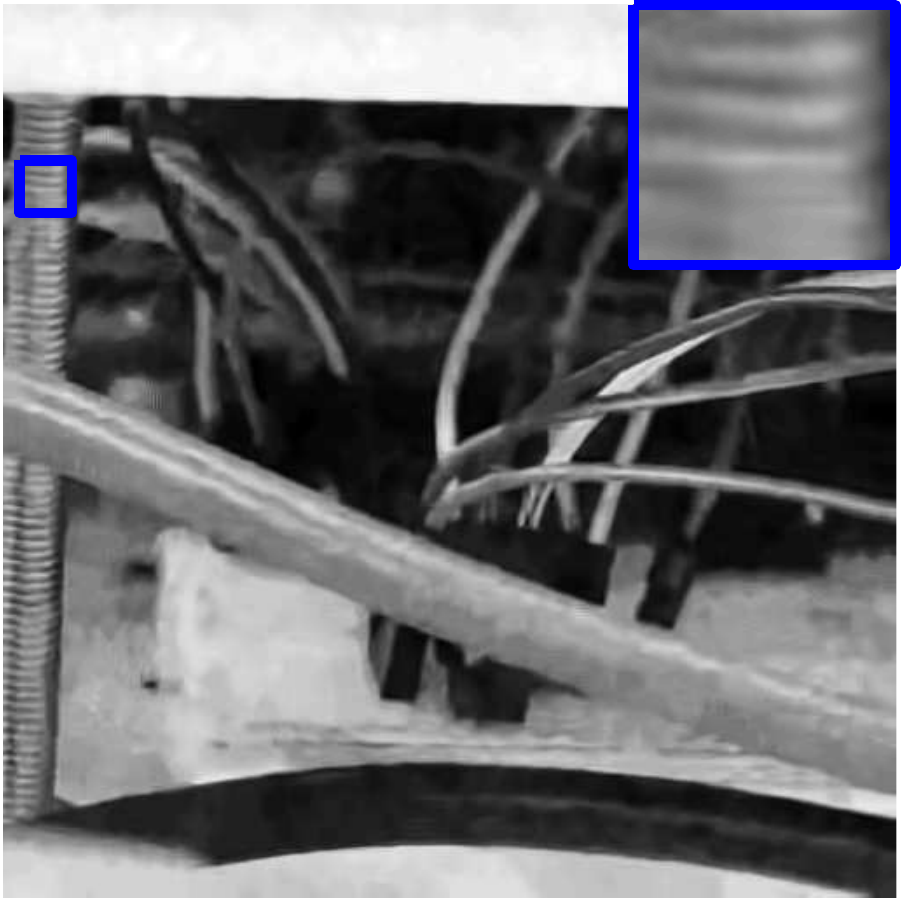}
{\footnotesize{BM3D (28.49dB)} }
\end{minipage}
\begin{minipage}[htbp]{0.137\linewidth}
\centering
\includegraphics[width=2.27cm]{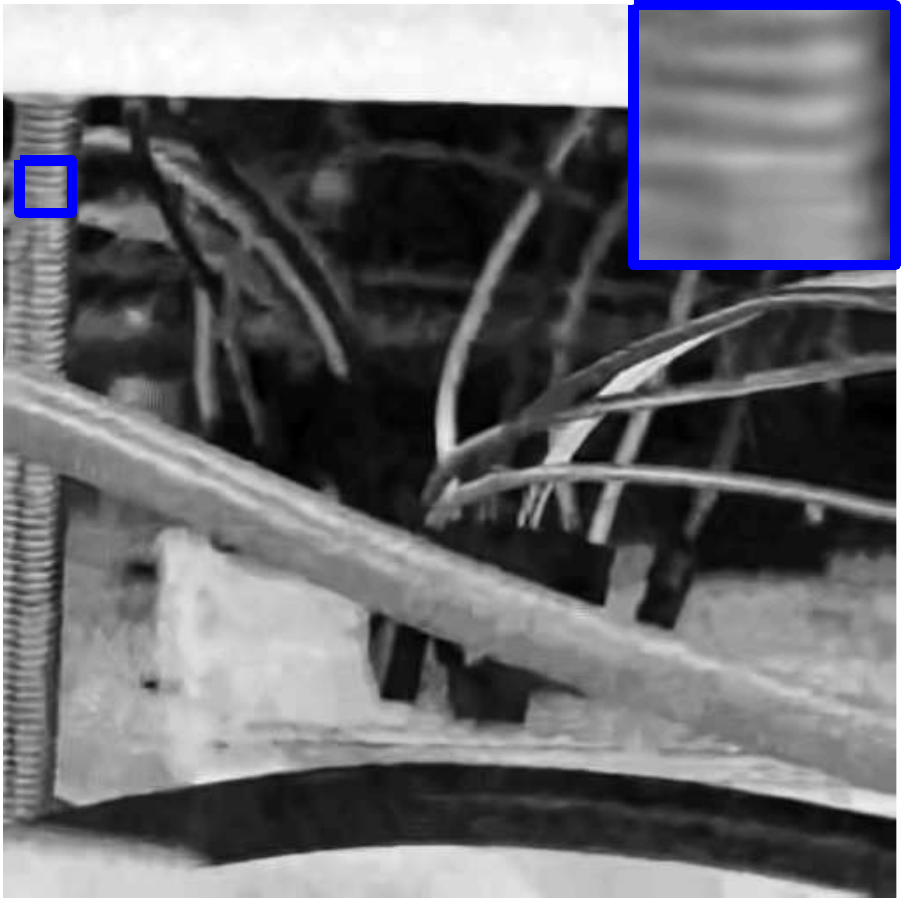}
{ \tiny{NN+BM3D} \footnotesize{(29.16dB)} }
\end{minipage}
\begin{minipage}[htbp]{0.137\linewidth}
\centering
\includegraphics[width=2.27cm]{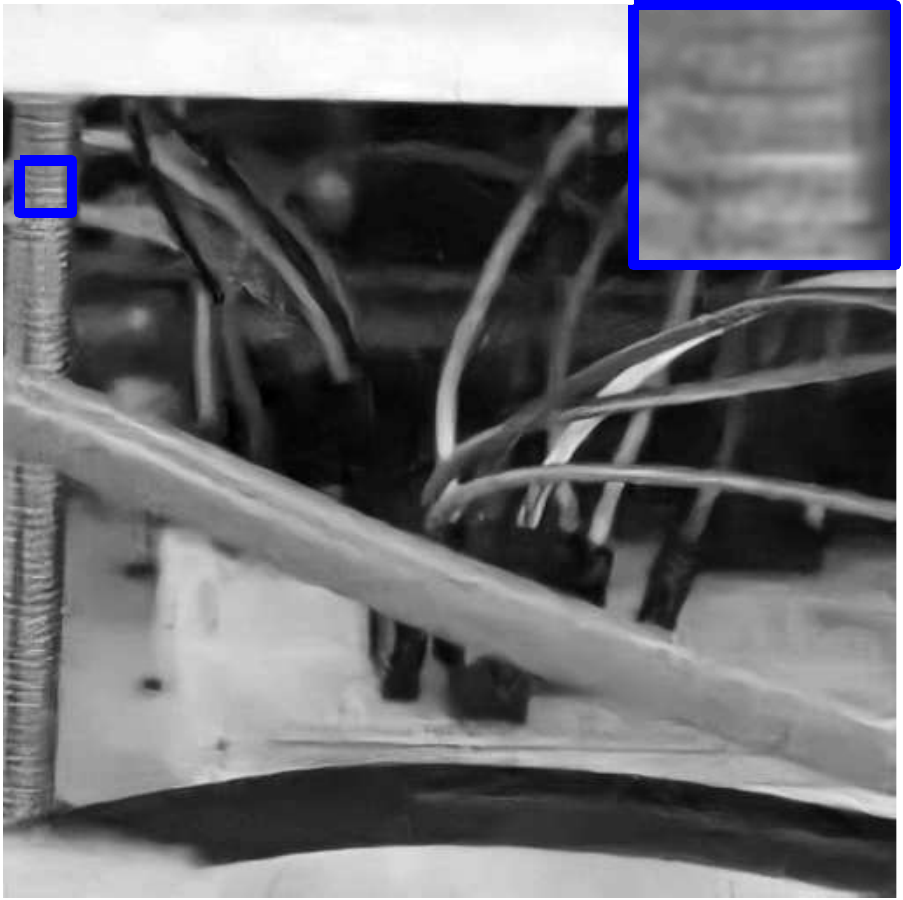}
{ \footnotesize{IRCNN (30.54dB)} }
\end{minipage}
\begin{minipage}[htbp]{0.137\linewidth}
\centering
\includegraphics[width=2.27cm]{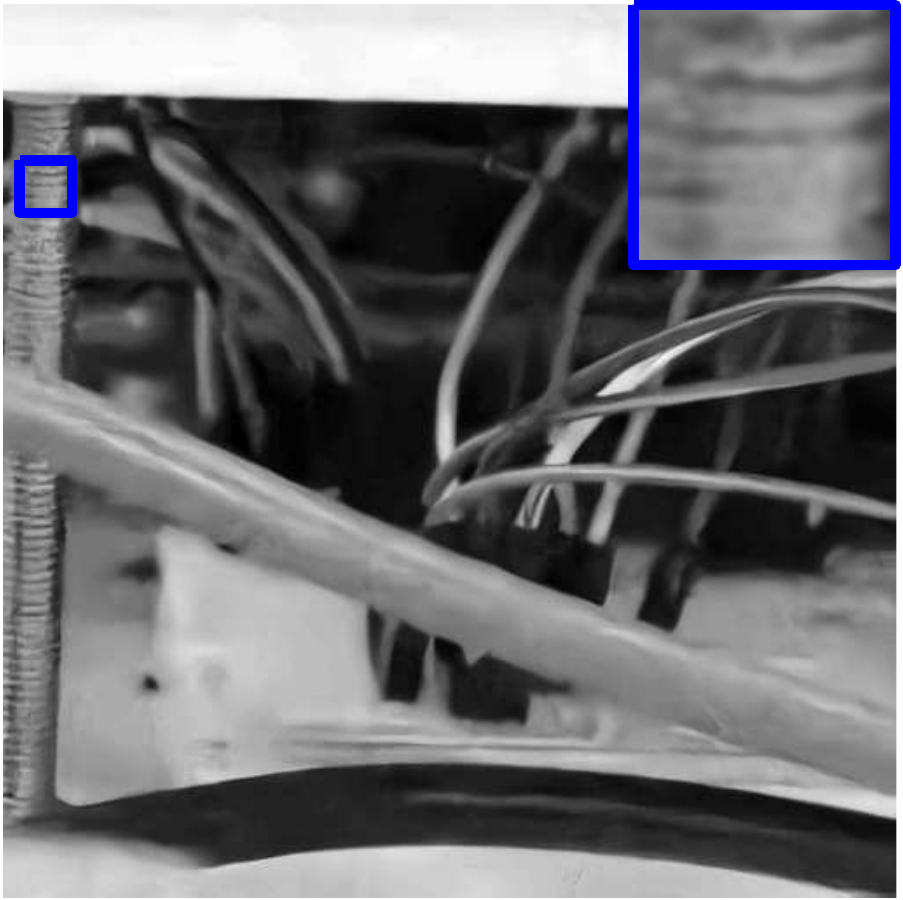}
{ \footnotesize{FDnCNN (30.92dB)} }
\end{minipage}
\begin{minipage}[htbp]{0.137\linewidth}
\centering
\includegraphics[width=2.27cm]{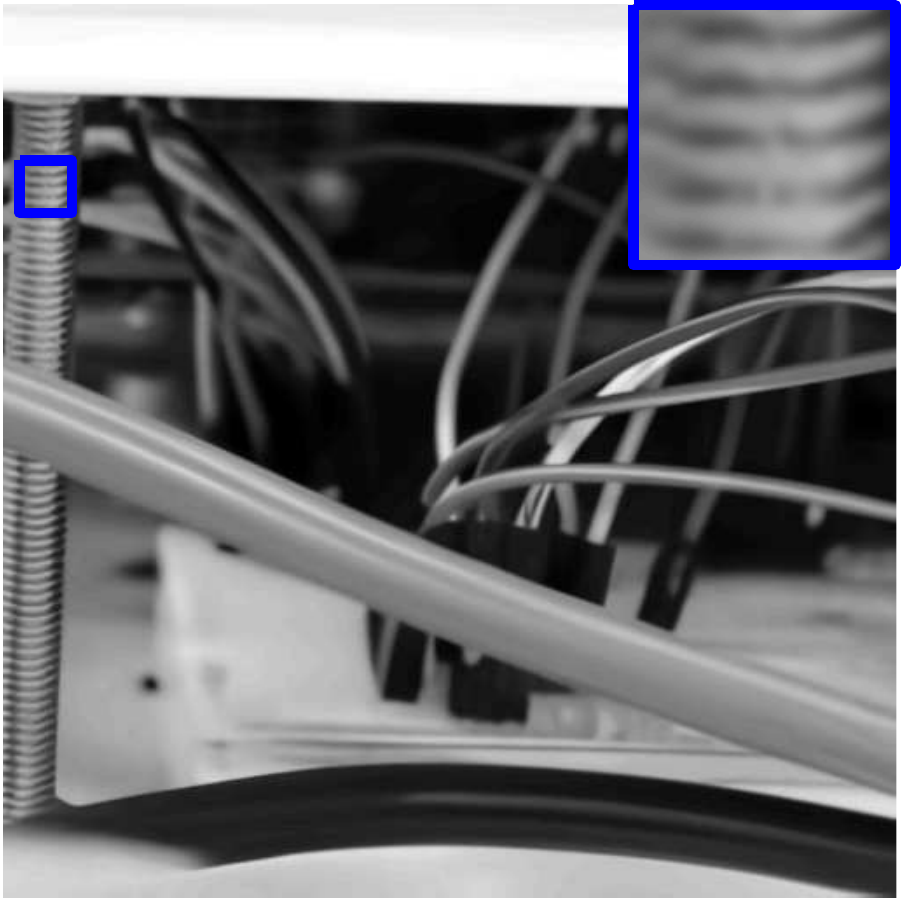}
{ \footnotesize{DRUNet (32.89dB)} }
\end{minipage}
\begin{minipage}[htbp]{0.137\linewidth}
\centering
\includegraphics[width=2.27cm]{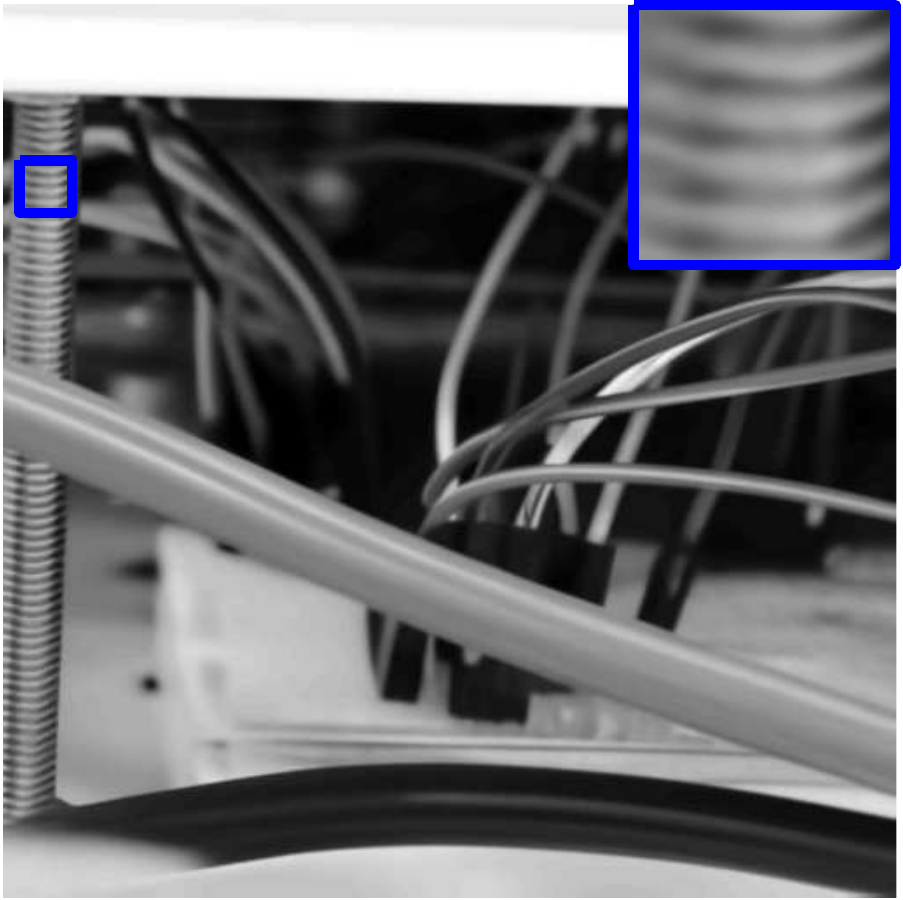}
{ \footnotesize{C-UNet (33.17dB)} }
\end{minipage}
\caption{Grayscale image denoising results of different methods on the image $``A0025''$ from PIPAL dataset and image $``Canon5D2\_5\_160\_6400\_circuit\_11''$ from PolyU dataset with noise level of 50.}
\label{figgraydenoise}
\end{figure*}

\begin{table*}[h]
\begin{center}
\begin{minipage}{\textwidth}
\caption{Average PSNR(dB)/SSIM results of different methods with noise levels 20, 35 and 50 for grayscale image denoising on Set9, Urban100, PIPAL, CC and PolyU datasets. The best and second best scores are \textbf{highlighted} and \underline{underline}, respectively.}\label{tabdenoiseg}
\small
\begin{tabular*}{\textwidth}
{@{\extracolsep{\fill}}llllllll@{\extracolsep{\fill}}}
\toprule
Datasets & $\sigma_s$  & BM3D  & NN+BM3D & IRCNN & FDnCNN & DRUNet & C-UNet\\
\midrule
 Set9   & 20 & 31.73/0.8498   & 31.70/0.8501  & 32.11/0.8550  & 32.20/0.8601 & \underline{32.60}/\underline{0.8703}  & \textbf{32.64}/\textbf{0.8710}  \\
   &  35  & 29.32/0.7851  & 29.08/0.7863   & 29.84/0.8033  & 29.90/0.8059 & \underline{30.41}/\underline{0.8215}  & \textbf{30.46}/\textbf{0.8221}  \\
   &  50  & 27.59/0.7347  & 27.65/0.7382   & 28.36/0.7620  & 28.45/0.7660 & \underline{29.08}/\underline{0.7877}  & \textbf{29.15}/\textbf{0.7887}  \\

 \cline{2-8}

 Urban100  & 20 & 30.54/0.8878   & 30.48/0.8872  & 30.95/0.9014  & 31.25/0.9080 & \underline{32.12}/\underline{0.9222}  & \textbf{32.18}/\textbf{0.9223}  \\
    &  35  & 27.21/0.8151  &  27.11/0.8153  & 28.11/0.8453  & 28.38/0.8519 & \underline{29.58}/\underline{0.8825}  & \textbf{29.69}/\textbf{0.8841}  \\
   &  50  & 24.83/0.7465  &  24.79/0.7485  & 26.23/0.7919  & 26.52/0.8011 & \underline{27.96}/\underline{0.8483}  & \textbf{28.12}/\textbf{0.8517}  \\

 \cline{2-8}

 PIPAL  & 20 & 29.41/0.8811   & 29.37/0.8802  & 29.92/0.8859  & 30.04/0.8900 & \underline{30.42}/\underline{0.8985}  & \textbf{30.49}/\textbf{0.8998}  \\
    &  35  & 26.33/0.7923  & 26.35/0.7989   & 27.08/0.8133  & 27.20/0.8171 &  \underline{27.66}/\underline{0.8340}  & \textbf{27.74}/\textbf{0.8369}  \\
   &  50  & 24.29/0.7138  &  24.43/0.7251  & 25.37/0.7507  & 25.49/0.7557 & \underline{26.00}/\underline{0.7806}  & \textbf{26.10}/\textbf{0.7851}  \\

 \cline{2-8}

 CC  & 20 & 36.74/0.9399   & 36.69/0.9407  & 37.30/0.9477  & 37.58/0.9503 & \underline{38.57}/\underline{0.9567}  & \textbf{38.73}/\textbf{0.9575}  \\
    &  35  & 33.08/0.8981  & 32.80/0.8985   & 34.22/0.9159  & 34.57/0.9216 & \underline{35.89}/\underline{0.9345}  & \textbf{36.09}/\textbf{0.9358}  \\
   &  50  & 30.22/0.8585  & 30.16/0.8581   & 32.17/0.8832  & 32.61/0.8964 & \underline{34.10}/\underline{0.9159}  & \textbf{34.36}/\textbf{0.9180}  \\

 \cline{2-8}

 PolyU  & 20 & 37.83/0.9474   & 37.76/0.9479  & 38.81/0.9593  & 39.11/0.9614 & \underline{40.43}/\underline{0.9698}  & \textbf{40.56}/\textbf{0.9703}  \\
    &  35  & 33.94/0.9093  & 33.57/0.9088   & 35.94/0.9375  & 36.29/0.9422 & \underline{38.05}/\underline{0.9577}  & \textbf{38.26}/\textbf{0.9590}  \\
   &  50  & 30.84/0.8747  & 30.65/0.8742   & 33.87/0.9138  & 34.36/0.9253 & \underline{36.37}/\underline{0.9473}  & \textbf{36.62}/\textbf{0.9491}  \\
\botrule
\end{tabular*}
\end{minipage}
\end{center}
\end{table*}

\subsection{Parameter Selection}
There are two parameters in our model, i.e., the regularization parameter $\lambda$ and noise level control parameter $\rho$. Our method is not sensitive to the choices of $\lambda$, which is fixed as $\lambda=0.37$ for image deblurring and super-resolution tasks. Because the noise distribution changes with different image restoration tasks, the value of $\rho$ varies for different degradation operations. We choose $\rho=1.2$ for image deblurring and $\rho=5$, $25$ and $50$ for SISR with scale factor $2$, $3$ and $4$, respectively. For both image deblurring and super-resolution tasks, the maximum iteration number $K$ is set to 15. Fig. \ref{surfdeblur} displays the PSNR results on image $Church$ from  PIPAL dataset in Fig. \ref{figdeblur} with different combinations of $\lambda$ and $\rho$, where $\lambda$ and $\rho$ are chosen from $(\lambda, \rho)\in\{2^{-5}, 2^{-4}, \ldots, 2^5\}\times\{2^{-1}, 2^0, \ldots, 2^9\}$. As shown, PSNR values are relatively stable such that there are reasonably large internals for parameters to generate good restoration results. Note that $\rho$ should be chosen larger than $1$ to obtain high quality restoration results.

\subsection{Image Denoising}

\begin{table*}[h]
\begin{center}
\begin{minipage}{\textwidth}
\caption{Average PSNR(dB)/SSIM results of different methods with noise levels 20, 35 and 50 for color image denoising on Kodak24, Urban100, PIPAL, CC and PolyU datasets. The best and second best scores are \textbf{highlighted} and \underline{underline}, respectively.}\label{tabdenoisec}%
\small
\begin{tabular*}{\textwidth}
{p{1.0cm}p{0.4cm}p{1.55cm}p{1.55cm}p{1.55cm}p{1.55cm}p{1.55cm}p{1.55cm}p{1.55cm}p{1.55cm}}
\toprule
Datasets & $\sigma_s$  & BM3D & NN+BM3D  & IRCNN & CBDNet & FDnCNN & DRUNet & C-UNet\\
\midrule
 Kodak24  & 20 & 30.53/0.8912  & 30.51/0.8913 & 33.08/0.8956  & 33.10/0.8965  & 33.20/0.8995  &  \underline{33.83}/\underline{0.9094} & \textbf{33.89}/\textbf{0.9102} \\
    &  35  & 29.46/0.8190  & 29.49/0.8232   & 30.43/0.8386  & 30.45/0.8371  &  30.57/0.8423  & \underline{31.28}/\underline{0.8599}  & \textbf{31.36}/\textbf{0.8614}  \\
   &  50  & 27.28/0.7579  & 27.47/0.7652   & 28.81/0.7932  & 28.85/0.7894  &  28.97/0.7965  &  \underline{29.74}/\underline{0.8202}  & \textbf{29.82}/\textbf{0.8224}  \\

 \cline{2-9}

 Urban100  & 20 & 31.97/0.9081 & 31.92/0.9075 & 32.30/0.9222  & 32.28/0.9226 & 32.57/0.9275  &  \underline{33.56}/\underline{0.9385} & \textbf{33.63}/\textbf{0.9392} \\
    &  35  & 28.55/0.8541  & 28.52/0.8519   & 29.49/0.8788  & 29.59/0.8784  & 29.87/0.8862   &  \underline{31.15}/\underline{0.9088}  & \textbf{31.27}/\textbf{0.9103}  \\
   &  50  & 25.96/0.8008  &  26.09/0.8003  & 27.70/0.8393  & 27.85/0.8381  &  28.10/0.8482  &  \underline{29.61}/\underline{0.8832}  & \textbf{29.76}/\textbf{0.8857}  \\

 \cline{2-9}

 PIPAL  & 20 & 31.17/0.9149  & 31.18/0.9151  & 31.72/0.9211  &  31.73/0.9213 & 31.89/0.9241  &  \underline{32.48}/\underline{0.9323} & \textbf{32.56}/\textbf{0.9332} \\
    &  35  & 27.85/0.8479  &  27.98/0.8547  & 28.83/0.8673  & 28.89/0.8670  &  29.02/0.8720  &  \underline{29.69}/\underline{0.8871}  & \textbf{29.79}/\textbf{0.8890}  \\
   &  50  & 25.45/0.7807  &  25.75/0.7937  & 27.08/0.8199  & 27.13/0.8180  &  27.26/0.8252  &  \underline{27.98}/\underline{0.8470}  & \textbf{28.09}/\textbf{0.8499}  \\

 \cline{2-9}

  CC & 20 & 36.89/0.9276   & 36.87/0.9284   & 37.92/0.9552  & 38.27/0.9549  & 38.52/0.9571  &  \underline{39.92}/\underline{0.9663}  & \textbf{40.10}/\textbf{0.9672} \\
    &  35  & 32.55/0.8843  & 32.77/0.8882   & 35.10/0.9266  & 35.33/0.9267  &  35.58/0.9319  &  \underline{37.34}\underline{/0.9486}  & \textbf{37.57}/\textbf{0.9500}  \\
   &  50  & 29.24/0.8434  & 29.42/0.8465   & 33.13/0.8978  & 33.36/0.8993  &  33.65/0.9086  &  \underline{35.63}/\underline{0.9332}  & \textbf{35.91}/\textbf{0.9353}  \\

 \cline{2-9}

 PolyU  & 20 & 38.55/0.9492  & 38.59/0.9502   & 39.49/0.9638  & 39.82/0.9644 & 40.07/0.9663  &  \underline{41.21}/\underline{0.9724}  & \textbf{41.35}/\textbf{0.9729} \\
    &  35  & 34.17/0.9139  & 34.53/0.9179   & 36.98/0.9440  & 37.12/0.9457  &  37.45/0.9496  &  \underline{39.19/0.9624}  & \textbf{39.38}/\textbf{0.9634}  \\
   &  50  & 30.65/0.8811  & 31.19/0.8864   & 34.96/0.9249  & 35.22/0.9280  &  35.63/0.9347  &  \underline{37.78}/\underline{0.9539}  & \textbf{38.02}/\textbf{0.9556}  \\
\botrule
\end{tabular*}
\end{minipage}
\end{center}
\end{table*}

\begin{figure*}[h]
\centering
\begin{minipage}[htbp]{0.24\linewidth}
\centering
\includegraphics[width=3.9cm]{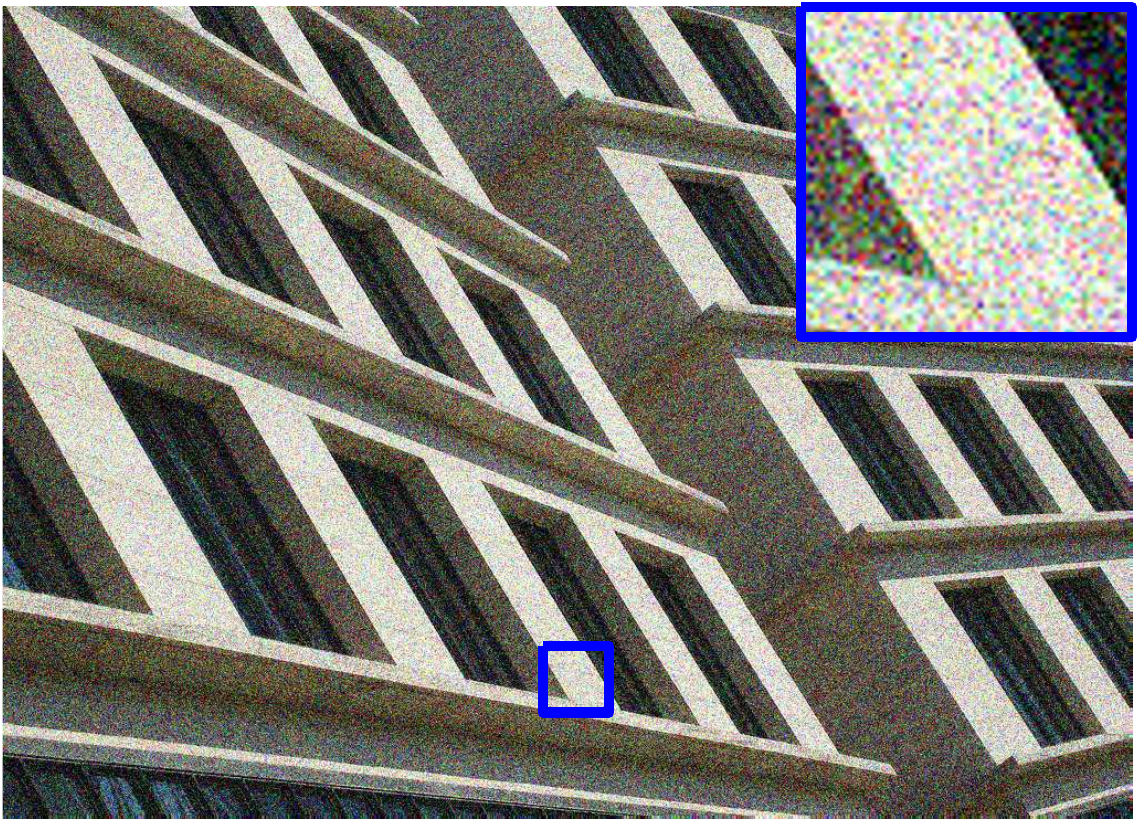}
{ \footnotesize{Noisy Image} }
\end{minipage}
\begin{minipage}[htbp]{0.24\linewidth}
\centering
\includegraphics[width=3.9cm]{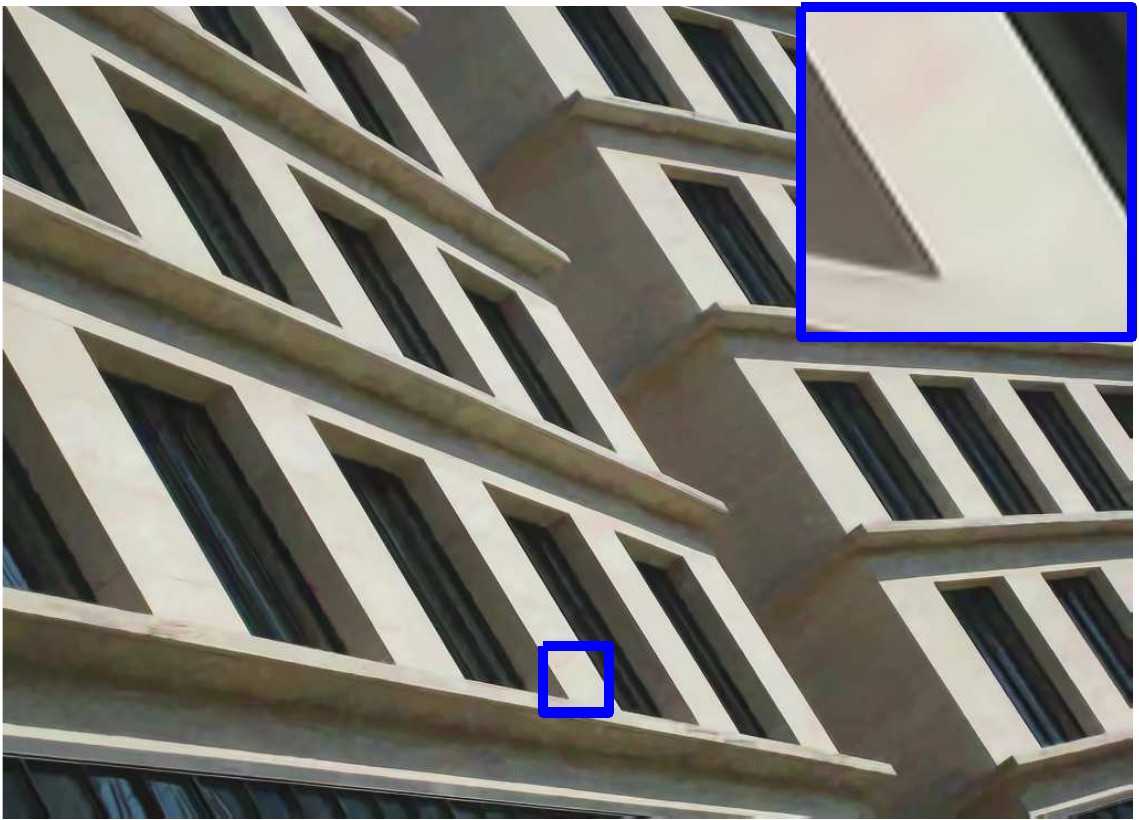}
{\footnotesize{BM3D (29.06dB)} }
\end{minipage}
\begin{minipage}[htbp]{0.24\linewidth}
\centering
\includegraphics[width=3.9cm]{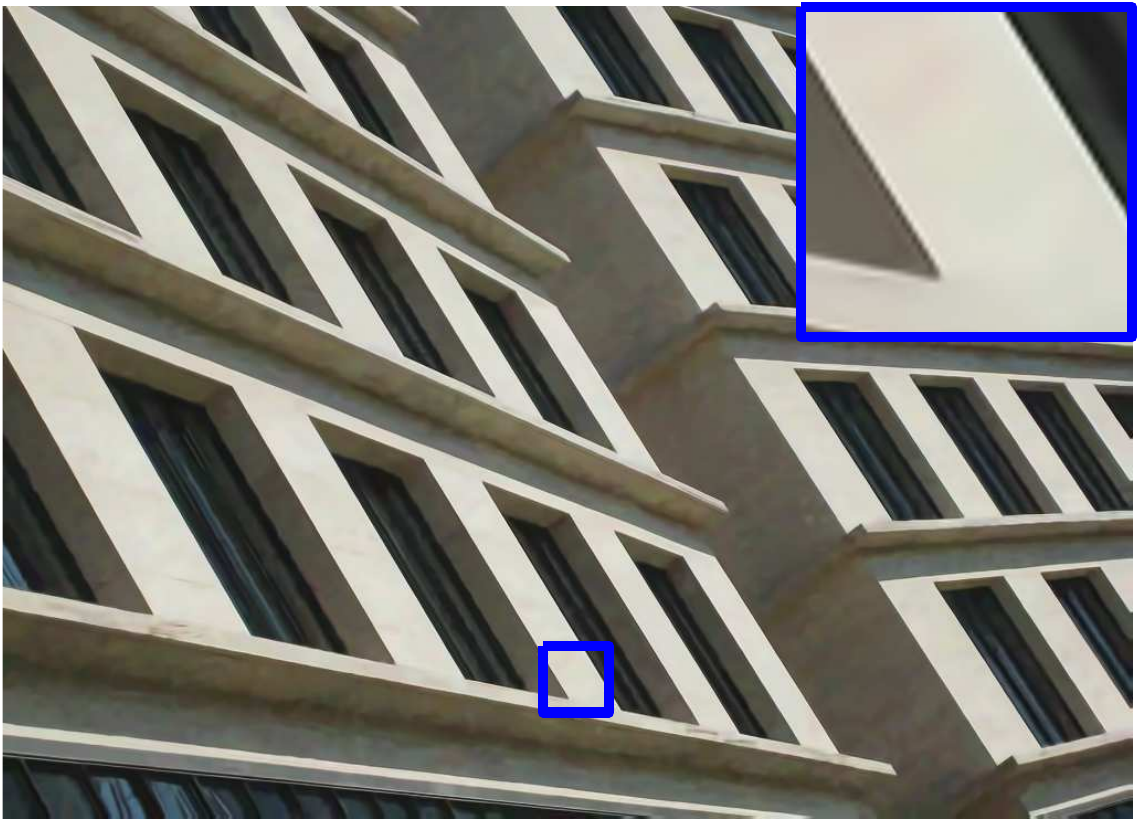}
{\footnotesize{NN+BM3D (29.30dB)} }
\end{minipage}
\begin{minipage}[htbp]{0.24\linewidth}
\centering
\includegraphics[width=3.9cm]{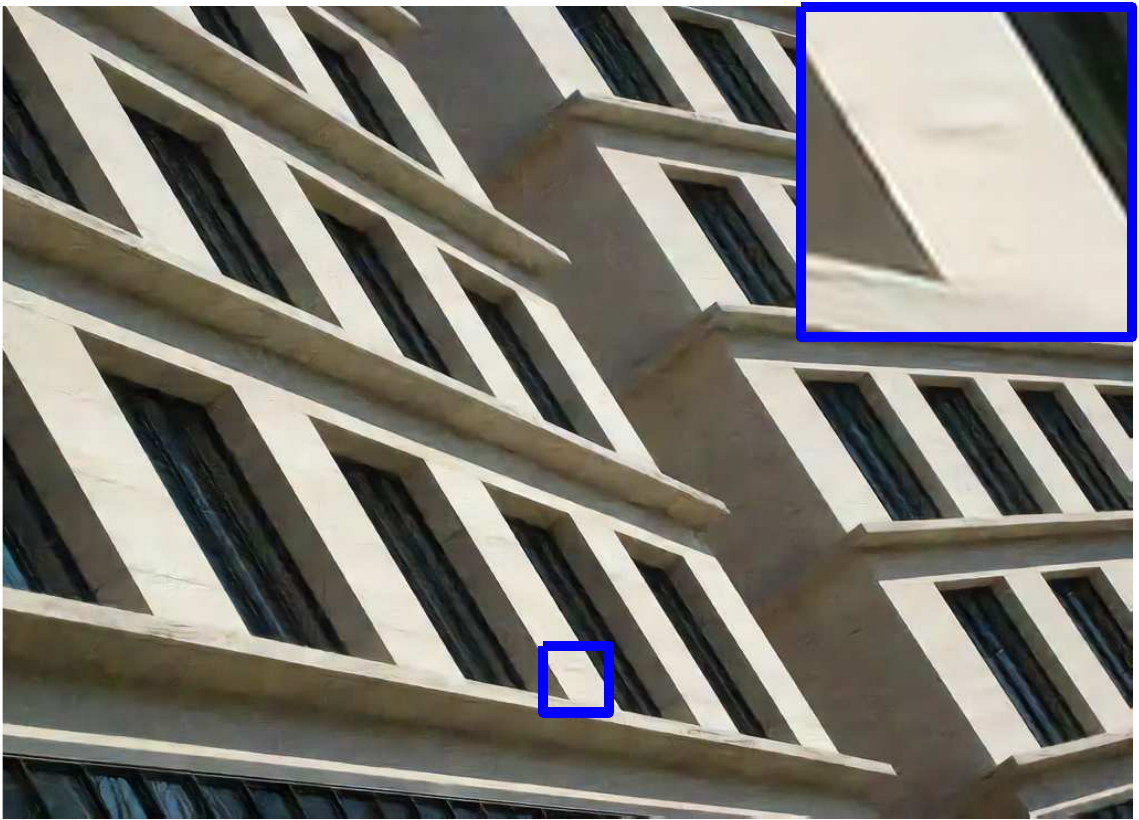}
{ \footnotesize{IRCNN (30.44dB)} }
\end{minipage}\\
\begin{minipage}[htbp]{0.24\linewidth}
\centering
\includegraphics[width=3.9cm]{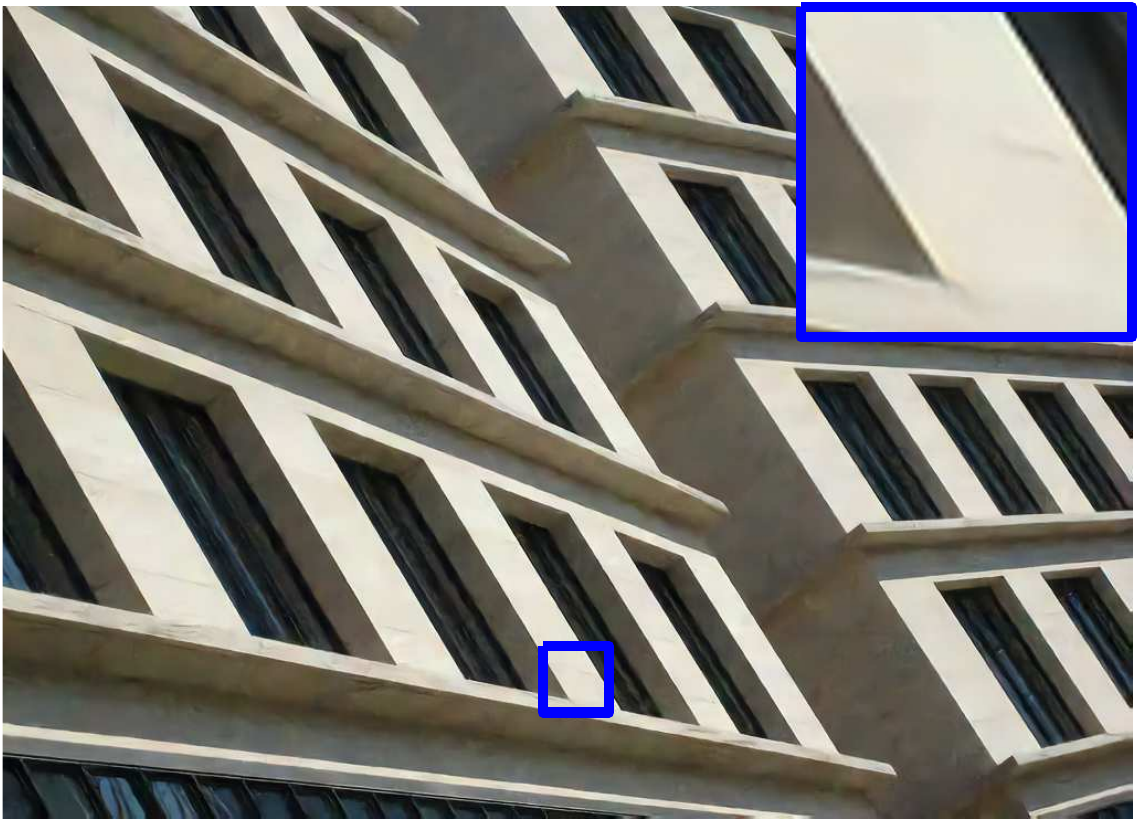}
{ \footnotesize{CBDNet (30.80dB)} }
\end{minipage}
\begin{minipage}[htbp]{0.24\linewidth}
\centering
\includegraphics[width=3.9cm]{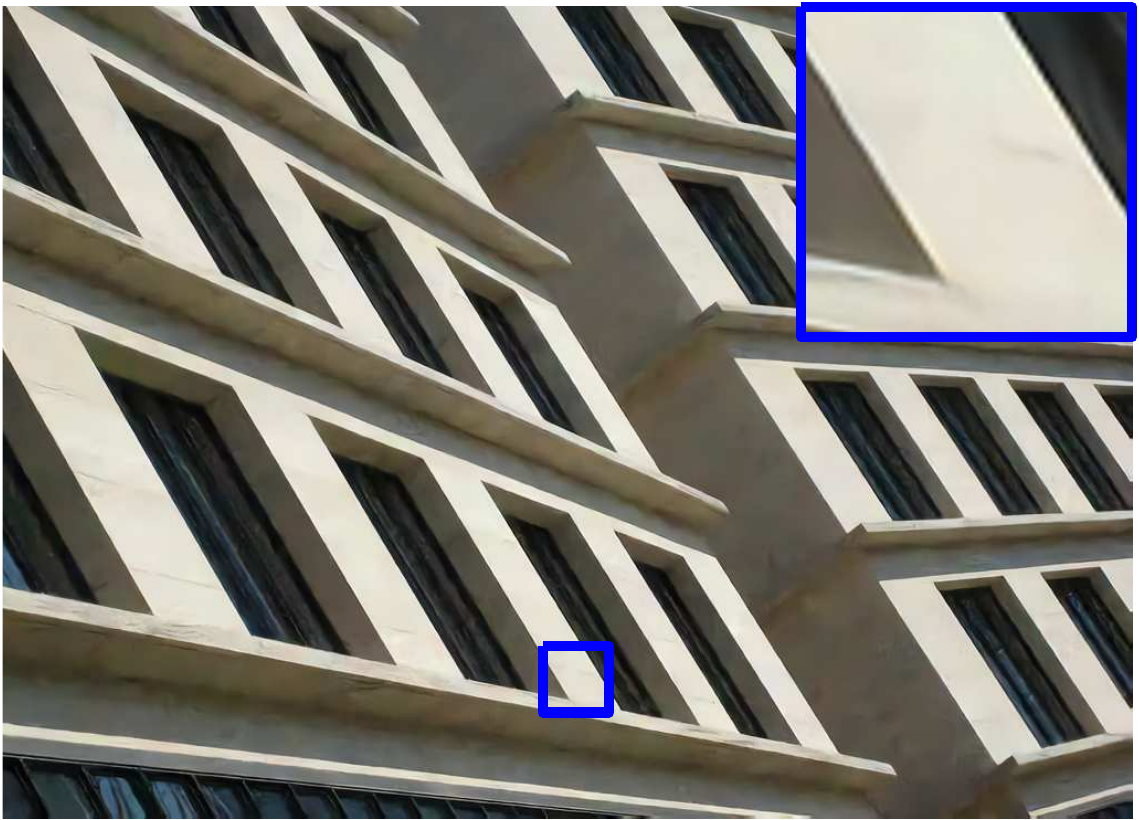}
{ \footnotesize{FDnCNN (30.99dB)} }
\end{minipage}
\begin{minipage}[htbp]{0.24\linewidth}
\centering
\includegraphics[width=3.9cm]{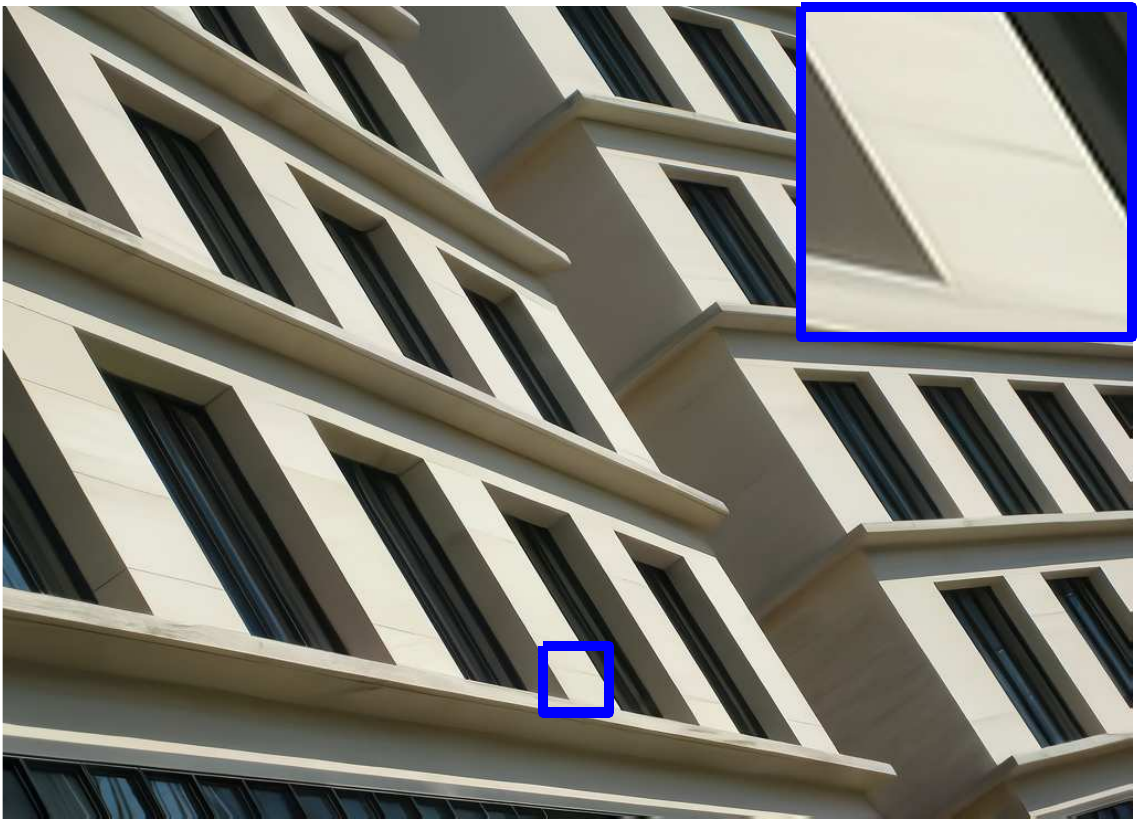}
{ \footnotesize{DRUNet (32.96dB)} }
\end{minipage}
\begin{minipage}[htbp]{0.24\linewidth}
\centering
\includegraphics[width=3.9cm]{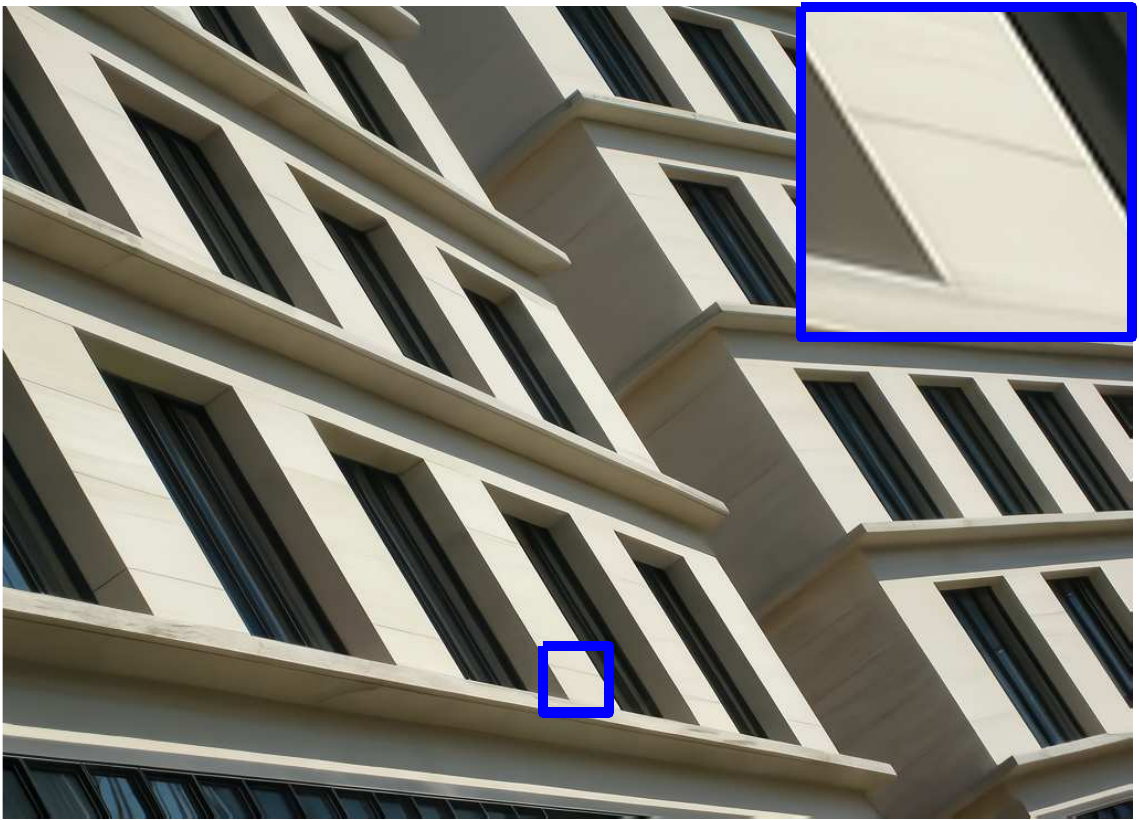}
{ \footnotesize{C-UNet (33.25dB)} }
\end{minipage}\\
\vspace{0.2cm}
\begin{minipage}[htbp]{0.24\linewidth}
\centering
\includegraphics[width=3.9cm]{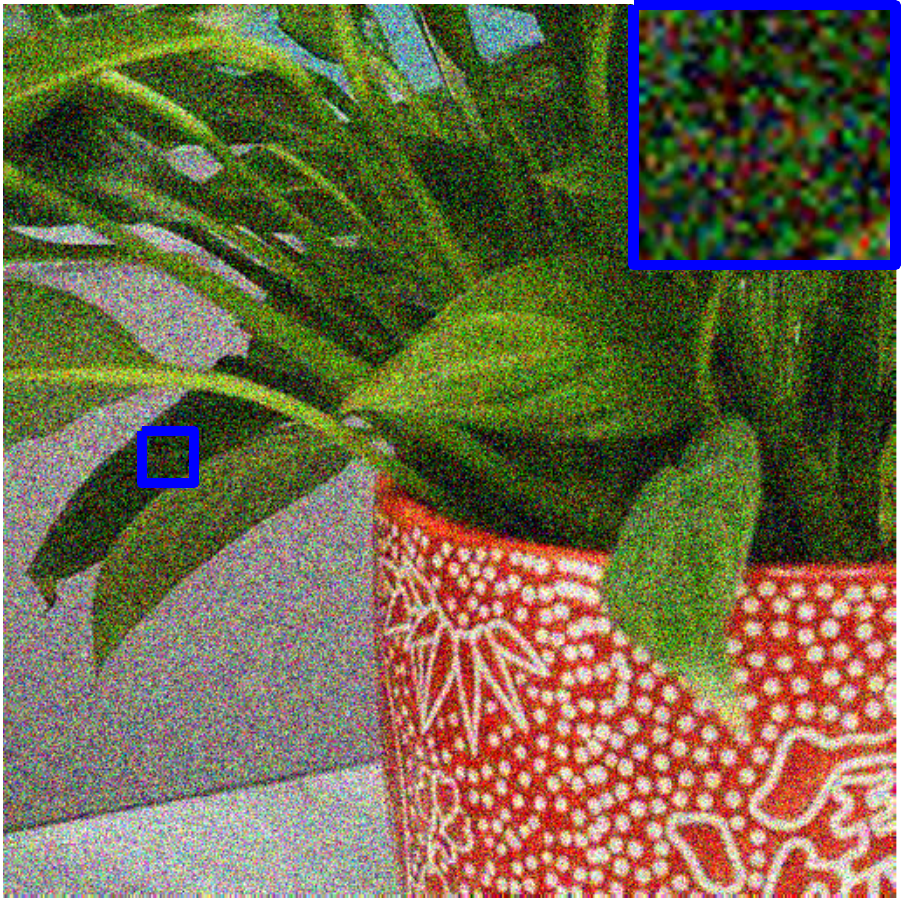}
{ \footnotesize{Noisy Image} }
\end{minipage}
\begin{minipage}[htbp]{0.24\linewidth}
\centering
\includegraphics[width=3.9cm]{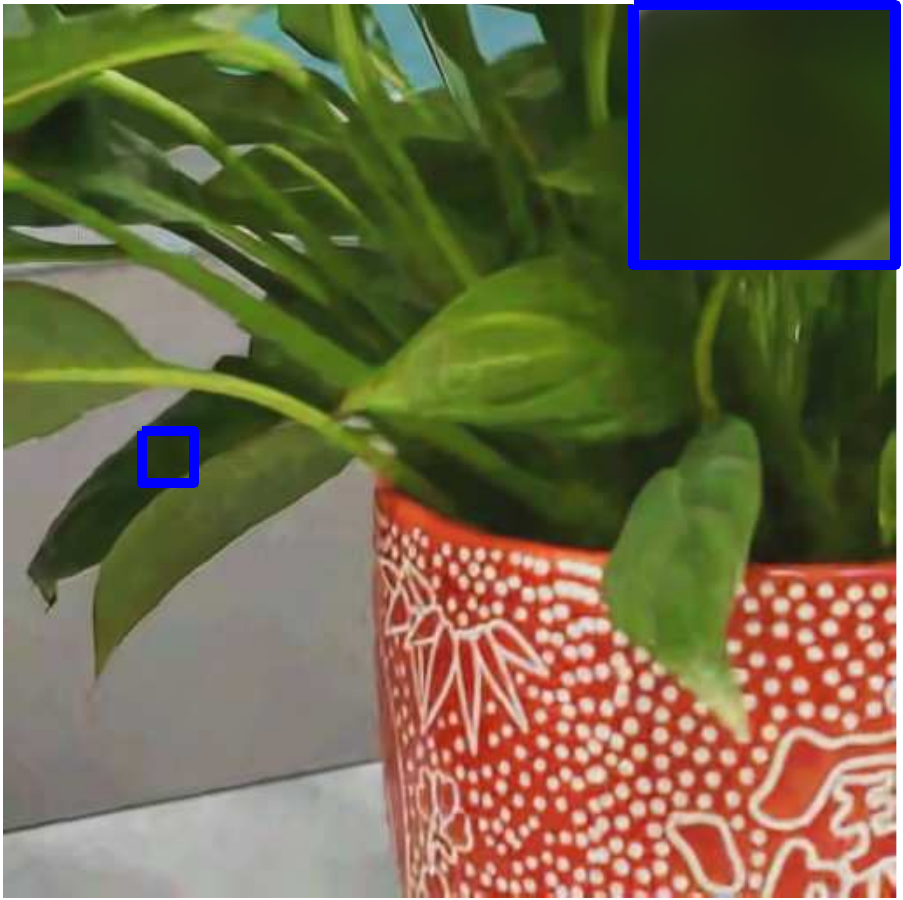}
{ \footnotesize{BM3D (27.92dB)} }
\end{minipage}
\begin{minipage}[htbp]{0.24\linewidth}
\centering
\includegraphics[width=3.9cm]{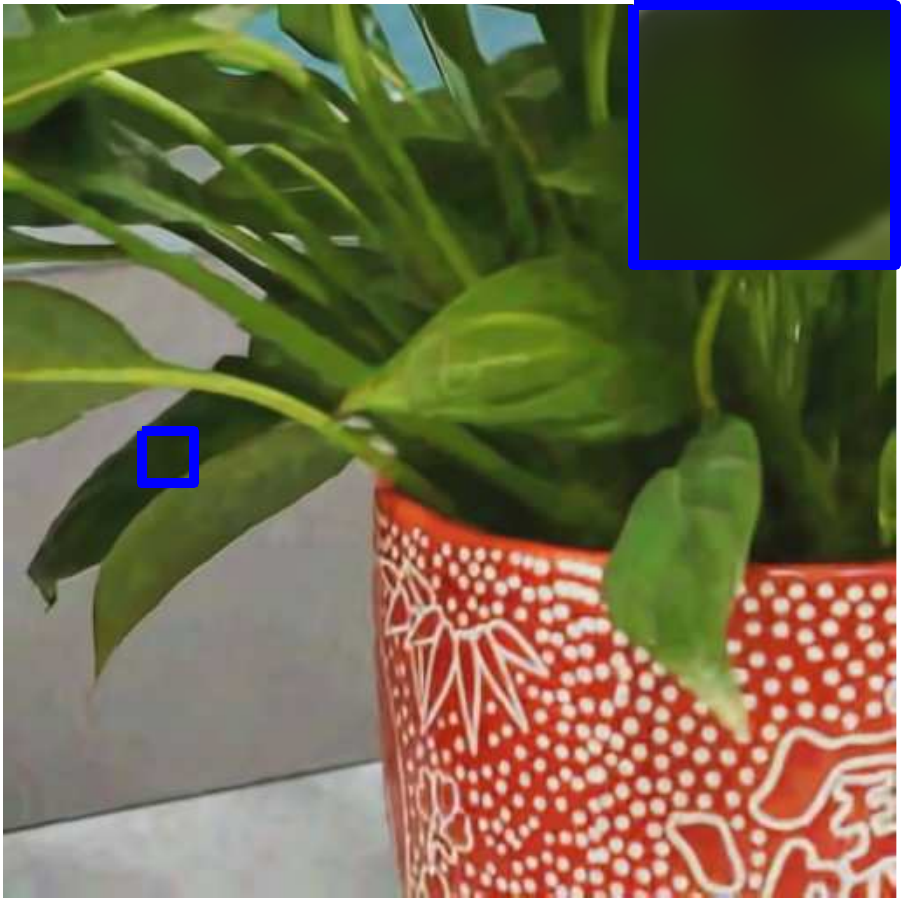}
{ \footnotesize{NN+BM3D (28.26dB)} }
\end{minipage}
\begin{minipage}[htbp]{0.24\linewidth}
\centering
\includegraphics[width=3.9cm]{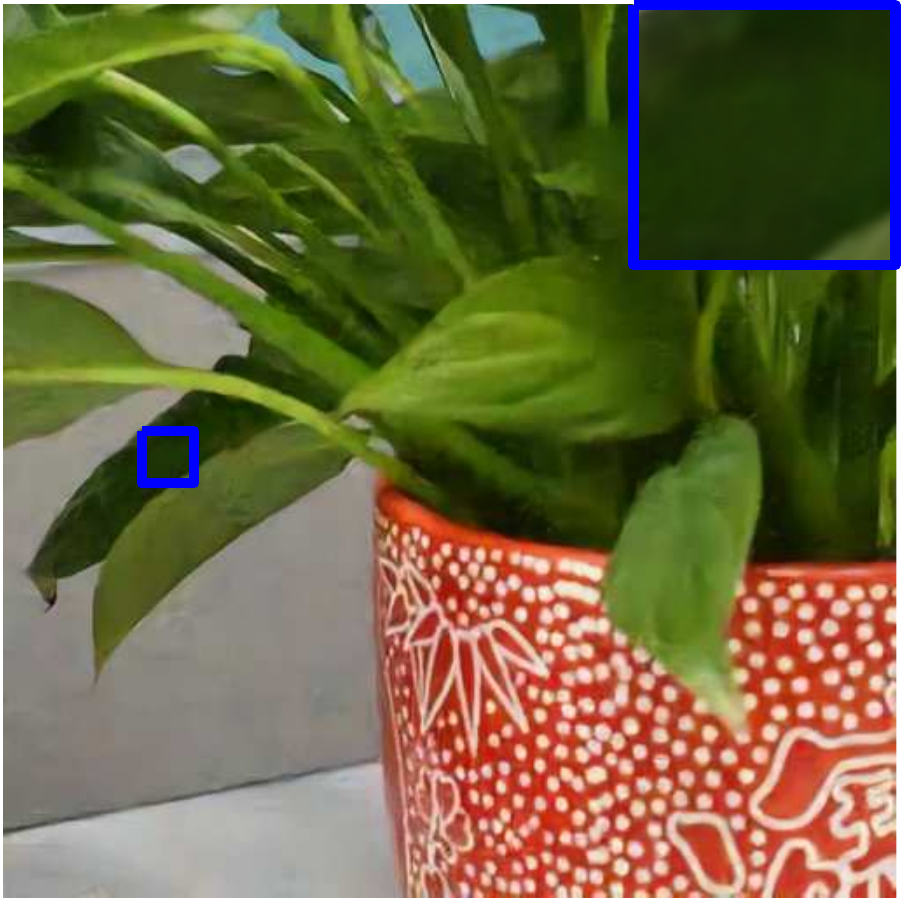}
{ \footnotesize{IRCNN (29.93dB)} }
\end{minipage}\\
\begin{minipage}[htbp]{0.24\linewidth}
\centering
\includegraphics[width=3.9cm]{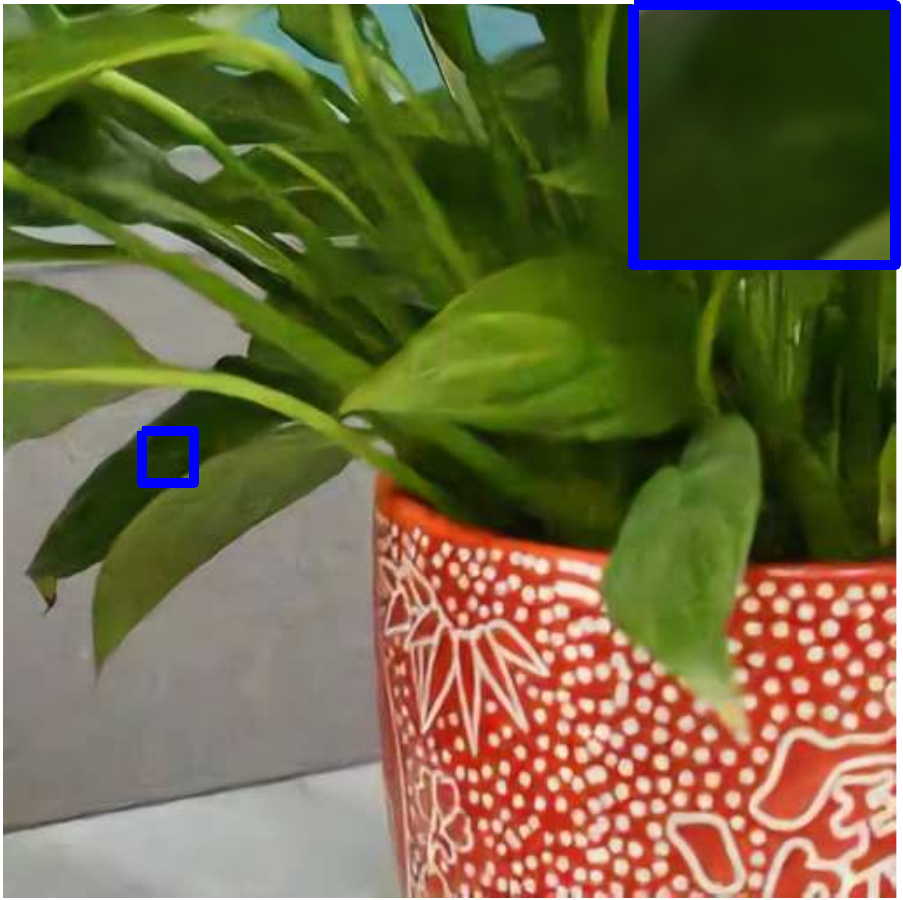}
{ \footnotesize{CBDNet (30.19dB)} }
\end{minipage}
\begin{minipage}[htbp]{0.24\linewidth}
\centering
\includegraphics[width=3.9cm]{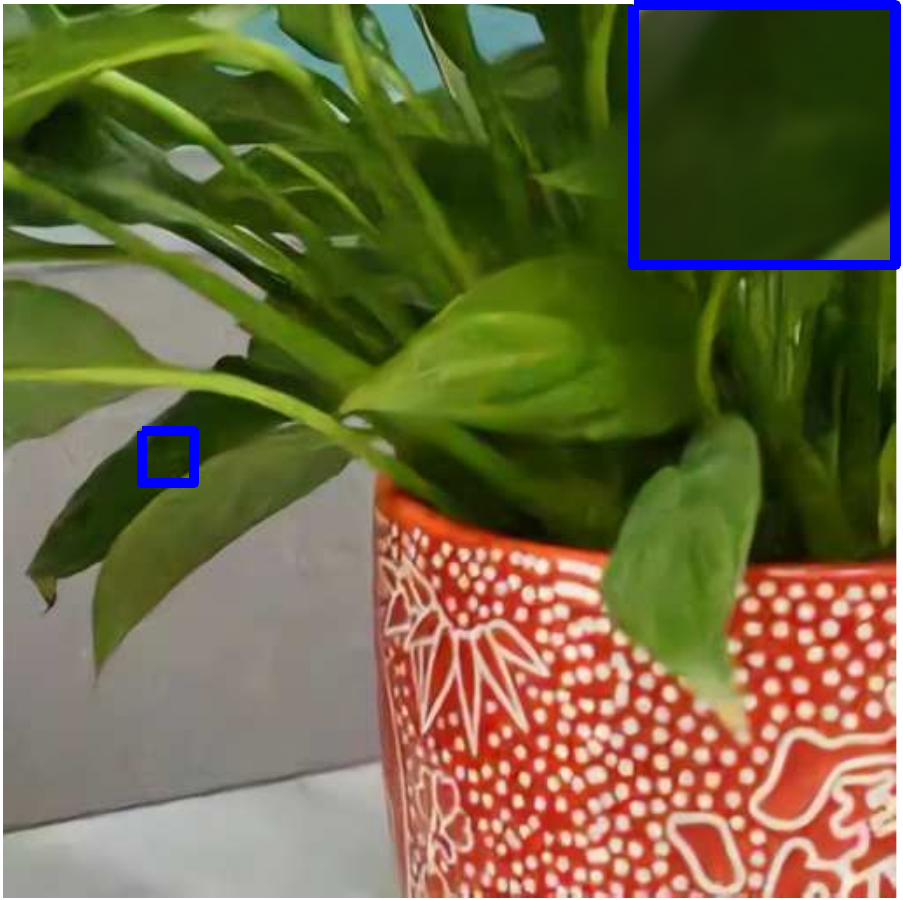}
{ \footnotesize{FDnCNN (30.30dB)} }
\end{minipage}
\begin{minipage}[htbp]{0.24\linewidth}
\centering
\includegraphics[width=3.9cm]{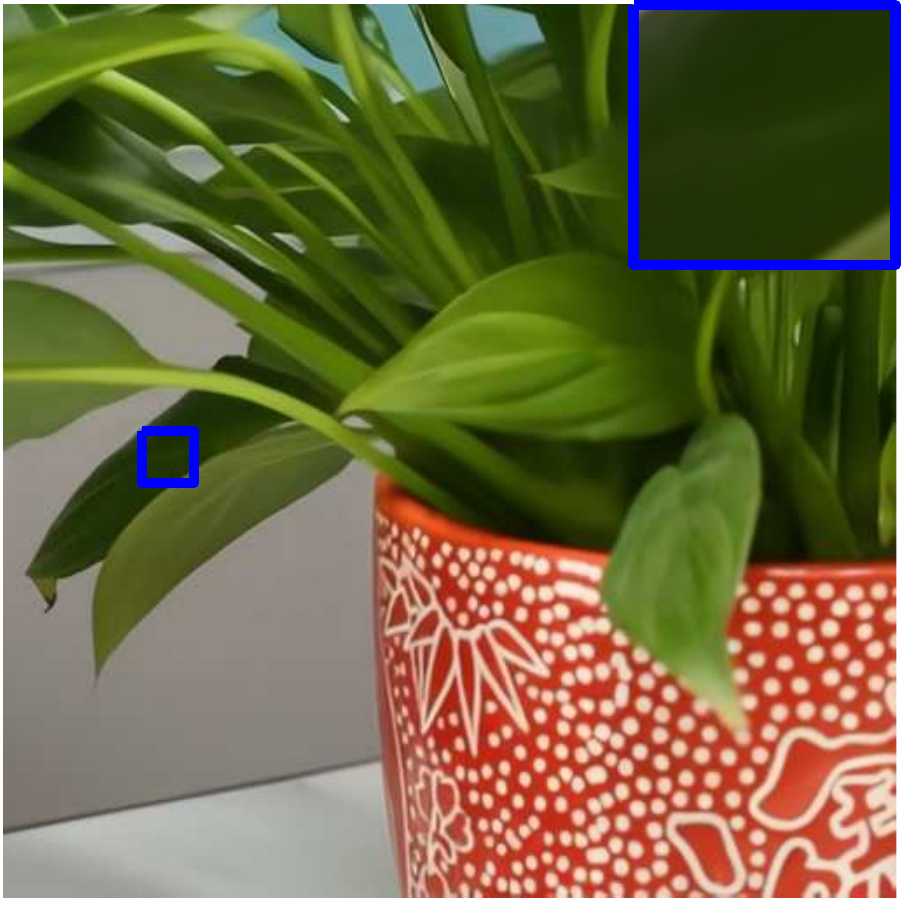}
{ \footnotesize{DRUNet (31.86dB)} }
\end{minipage}
\begin{minipage}[htbp]{0.24\linewidth}
\centering
\includegraphics[width=3.9cm]{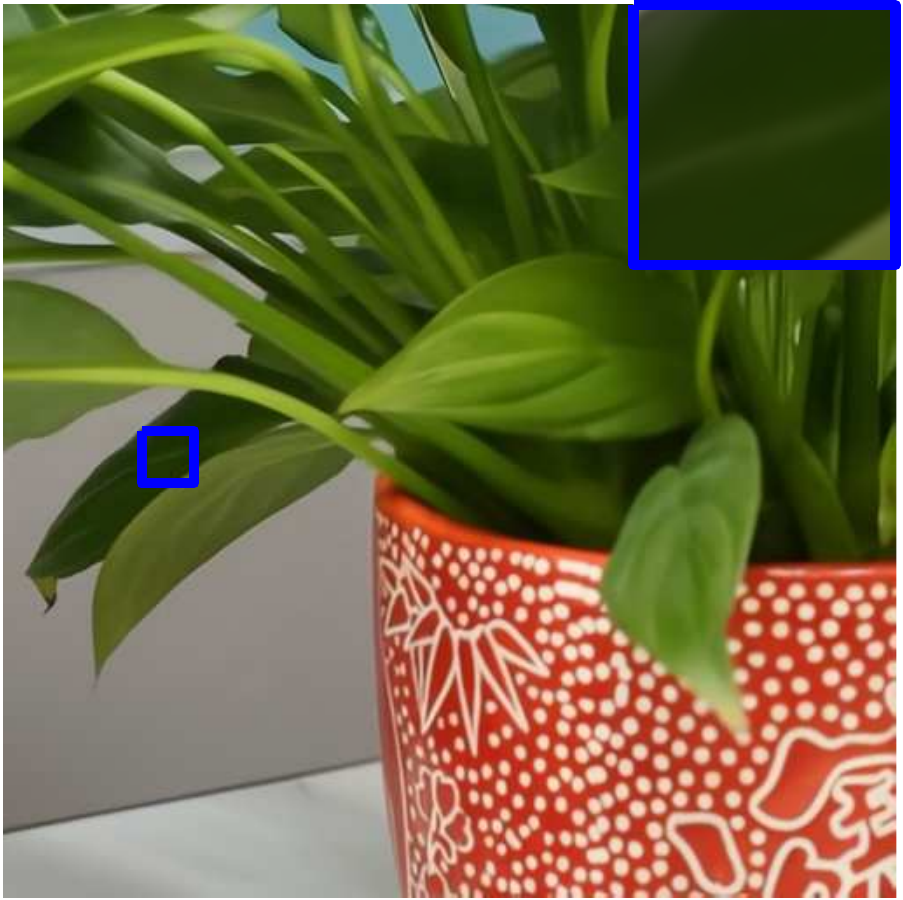}
{ \footnotesize{C-UNet (32.06dB)} }
\end{minipage}
\caption{Color image denoising results of different methods on the image $``img025''$ from Urban100 dataset and image $``Sony\_4-5\_125\_3200\_plant\_10''$from PolyU dataset with noise level of 50.}
\label{figcolordenoise}
\end{figure*}

\subsubsection{Comparison methods}
We compare the proposed C-UNet with the well-known traditional and learning-based denosing methods, the details of which are summarized as follows
\begin{itemize}
    \item BM3D \citep{Dabov2007Image}: The Block Matching and 3D filtering (BM3D) is an image denoising method which integrates 3-D transformation of a group, shrinkage of the transform spectrum and inverse 3-D transformation and demands the noise level as input.
    \item NN+BM3D \citep{Zheng2021an}: The unsupervised denoising method NN+BM3D combines the encoder-decoder convolutional neural network with the Gaussian denoiser BM3D, which does not require any training samples, but needs the noise level as the input.
    \item IRCNN \citep{zhang2017learning}: The Image Restoration Convolutional Neural Network (IRCNN) involves 25 separate 7-layer denoisers, which were trained on a certain noise level from the range of [0, 50]. The training dataset consists of 400 BSD images, 400 selected ImageNet database images and 4744 Waterloo Exploration Database images.
    \item CBDNet \citep{guo2019toward}: The Convolutional Blind Denoising Network (CBDNet) model includes a noise estimation sub-network to obtain the noise level map and a non-blind denoising subnetwork to estimate the denoising results. For a fair comparison, we retrained the CBDNet using the same training dataset as ours, i.e., 4744 Waterloo Exploration Database, 900 DIV2K and 2650 Flick2K images corrupted by additive Gaussian noises of noise level [0, 50].
    \item FDnCNN \citep{zhang2017beyond}: By taking a noise level map as input, the Flexible Denoising Convolutional Neural Network (FDnCNN) was trained on 400 BSD images corrupted by Gaussian noises with noise level ranging form 0 to 75.
    \item DRUNet \citep{zhang2021plug}: The Denoising Residual block based U-Net (DRUNet) adopts the U-Net architecture with ResNet blocks, which takes the noise level map as input. It was trained on BSD400, Waterloo Exploration Database, DIV2K and Flick2K datasets degraded by random Gaussian noises of level [0, 50].
\end{itemize}

\subsubsection{Quantitative and qualitative comparison results}
Table \ref{tabdenoiseg} presents the grayscale image denoising results of the noise level $\sigma_s=20$, $35$ and $50$ on different datasets. As shown, our C-UNet achieves the best restoration results for all datasets, especially for images with big noises. The visual comparisions on the images $``A0025''$ from PIPAL dataset and $``Canon5D2\_5\_160\_6400\_circuit\_11''$ from PolyU dataset with noise level $\sigma_s=50$ are displayed in Fig. \ref{figgraydenoise}. As can be observed by the magnified portions, our C-UNet is very effective in preserving edges and texture structures. Concretely speaking, our C-UNet can preserve the texture of the wall and the fringe of the device.

Similarly, we conduct the comparison experiments on color image restoration problems. The denoising results on different datasets are presented in Table \ref{tabdenoisec}. Once again, we observe that our C-UNet gives the best performance for all noise levels, which demonstrate the superiority of our C-UNet in dealing with noises in a wide range. Fig. \ref{figcolordenoise} shows the visual results of different methods on the images $``img025''$ from Urban100 dataset and $``Sony\_4-5\_125\_3200\_plant\_10''$ from PolyU dataset with noise level $\sigma_s=50$. It can be observed that BM3D, NN+BM3D, IRCNN, CBDNet, FDnCNN and DRUNet can remove the noises but cannot recover the sharp structures of images. Obviously, our C-UNet performs better in preserving fine structures such as the stripes on the wall and the leaf vein, proving the effect of the curvature information.

\begin{table*}[h]
\begin{center}
\begin{minipage}{\textwidth}
\caption{Average runtime comparison (in seconds) among different methods on color image denoising datasets.} \label{denoisetime}
\small
\begin{tabular*}{\textwidth}
{@{\extracolsep{\fill}}llllllll@{\extracolsep{\fill}}}
\toprule
Datasets & BM3D & NN+BM3D  & IRCNN & CBDNet & FDnCNN & DRUNet & C-UNet\\
\midrule
 Kodak24  & 7.3183  & 1134.26 & 0.0383  & 0.0472  &  0.1846 &  0.0867 & 0.3417 \\
 Urban100  & 23.0133  & 1468.20 & 0.1153  & 0.1159  & 0.5239  & 0.2660  & 1.2084 \\
 PIPAL  & 2.3057  & 949.20  & 0.0144  & 0.0161  & 0.0712  & 0.0462  & 0.1190 \\
 CC  & 7.9339  & 1238.95 & 0.0437  & 0.0499  &  0.1887 & 0.1063  & 0.3844 \\
 PolyU  & 8.1621  & 1218.75 & 0.0387  & 0.0458  & 0.1849 & 0.1015  & 0.4055 \\
\botrule
\end{tabular*}
\end{minipage}
\end{center}
\end{table*}

\begin{figure*}[t]
\centering
\subfigure[CBDNet]{\includegraphics[width=7.5cm]{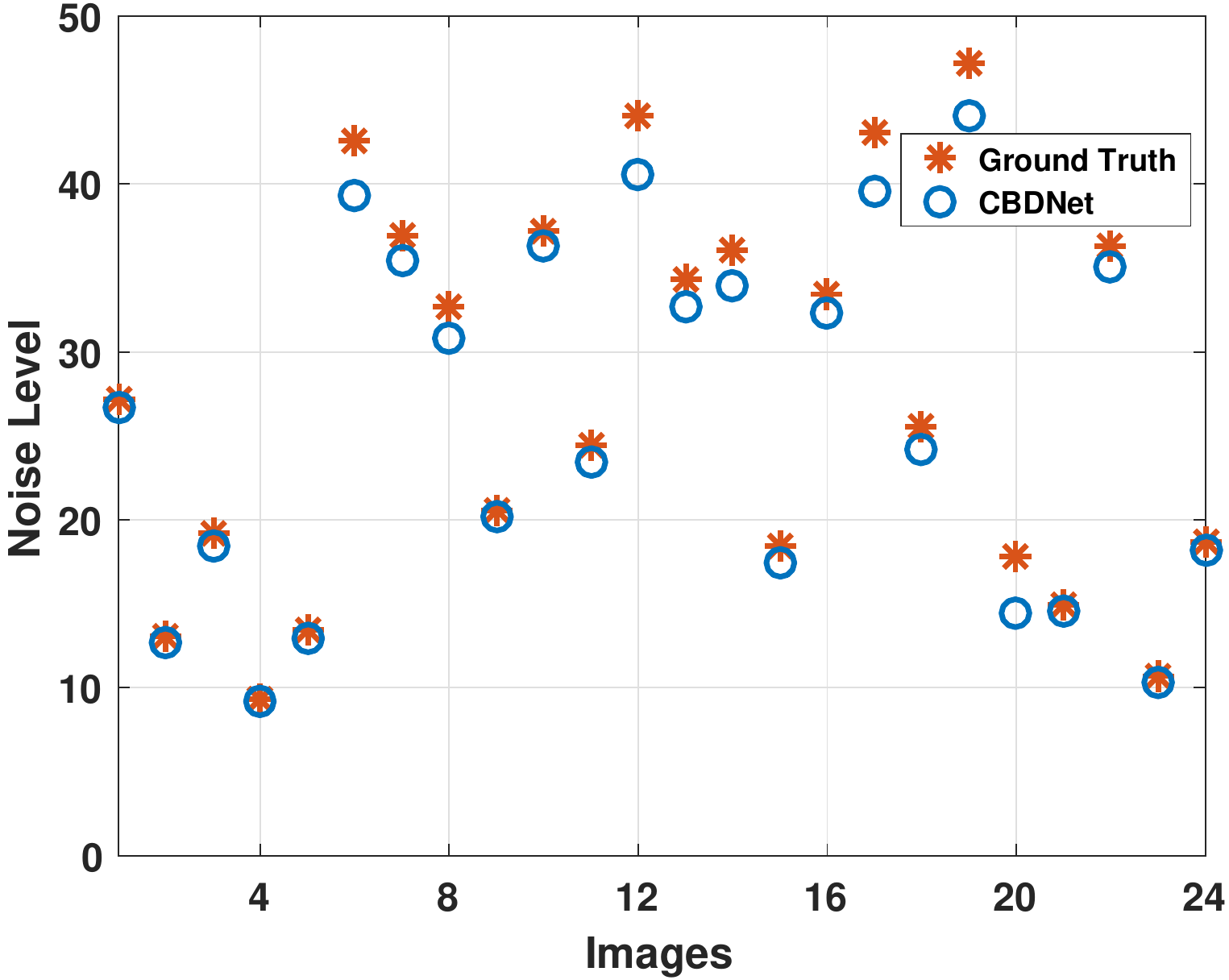}}\qquad
\subfigure[C-UNet]{\includegraphics[width=7.5cm]{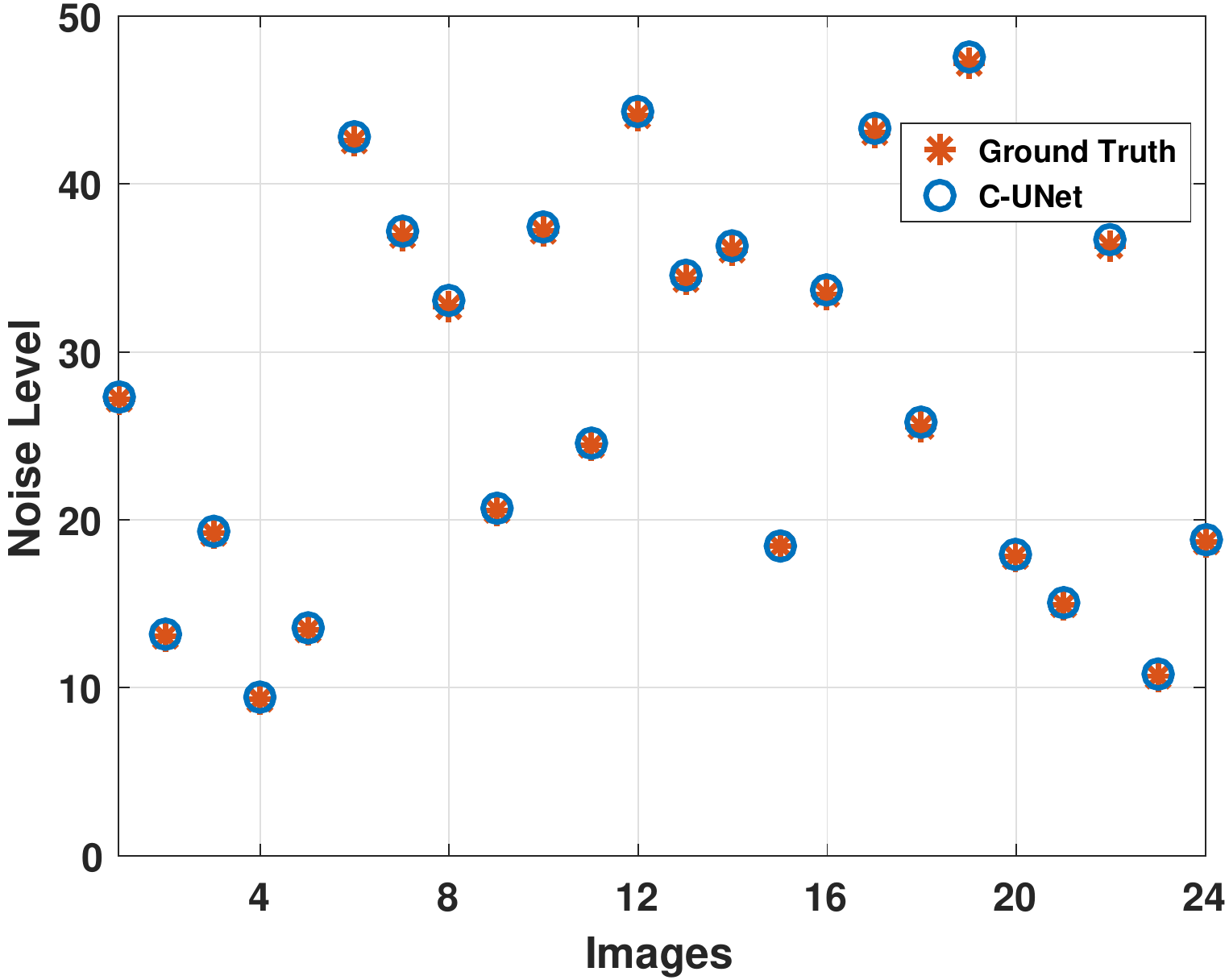}}
\caption{The noise level obtained by CBDNet and C-UNet on Kodak24 dataset, which are degraded by random Gaussian noises of level $\sigma\in[0,50]$.}
\label{comparenoise}
\end{figure*}

\subsubsection{Computational efficiency}

Table \ref{denoisetime} reports the comparison results on the average run time  of the aforementioned denoising methods on color image datasets with noise level 50. Compared with CBDNet, IRCNN, FDnCNN and DRUNet,  although our C-UNet consumes a little bit more time due to the curvature map, it provides much better PSNR values. Note that the unsupervised NN+BM3D costs the most computational time due to the on-line learning strategy.

\subsubsection{Noise estimation}
To verify the accuracy of the noise estimation subnetwork in our C-UNet, we compare the noise levels obtained by CBDNet and C-UNet on Kodak24 dataset, where the test images are corrupted by Gaussian noises of noise level $\sigma_s \in [0, 50]$. As can be observed in Fig. \ref{comparenoise}, the noise levels estimated by our C-UNet are close to the truth noise levels added into the test images. Since the noise estimation sub-network of CBDNet adopts a five-layer plain  fully connected convolutional network, which does not work as powerful as the UNet used in our C-UNet, it tends to underestimate the noise levels and the performance is also not as stable as ours.

\subsection{Image Deblurring}

\begin{figure*}[t]
\centering
\begin{minipage}[htbp]{0.195\linewidth}
\centering
\includegraphics[width=3.1cm]{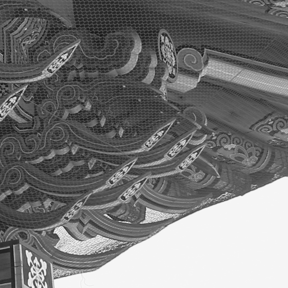}
{ \footnotesize{(a) Eave} }
\end{minipage}
\begin{minipage}[htbp]{0.195\linewidth}
\centering
\includegraphics[width=3.1cm]{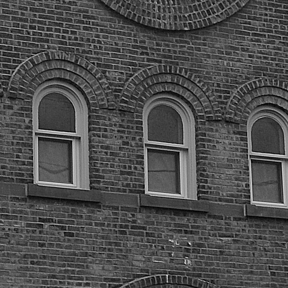}
{ \footnotesize{(b) Wall} }
\end{minipage}
\begin{minipage}[htbp]{0.195\linewidth}
\centering
\includegraphics[width=3.1cm]{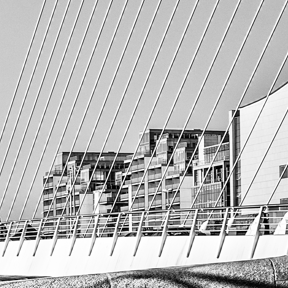}
{ \footnotesize{(c) Bridge} }
\end{minipage}
\begin{minipage}[htbp]{0.195\linewidth}
\centering
\includegraphics[width=3.1cm]{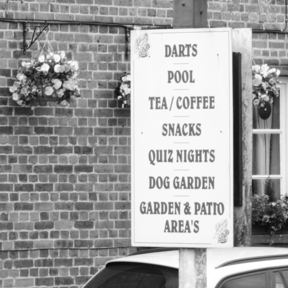}
{ \footnotesize{(d) Sign} }
\end{minipage}
\begin{minipage}[htbp]{0.195\linewidth}
\centering
\includegraphics[width=3.1cm]{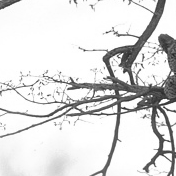}
{ \footnotesize{(e) Branch} }
\end{minipage}\\
\begin{minipage}[htbp]{0.195\linewidth}
\centering
\includegraphics[width=3.1cm]{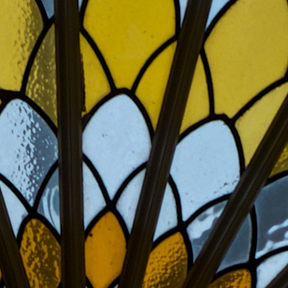}
{ \footnotesize{(f) Glass} }
\end{minipage}
\begin{minipage}[htbp]{0.195\linewidth}
\centering
\includegraphics[width=3.1cm]{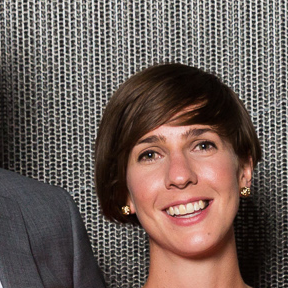}
{ \footnotesize{(g) Woman} }
\end{minipage}
\begin{minipage}[htbp]{0.195\linewidth}
\centering
\includegraphics[width=3.1cm]{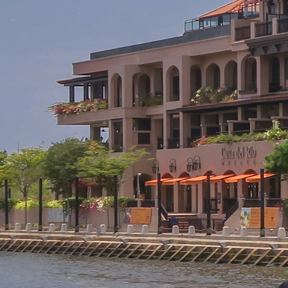}
{ \footnotesize{(h) House} }
\end{minipage}
\begin{minipage}[htbp]{0.195\linewidth}
\centering
\includegraphics[width=3.1cm]{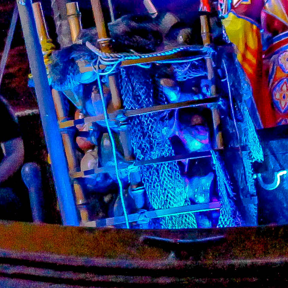}
{ \footnotesize{(i) Blue} }
\end{minipage}
\begin{minipage}[htbp]{0.195\linewidth}
\centering
\includegraphics[width=3.1cm]{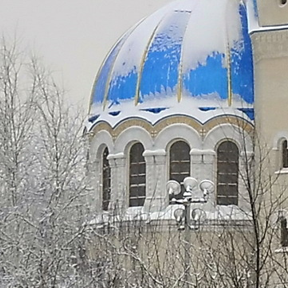}
{ \footnotesize{(j) Church} }
\end{minipage}\\
\caption{Ten test images for image deblurring experiments.}
\label{blurimg}
\end{figure*}

The degradation model of the blurry observation $y$ can be described by
\begin{equation}
y=Ax+n,
\end{equation}
where $Ax$ denotes the convolution of the clean image $x$ with the blur operator $A$.

Our CurvPnP method focuses on the non-blind deblurring with the blur kernel being assumed known. For image deblurring, the $x$ subproblem  \eqref{subprobx} can be defined as follows
\begin{equation}
x_{k+1}=\arg\min_{x}~\frac{1}{2\lambda\sigma_s^2}\vert\vert y-Ax\vert\vert^2+\frac{1}{\sigma_k^2}\vert\vert v_k-x\vert\vert^2,
\label{dbminx}
\end{equation}
the Euler-Lagrange equation of which gives
\begin{equation}
(\sigma_k^2A^*A+\lambda\sigma_s^2\mathcal{I})x_{k+1}=\sigma_k^2A^*y+\lambda\sigma_s^2v_k,
\end{equation}
with $A^*$ being the adjoint of $A$ and $\mathcal{I}$ being the identity operator. The above equation can be efficiently solved by the discrete fast Fourier transform (FFT), that is
\begin{equation}
x_{k+1}=\mathcal{F}^{-1}\Bigg(\frac{\sigma_k^2\overline{\mathcal{F}(A)}\mathcal{F}(y)+\lambda\sigma_s^2\mathcal{F}(v_k)}{\sigma_k^2\overline{\mathcal{F}(A)}\mathcal{F}(A)+\lambda\sigma_s^2\mathcal{I}}\Bigg).
\label{solvxdeb}
\end{equation}

\subsubsection{Comparison methods}
We compare the deblurring results of our CurvPnP with several state-of-the-art deblurring approaches, including the learning-based blind methods, non-blind method DWDN and PnP approaches. The details of deblurring methods are expressed as follows
\begin{itemize}
    \item DMPHN \citep{Zhang_2019_CVPR}: The Deep Multi-Patch Hierarchical Network (DMPHN) deals with blurry images via a fine-to-coarse hierarchical representation. The DMPHN was trained on the dataset consists of 2103 image pairs of GoPro dataset and 61 videos from VideoDeblurring dataset.
    \item MPRNet \citep{Zamir2021MPRNet}: The Multi-stage Progressive Restoration Network (MPRNet) model injects supervision at each stage to progressively improve degraded inputs. It was trained end-to-end on 2103 image pairs from GoPro dataset.
    \item DWDN \citep{dong2020deep}: The Deep Wiener Deconvolution Network (DWDN) is a domain-specific network that integrates the feature-based Wiener deconvolution into a deep neural network for non-blind image deblurring. It was trained on BSD400 and 4744 Waterloo Exploration datasets with synthetic realistic blur kernels of random sizes in the range from 13$\times$13 to 15$\times$15 pixels and Gaussian noises with noise levels in between [0, 12.75].
    \item IRCNN \citep{zhang2017learning}: It is a CNN denoiser prior based PnP image restoration method, for which the number of iteration was fixed as 30 for image deblurring problems.
    \item DPIR \citep{zhang2021plug}: The Deep PnP Image Restoration (DPIR) is built up in the PnP framework using DRUNet as the denoiser, for which the iteration number of is fixed as 8 for image deblurring problems.
    \item DPIR+: We introduce the noise estimation method \citep{chen2015efficient} into the denoiser DRUNet for dealing with the blind image restoration problem, which is shorted as DPIR+.
\end{itemize}

\begin{table*}[h]
\begin{center}
\begin{minipage}{\textwidth}
\caption{PSNR(dB) results of different methods for image deblurring on GSet5 and CSet5 datasets. The best and second best scores are \textbf{highlighted} and \underline{underline}, respectively.}\label{tabdeblur}
\small
\begin{tabular*}{\textwidth}
{@{\extracolsep{\fill}}lllllllllllll@{\extracolsep{\fill}}}
\toprule
Kernel & $\sigma_s$  & Methods & Eave & Wall  & Bridge & Sign & Branch & Glass & Woman & House & Blue & Church\\
\midrule
First  &  4  & DMPHN & 19.88   & 18.62   & 16.39  & 17.99  & 16.97  & 20.30  & 13.73  & 16.63  & 18.52  & 14.84  \\
(19$\times$19) &   & MPRNet & -   & -   & -  & -  & -  & 22.87  & 15.80  & 18.61  & 19.37  & 17.64  \\
&    & DWDN & -   & -   & -  & -  & -  & 31.79  & 25.07  & 30.31  & 25.80  & 27.55  \\
&    & IRCNN & 28.08  &  26.00  &  27.69  & 28.87  & 29.47  & 34.86  & 27.33  & 31.26  & 28.22  & 28.57  \\
&    & DPIR & \underline{28.70}  & \underline{27.42}   & \underline{28.65}   & 29.98  & \underline{29.89}  & \underline{36.26}  & \underline{28.75}  & 32.26  & 29.09  & \underline{29.34}  \\
&    & DPIR+ & 28.43  & 26.63   & 27.72   & \underline{29.99}  & 28.53  & \underline{36.26}  & 28.43  & \underline{32.27}  & \underline{29.19}  & 29.26  \\
&    & CurvPnP & \textbf{28.88}  & \textbf{27.56}   & \textbf{28.67}   & \textbf{30.12}  & \textbf{30.23}  & \textbf{36.67}  & \textbf{29.09}  & \textbf{32.33}  & \textbf{29.30}  & \textbf{29.36}  \\

 \cline{2-13}

& 8  & DMPHN & 19.03   & 17.74   & 15.57  & 17.96  & 16.92  & 19.71  & 13.80  & 16.20  & 18.26  & 14.82  \\
&    & MPRNet & -   & -   & -  & -  & -  & 21.55  & 16.46  & 17.62  & 18.89  & 16.54  \\
&    & DWDN & -   & -   & -  & -  & -  & 30.18  & 23.77  & 28.13  & 24.32  & 25.27  \\
&    & IRCNN & 26.05 &  23.40  &  24.34  & 25.56  & 26.24  & 32.33  & 23.85  & 28.48  & 25.67  & 25.72  \\
&    & DPIR & 26.42  & \underline{24.68}   & \underline{25.35}   & \underline{27.08}  & \underline{26.25}  & \underline{33.18}  & \underline{25.96}  & 29.32  & 26.31  & 26.17  \\
&    & DPIR+ & \underline{26.49}  & 24.67   &  25.06  & 27.06  & 26.04  & 32.55  & 25.78  & \underline{29.37}  &  \underline{26.37} & \underline{26.26}  \\
&    & CurvPnP & \textbf{26.56}  & \textbf{25.09}   & \textbf{25.62}   & \textbf{27.13}  & \textbf{26.64}  & \textbf{33.89}  & \textbf{26.60}  & \textbf{29.63}  & \textbf{26.71}  & \textbf{26.52}  \\
 \midrule
Fourth &  4   & DMPHN & 17.62   & 18.26   & 13.56  & 13.64  & 12.68  & 14.14  & 14.33  & 17.66  & 13.71  & 15.37  \\
(27$\times$27) &    & MPRNet & -   & -   & -  & -  & -  & 14.32  & 14.57  & 17.82  & 13.64  & 16.54  \\
&   & DWDN & -   & -   & -  & -  & -  & 29.46  & 23.54  & 28.40  & 23.62  & 26.04  \\
&    & IRCNN & 27.60  &  25.78  &  26.98  & 28.32  & 28.68  & 34.60  & 26.71  & 30.62  & 27.85  & 27.85  \\
&    & DPIR & \underline{28.24}  & \underline{27.00}   & \underline{28.08}   & 29.51  & \underline{28.95}  & \underline{35.91}  & \underline{28.06}  & \underline{31.65}  & 28.56  & \underline{28.70}  \\
&    & DPIR+ & 28.08  &  26.86  &  27.64  & \underline{29.55}  & 27.77  & 35.77  & 27.98  & 31.56  & \underline{28.59}  & 28.60  \\
&    & CurvPnP & \textbf{28.36}  & \textbf{27.22}   & \textbf{28.14}   & \textbf{29.72}  & \textbf{29.19}  & \textbf{36.29}  & \textbf{28.60}  & \textbf{31.71}  & \textbf{28.82}  & \textbf{28.76}  \\

 \cline{2-13}

&  8  & DMPHN & 17.02   & 17.50   & 13.17  & 13.39  & 12.82  & 13.40  & 13.77  & 16.24  & 13.37  & 14.85  \\
&    & MPRNet & -   & -   & -  & -  & -  & 13.95  & 14.35  & 17.00  & 12.91  & 16.18  \\
&    & DWDN & -   & -   & -  & -  & -  & 28.52  & 22.14  & 26.70  & 22.54  & 24.27  \\
&    & IRCNN & 25.53  &  22.66  &  23.61  & 24.86  & \underline{25.29}  & 32.05  & 23.33  & 27.72  & 25.29  & 25.17  \\
&    & DPIR & 25.81  & 24.28   & \underline{24.77}   & 26.47  & 25.27  & \underline{33.12}  & 25.26  & 28.68  & 25.85  & 25.55  \\
&    &  DPIR+ & \underline{25.98} &  \underline{24.30}  &  \underline{24.77}  & \underline{26.48} & 24.91  & 32.46  & \underline{25.32}  & \underline{28.73}  & \underline{25.88}  & \underline{25.69}  \\
&    & CurvPnP & \textbf{26.08}  & \textbf{24.92}   & \textbf{25.14}   & \textbf{26.88}  & \textbf{25.90}  & \textbf{33.62}  & \textbf{26.06}  & \textbf{28.99}  & \textbf{26.29}  & \textbf{25.93}  \\
\botrule
\end{tabular*}
\end{minipage}
\end{center}
\end{table*}

\begin{figure*}[h]
\centering
\begin{minipage}[htbp]{0.16\linewidth}
\centering
\includegraphics[width=2.6cm]{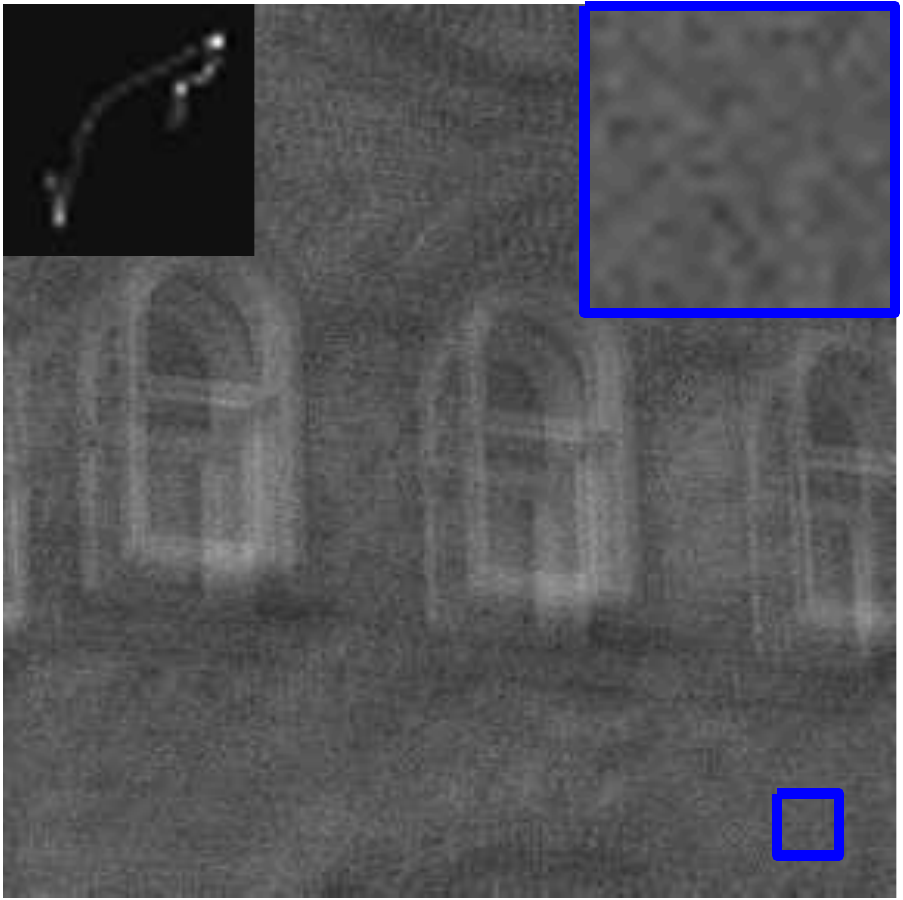}
{ \footnotesize{Blurry Image} }
\end{minipage}
\begin{minipage}[htbp]{0.16\linewidth}
\centering
\includegraphics[width=2.6cm]{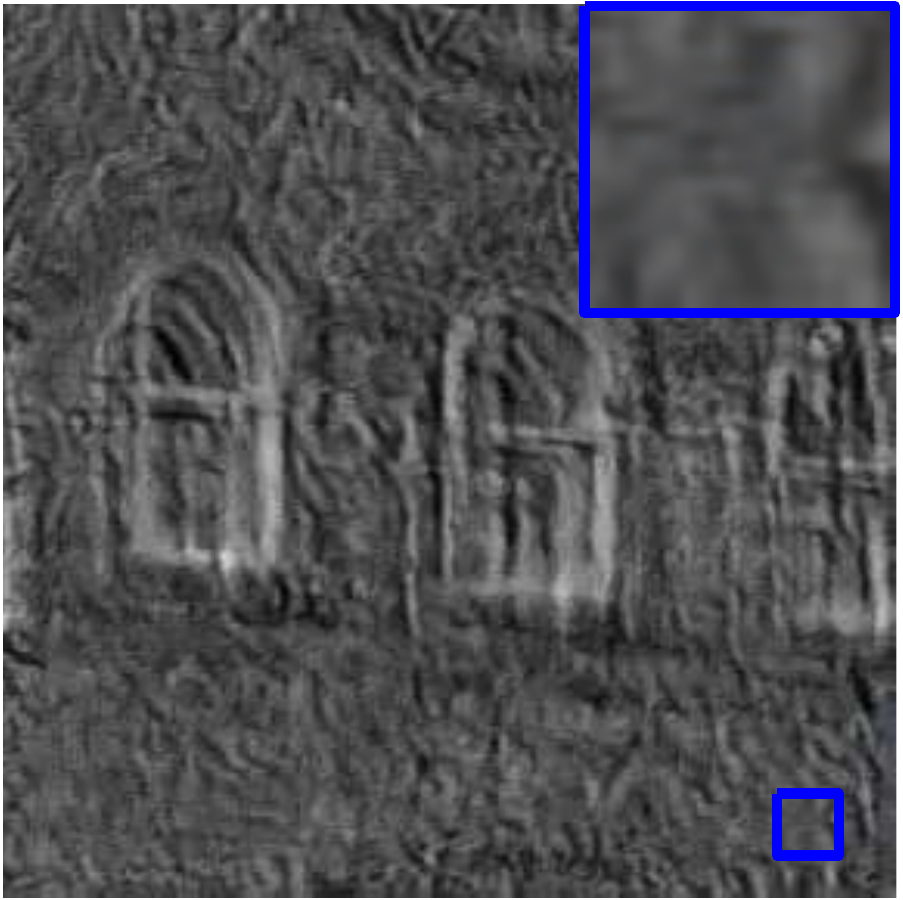}
{ \footnotesize{DMPHN (17.50dB)} }
\end{minipage}
\begin{minipage}[htbp]{0.16\linewidth}
\centering
\includegraphics[width=2.6cm]{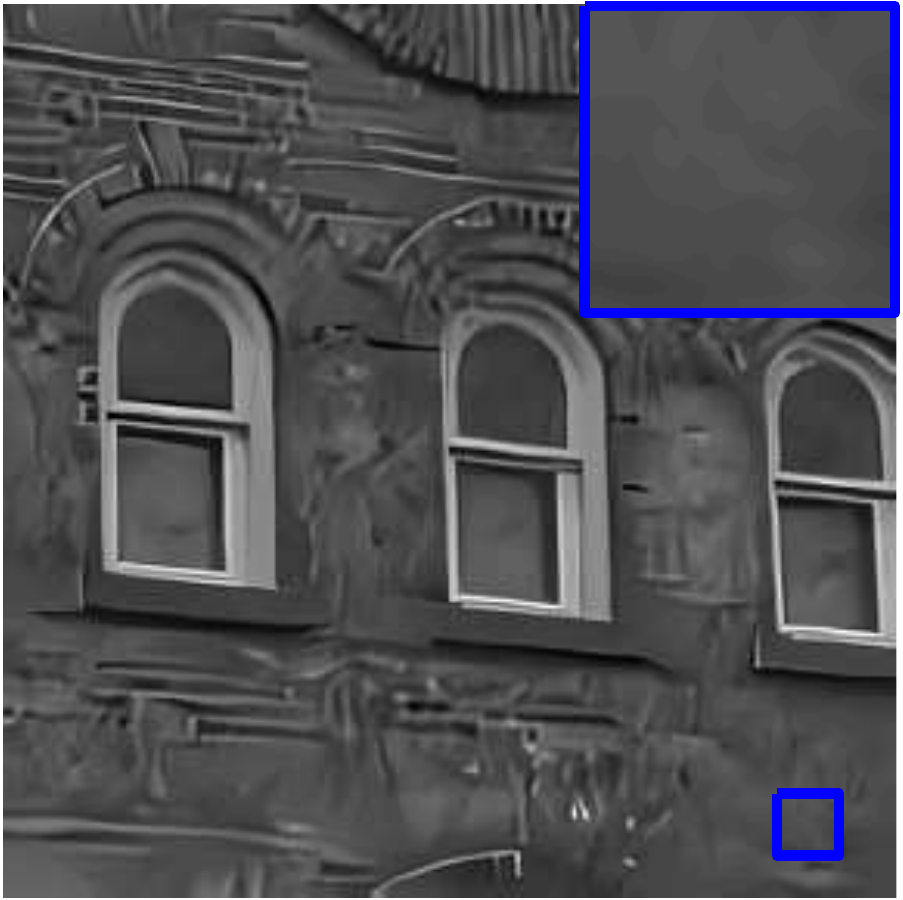}
{ \footnotesize{IRCNN (22.66dB)} }
\end{minipage}
\begin{minipage}[htbp]{0.16\linewidth}
\centering
\includegraphics[width=2.6cm]{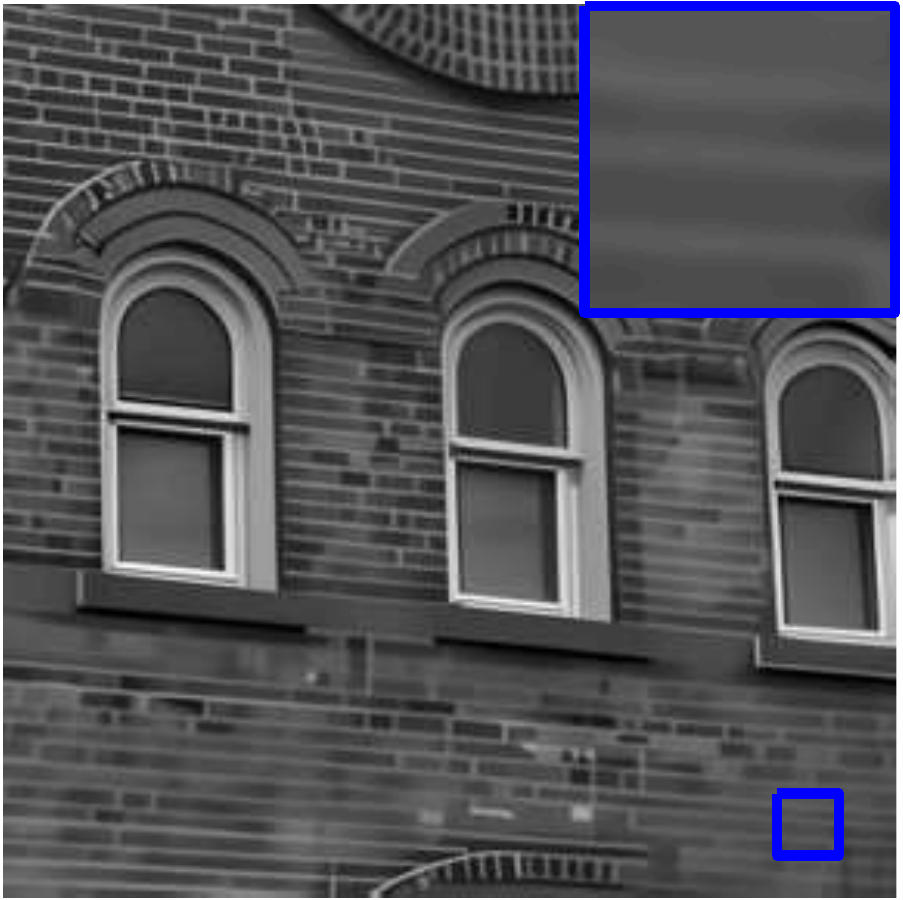}
{ \footnotesize{DPIR (24.28dB)} }
\end{minipage}
\begin{minipage}[htbp]{0.16\linewidth}
\centering
\includegraphics[width=2.6cm]{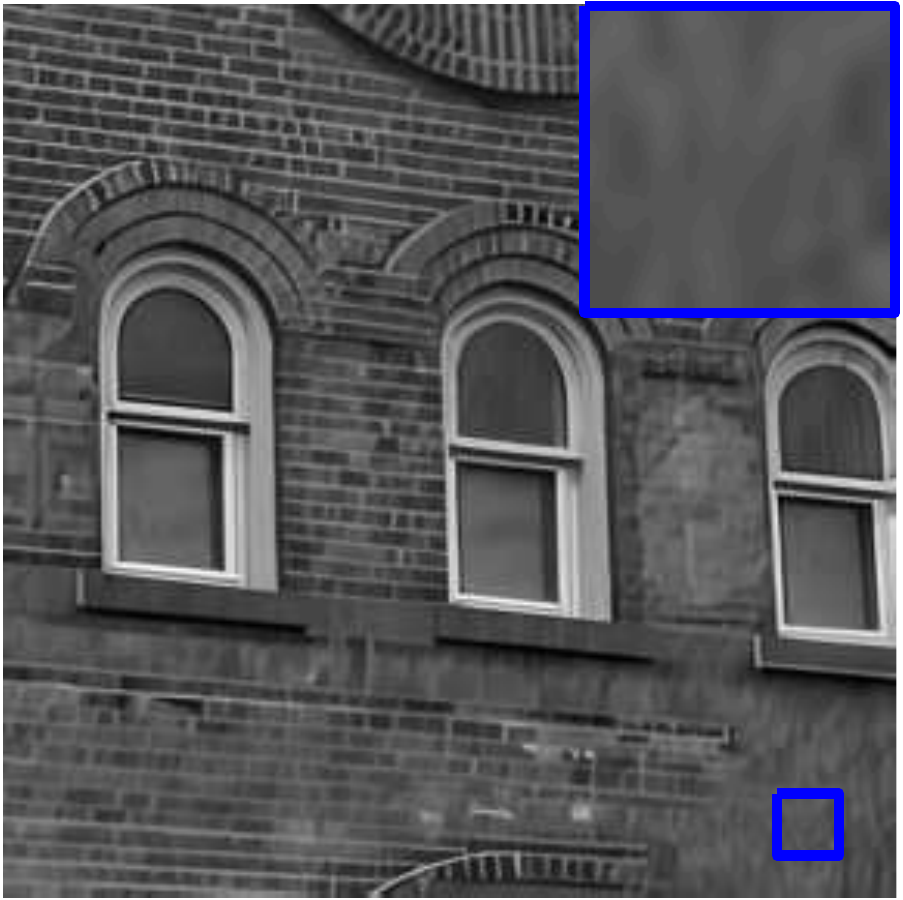}
{ \footnotesize{DPIR+ (24.30dB)} }
\end{minipage}
\begin{minipage}[htbp]{0.16\linewidth}
\centering
\includegraphics[width=2.6cm]{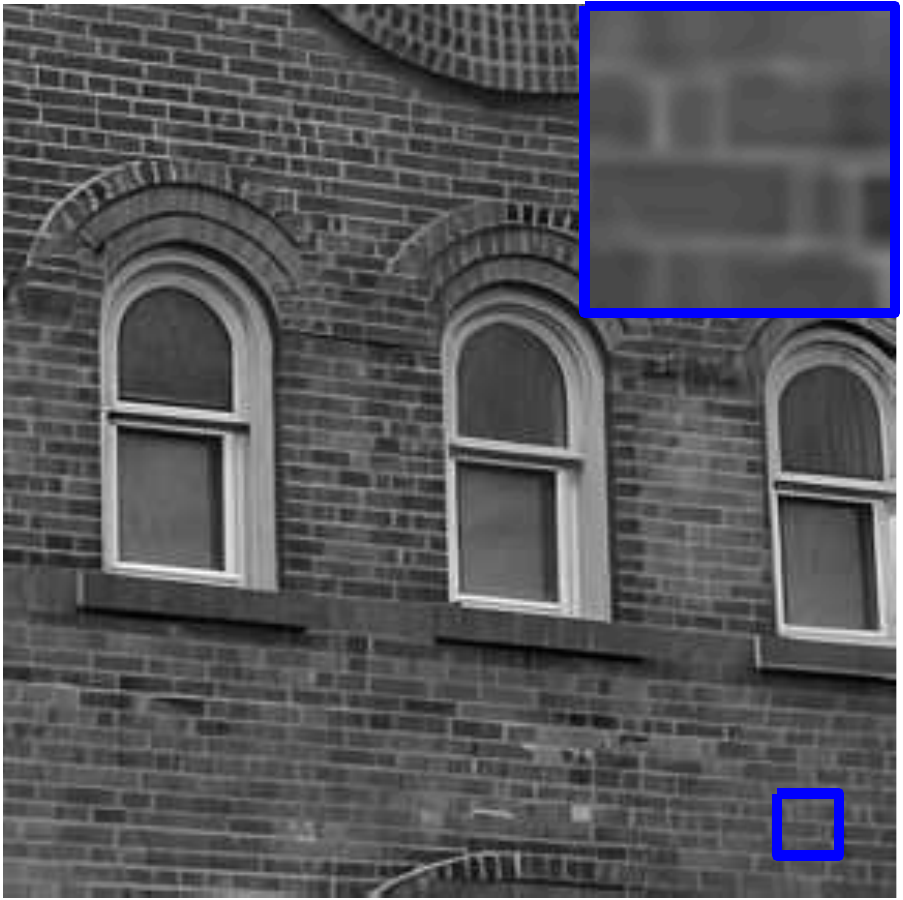}
{ \footnotesize{CurvPnP (24.92dB)} }
\end{minipage}\\
\vspace{0.2cm}
\begin{minipage}[htbp]{0.16\linewidth}
\centering
\includegraphics[width=2.6cm]{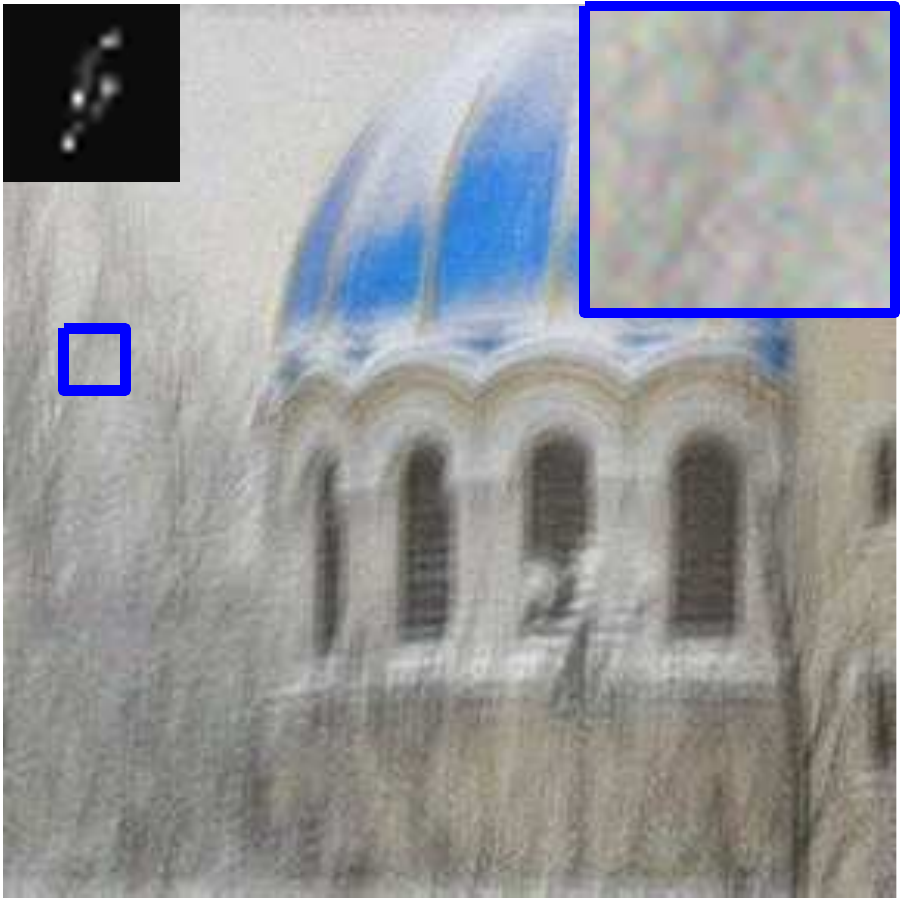}
{ \footnotesize{Blurry Image} }
\end{minipage}
\begin{minipage}[htbp]{0.16\linewidth}
\centering
\includegraphics[width=2.6cm]{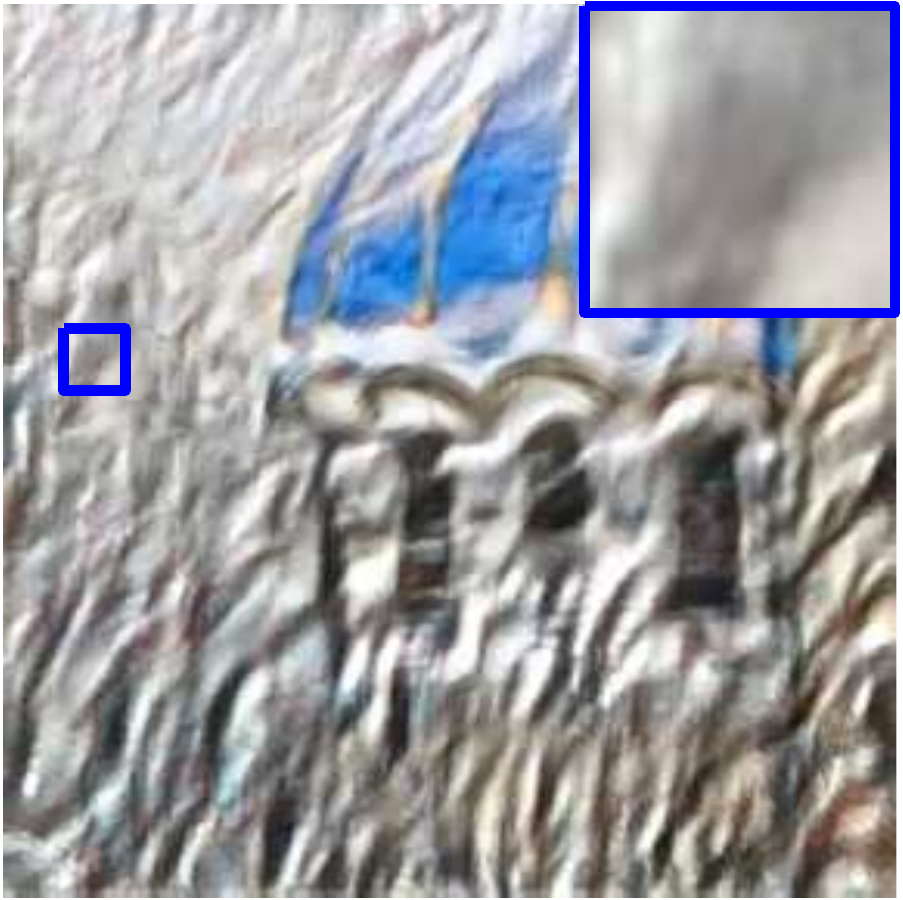}
{\footnotesize{DMPHN (14.82dB)} }
\end{minipage}
\begin{minipage}[htbp]{0.16\linewidth}
\centering
\includegraphics[width=2.6cm]{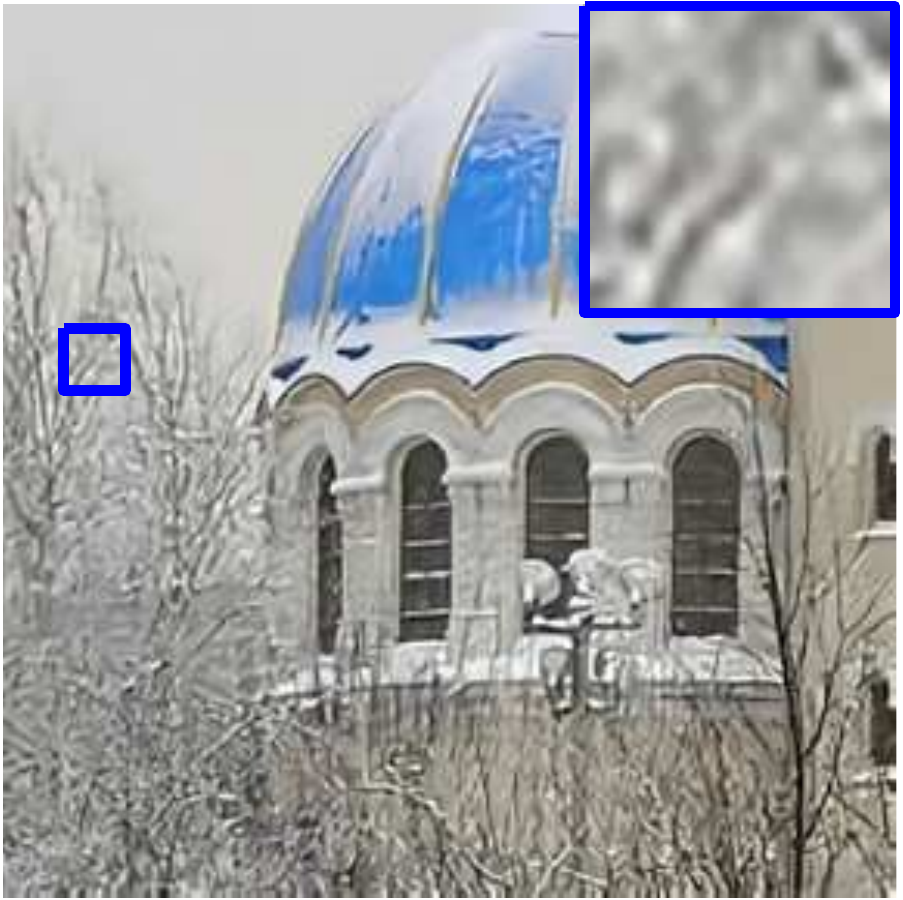}
{ \footnotesize{IRCNN (25.72dB)} }
\end{minipage}
\begin{minipage}[htbp]{0.16\linewidth}
\centering
\includegraphics[width=2.6cm]{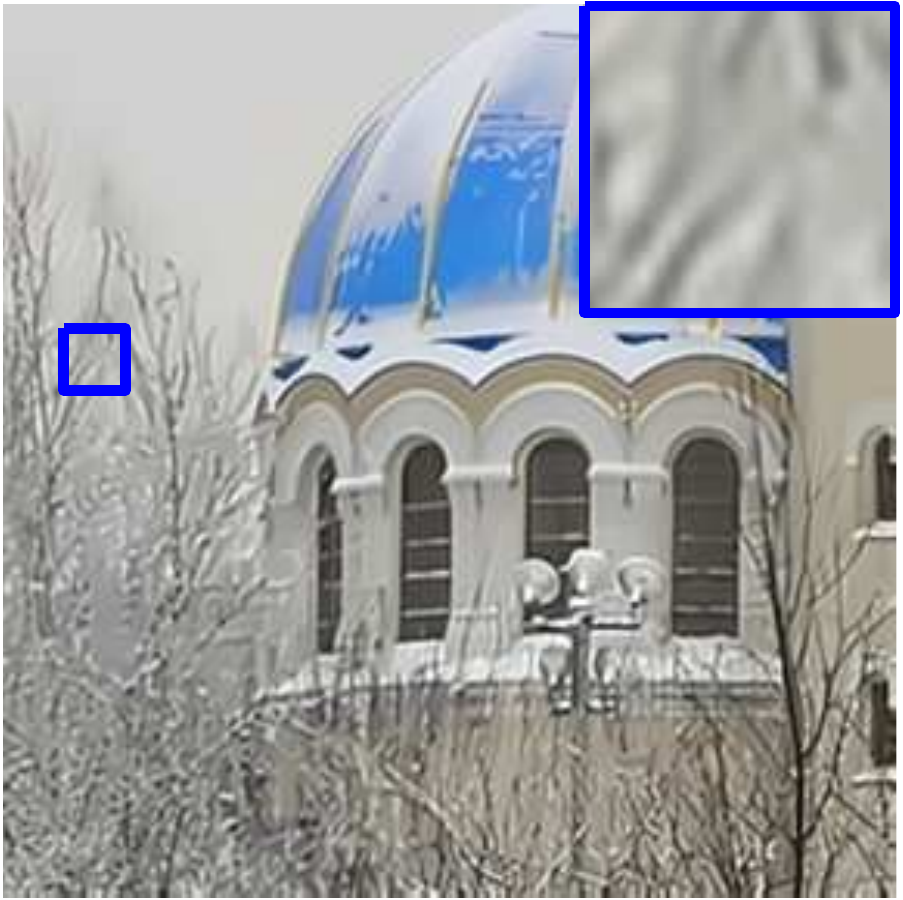}
{ \footnotesize{DPIR (26.17dB)} }
\end{minipage}
\begin{minipage}[htbp]{0.16\linewidth}
\centering
\includegraphics[width=2.6cm]{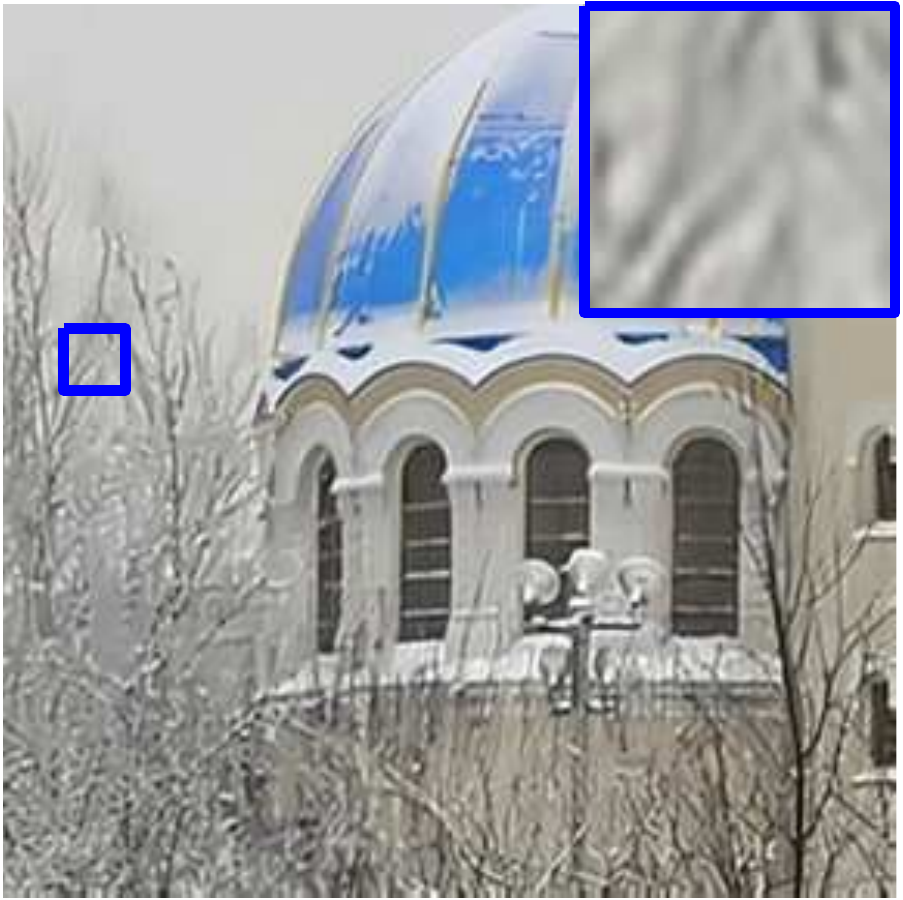}
{ \footnotesize{DPIR+ (26.26dB)} }
\end{minipage}
\begin{minipage}[htbp]{0.16\linewidth}
\centering
\includegraphics[width=2.6cm]{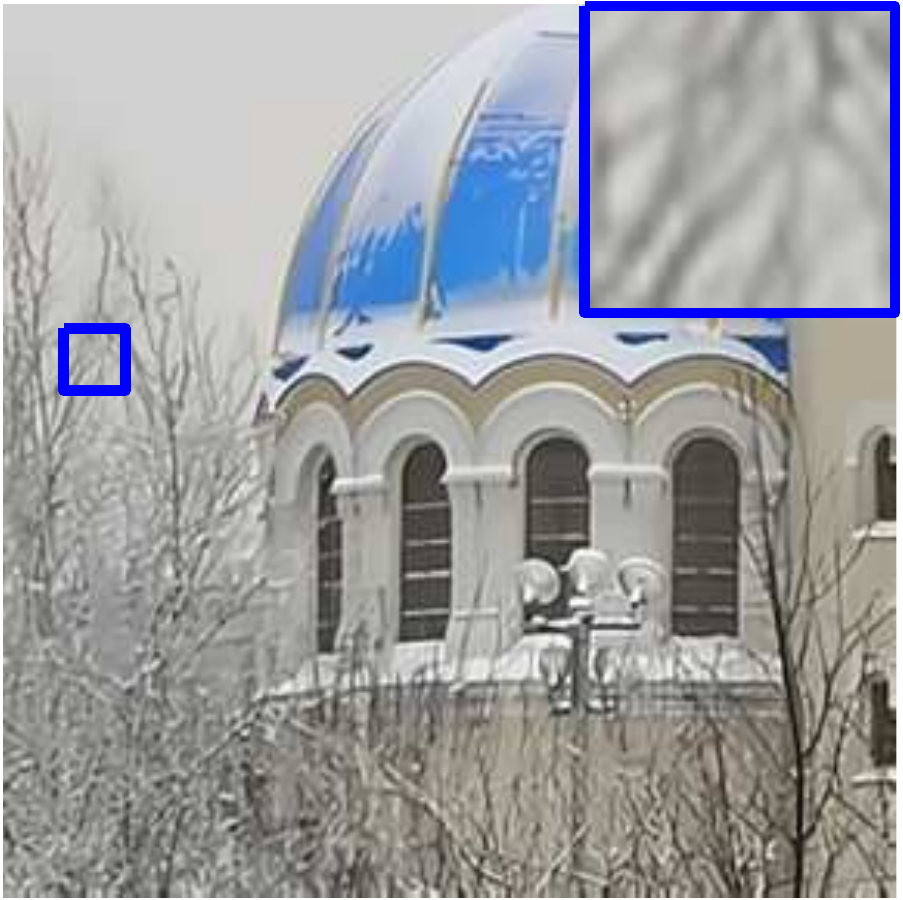}
{ \footnotesize{CurvPnP (26.52dB)} }
\end{minipage}
\caption{Image deblurring results of different methods on the image $``Wall''$ and $``Church''$ with noise level of 8. The blur kernels are shown on the upper left corner of the blurry images.}
\label{figdeblur}
\end{figure*}

\begin{table*}[h]
\begin{center}
\begin{minipage}{\textwidth}
\caption{Average runtime comparison (in seconds) among different methods on image deblurring datasets.} \label{deblurtime}
\small
\begin{tabular*}{\textwidth}
{@{\extracolsep{\fill}}llllllll@{\extracolsep{\fill}}}
\toprule
Datasets & DMPHN & DWDN  & MPRNet & IRCNN & DPIR & DPIR+ & CurvPnP \\
\midrule
 GSet5  & 0.0707  & - & -  & 0.0992  & 0.1442  & 0.1874  & 0.4490 \\
 CSet5  & 0.0782  &  0.3721 & 0.1184  & 0.1350  & 0.1844  & 0.4416  & 0.9762 \\
\botrule
\end{tabular*}
\end{minipage}
\end{center}
\end{table*}

\begin{figure*}[t]
\centering
\subfigure[Relative Error]{\includegraphics[width=7.5cm]{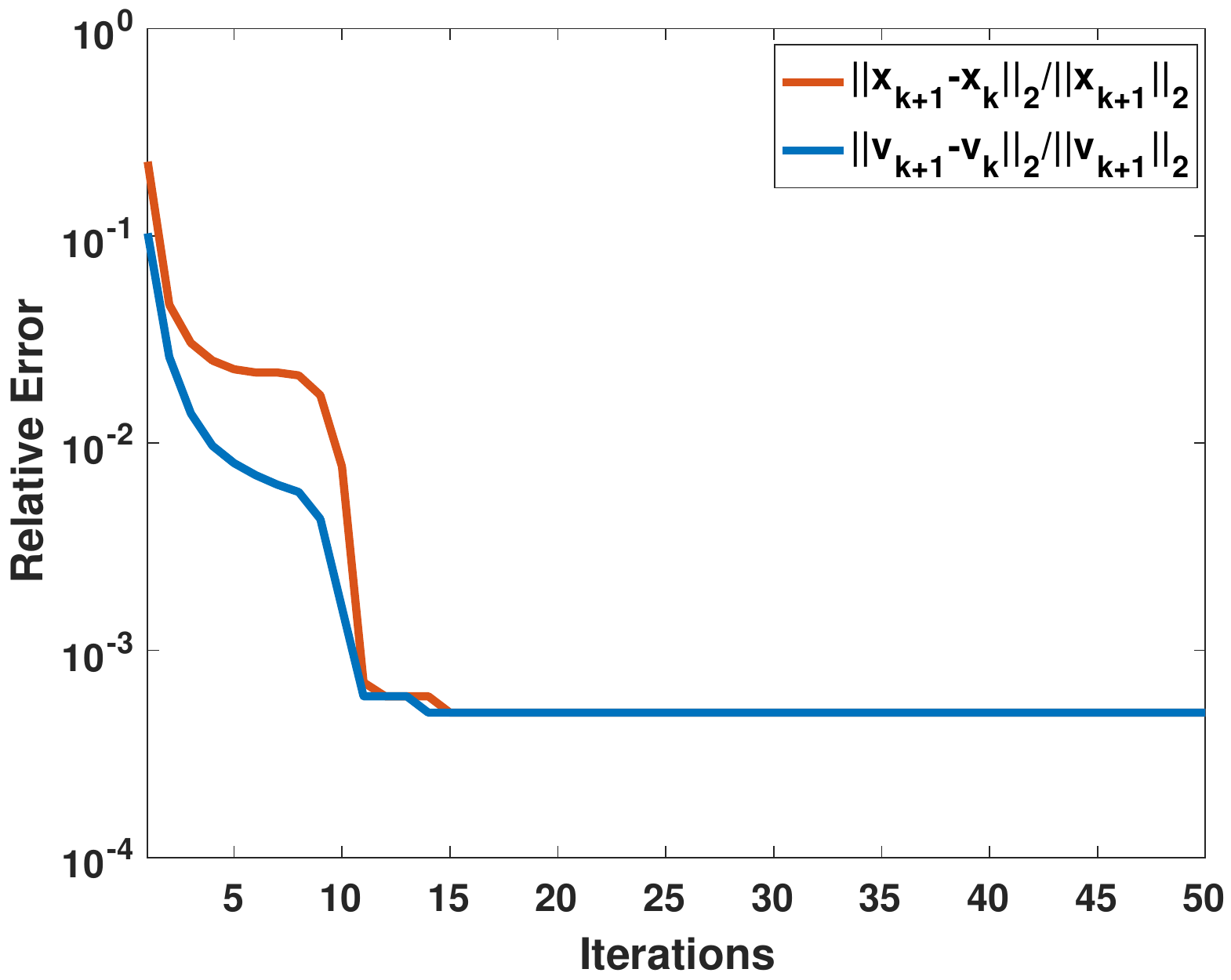}}\qquad
\subfigure[PSNR]{\includegraphics[width=7.5cm]{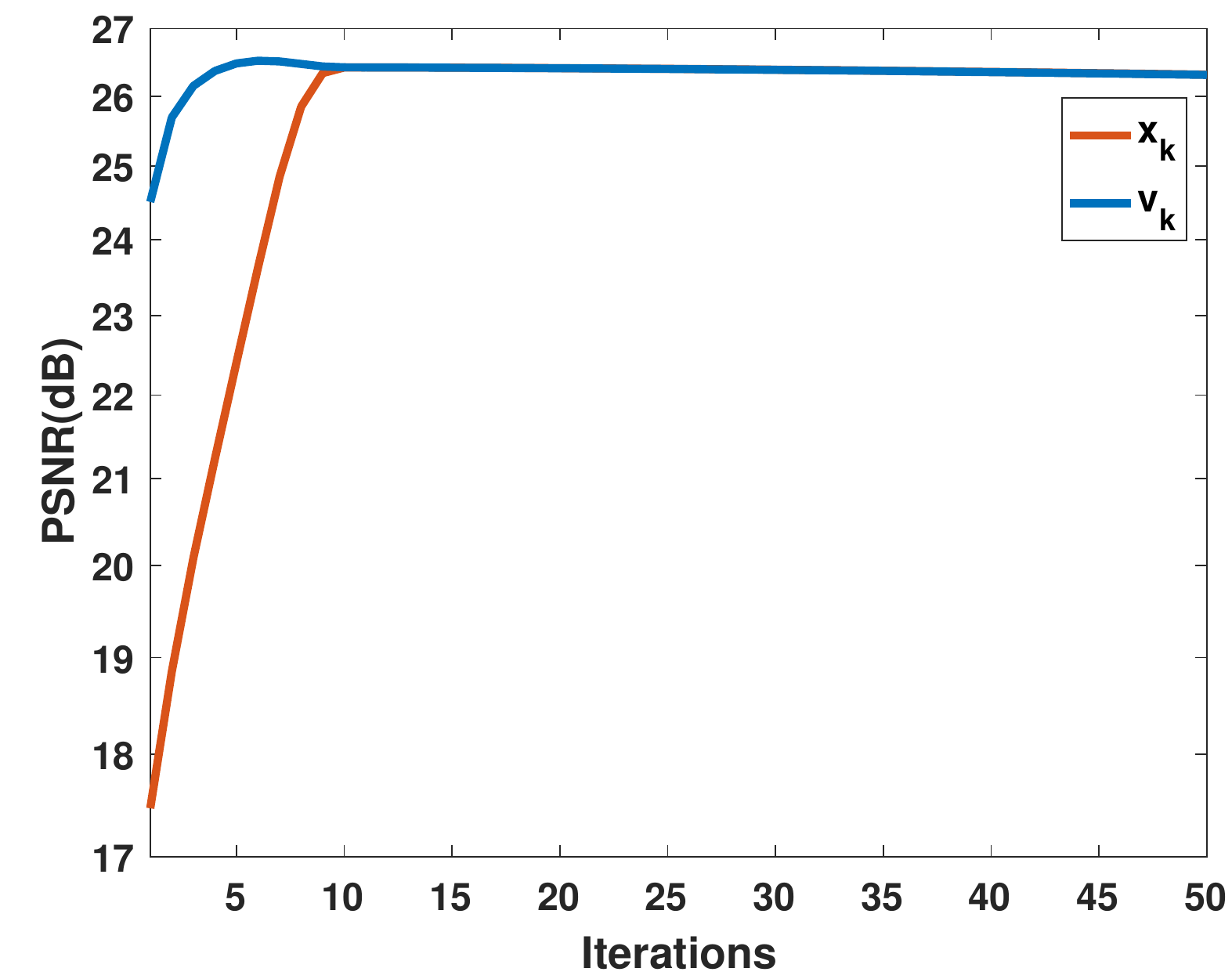}}
\caption{The evolution of relative errors $\frac{\|x_{k+1}-x_k\|_2}{\|x_{k+1}\|_2}$ and  $\frac{\|v_{k+1}-v_k\|_2}{\|v_{k+1}\|_2}$ and PSNR via iterations for the image $``Church''$ in Fig. \ref{figdeblur}.}
\label{errordeblur}
\end{figure*}

\subsubsection{Quantitative and qualitative comparison}
We choose ten testing images including five grayscale images (GSet5) and five color images (CSet5), for image deblurring experiments; see  Fig. \ref{blurimg}.  We quantitatively evaluate our method on the dataset with two real blur kernels of size $19\times 19$ and $27\times 27$ from \citet{Levin2009understanding} and  additive Gaussian noises of noise levels 4 and 8, respectively.

In Table \ref{tabdeblur}, it can be seen that our proposal is more stable and robust against different blur kernels and noises for both grayscale and color images. Since DMPHN and MPRNet are blind deblurring methods, they are not as powerful as other learning-based methods in removing the blurry contained in the images. More importantly, our CurvPnP outperforms the available most effective method DPIR by an average PSNR of 0.16dB$\sim$0.46dB on grayscale images and 0.21dB$\sim$0.49dB on color images. Note that the DPIR requires the ground truth noise level. When the noise level is approximated by the estimation method \citep{chen2015efficient}, the performance of the DPIR+ drop to a certain degree for some test images.

Fig. \ref{figdeblur} illustrates the deblurring results on the images $``Wall''$ and $``Church''$. As shown by both visual and quantitative results, DMPHN cannot handle the distortion of blur, while the IRCNN, DPIR and DPIR+ tend to smooth out the edges and details. Instead, the images restored by our CurvPnP are much sharper and closer to the ground-truth. Specifically, our CurvPnP can yield the crotch of the tree for the image $``Church''$, while other methods fail to provide the clear structures.

\subsubsection{Computational efficiency and numerical convergence}

In Table \ref{deblurtime}, we record the average running time of different restoration methods on image deblurring datasets with the first blur kernel and noise level 8. Although our CurvPnP is somehow slower than other learning based methods, the computational speed is still acceptable for consuming less than second to process a color image of size  288 $\times$ 288.

Fig. \ref{errordeblur} records the evolution of both the relative errors and PSNR on the image $``Church''$ in Fig. \ref{figdeblur} by 50 iterations for our CurvPnP method. Although the relative error curves vibrate at the beginning, they all converge to zero as the iteration number keeps increasing in Fig. \ref{errordeblur} (a). Fig. \ref{errordeblur} (b) also demonstrate that both PNSR of $x_k$ and $v_k$ converge quickly. As can be seen, since $v$ is the clean image estimated by the denoiser, it converges faster than $x$, and the two variables converge to the same steady values.

\subsection{Single Image Super-Resolution}
\begin{table*}[!htb]
\begin{center}
\begin{minipage}{\textwidth}
\caption{Average PSNR(dB) results of different methods for single image super-resolution on CC and PolyU datasets. The best and second best scores are \textbf{highlighted} and \underline{underline}, respectively.}\label{tabsisr}%
\small
\begin{tabular*}{\textwidth}
{@{\extracolsep{\fill}}lllllllllll@{\extracolsep{\fill}}}
\toprule
Datasets & Scale & $\sigma_s$  & Channel & TFOV & ZSSR  & SRFBN & IRCNN & DPIR & DPIR+ & CurvPnP\\
\midrule
 CC   & $\times$2 & 8   & RGB & 29.63  &  26.90  & 26.99 &  33.56  & \underline{39.97}  & 38.73  &  \textbf{40.11} \\
 &   &  & Y & 33.11  &  30.89 & 30.93 & 36.55  & \underline{41.98}  & 41.06  &  \textbf{42.08} \\
 & $\times$2 & 16   & RGB & 24.72  &  21.71  & 21.66 & 30.66  & \underline{37.15} & 36.09  &  \textbf{37.44} \\
 &   &  & Y & 28.89  & 26.05 & 25.97 & 33.41  & \underline{39.25} & 38.86  &  \textbf{39.48} \\
   & $\times$3  & 4   & RGB & 33.95  & 32.56   & 32.87 & 32.30  & 35.11  & \underline{35.22}  & \textbf{35.68}  \\
 &   &  & Y & 37.04  & 36.19  & 36.50 & 36.66  &  36.77 & \underline{37.03}  & \textbf{37.37}  \\
   & $\times$4  & 8   & RGB & 28.18  &  26.94  & 27.29 & 26.53  & \underline{30.46}  & 29.96  & \textbf{32.13}  \\
 &   &  & Y & 31.74  & 30.73 & 31.27 &  30.87 & 32.20 & \underline{32.52}  & \textbf{33.88}  \\

 \cline{2-11}

 PolyU   & $\times$2 & 8   & RGB & 29.75  &  27.03  & 27.03  & 33.56  & \underline{40.82}  &  40.12  & \textbf{40.88}  \\
 &   &  & Y & 33.39  & 31.08  & 31.14  & 36.53  & \underline{42.88}  &  42.29  & \textbf{42.91}  \\
 & $\times$2 & 16   & RGB & 24.70  &  21.68  & 21.64  & 31.04  &  \underline{38.78} & 37.92  & \textbf{38.97}  \\
 &   &  & Y & 29.00  & 26.16 & 26.09  & 33.77 & \underline{40.87}  & 40.13 &  \textbf{41.04} \\
   & $\times$3  & 4   & RGB & 33.84  &  32.56  &  32.36 & 32.22  & \underline{36.75}  & 36.41  & \textbf{37.04}  \\
 &   &  & Y & 37.05  & 36.34  & 35.91  &  36.62 & 38.33  &  \underline{38.34} &  \textbf{38.67} \\
   & $\times$4  & 8  & RGB & 28.29  &  26.98  &  27.35 & 26.59  & \underline{32.48}  & 30.54  & \textbf{33.41}  \\
 &   &  & Y & 32.08  & 30.91 & 31.55  & 31.06 & \underline{34.02}  & 33.48  &  \textbf{35.32} \\
\botrule
\end{tabular*}
\end{minipage}
\end{center}
\end{table*}

\begin{figure*}[h]
\centering
\begin{minipage}[htbp]{0.24\linewidth}
\centering
\includegraphics[width=3.9cm]{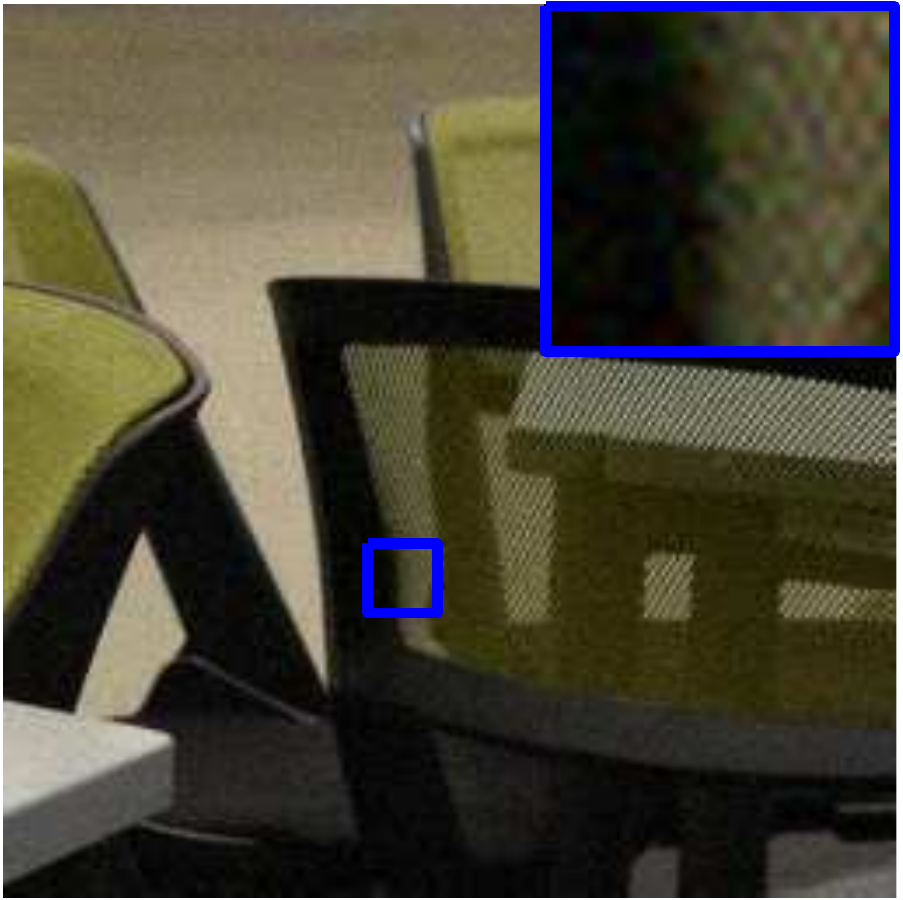}
{ \footnotesize{Low-resolution Image} }
\end{minipage}
\begin{minipage}[htbp]{0.24\linewidth}
\centering
\includegraphics[width=3.9cm]{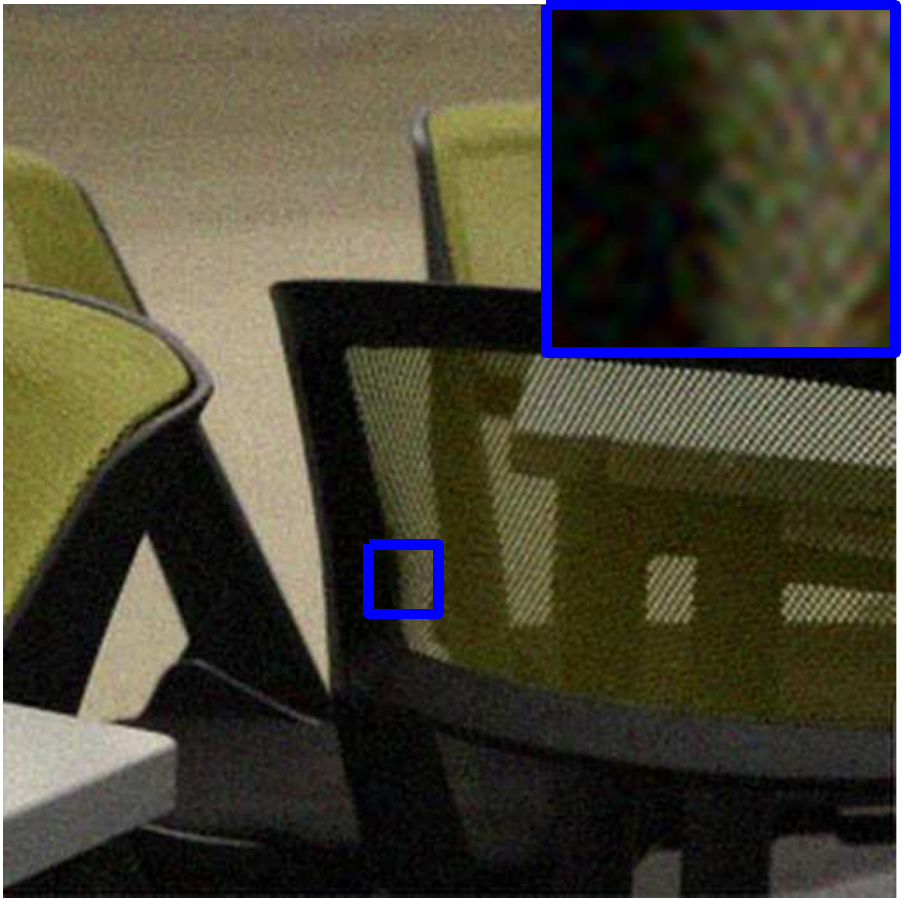}
{ \footnotesize{TFOV (26.29dB)} }
\end{minipage}
\begin{minipage}[htbp]{0.24\linewidth}
\centering
\includegraphics[width=3.9cm]{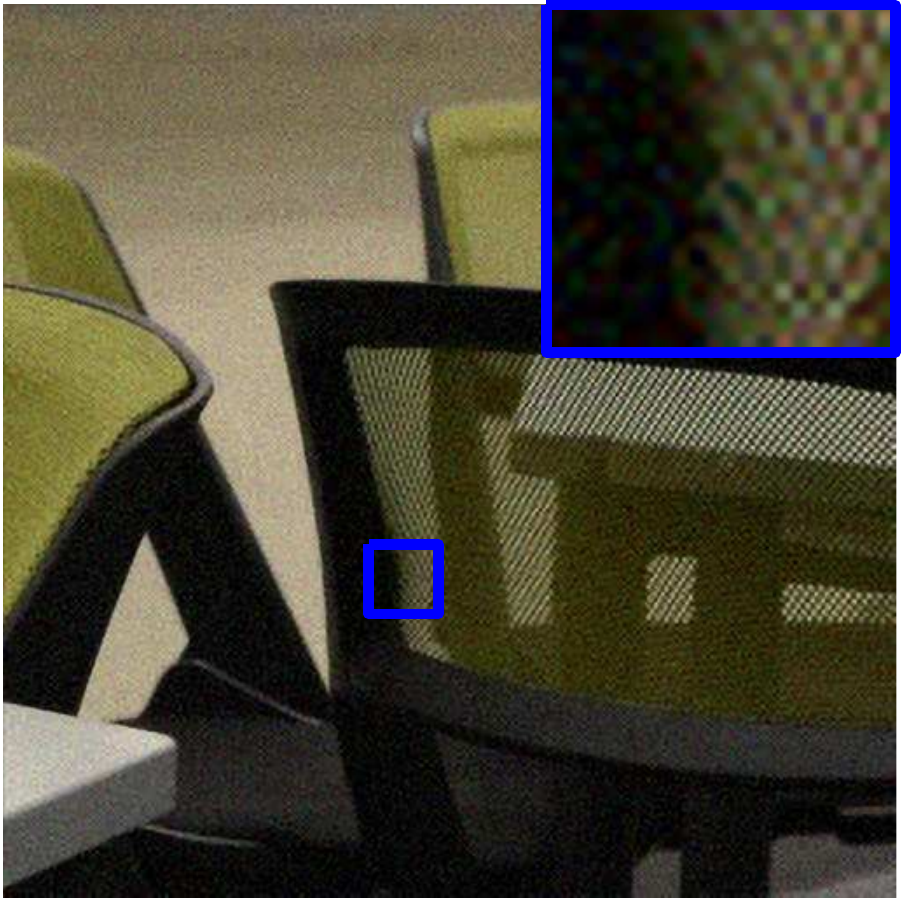}
{ \footnotesize{ZSSR (24.79dB)} }
\end{minipage}
\begin{minipage}[htbp]{0.24\linewidth}
\centering
\includegraphics[width=3.9cm]{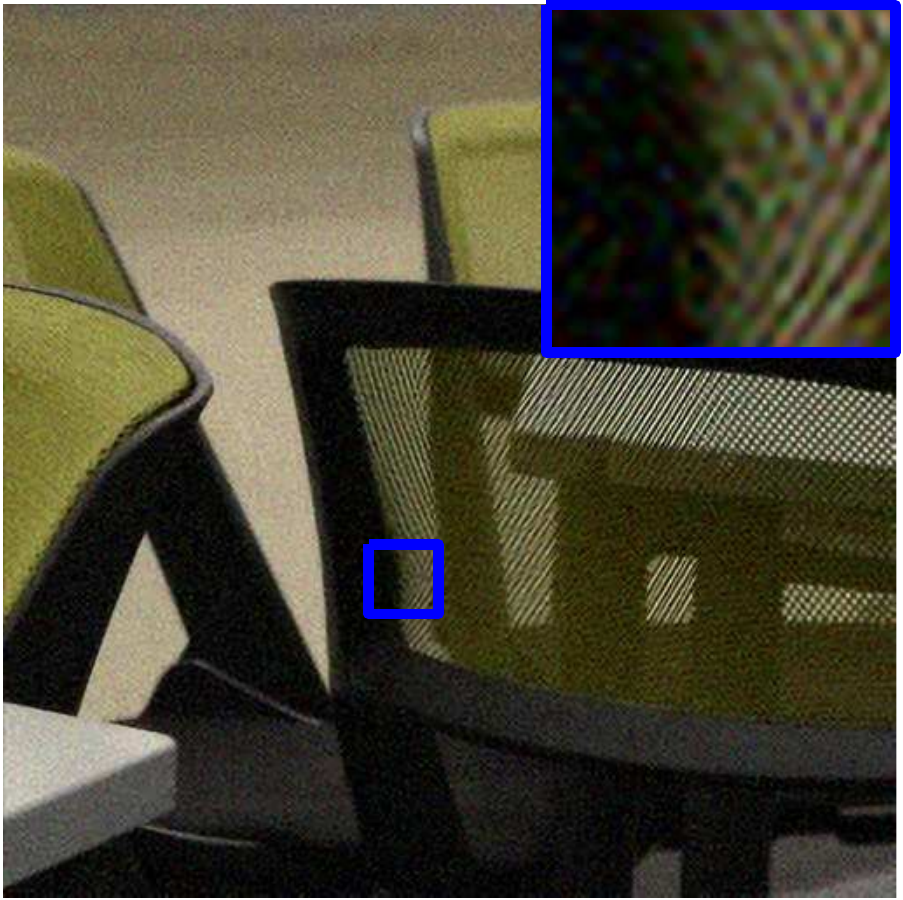}
{ \footnotesize{SRFBN (23.68dB)} }
\end{minipage}\\
\begin{minipage}[htbp]{0.24\linewidth}
\centering
\includegraphics[width=3.9cm]{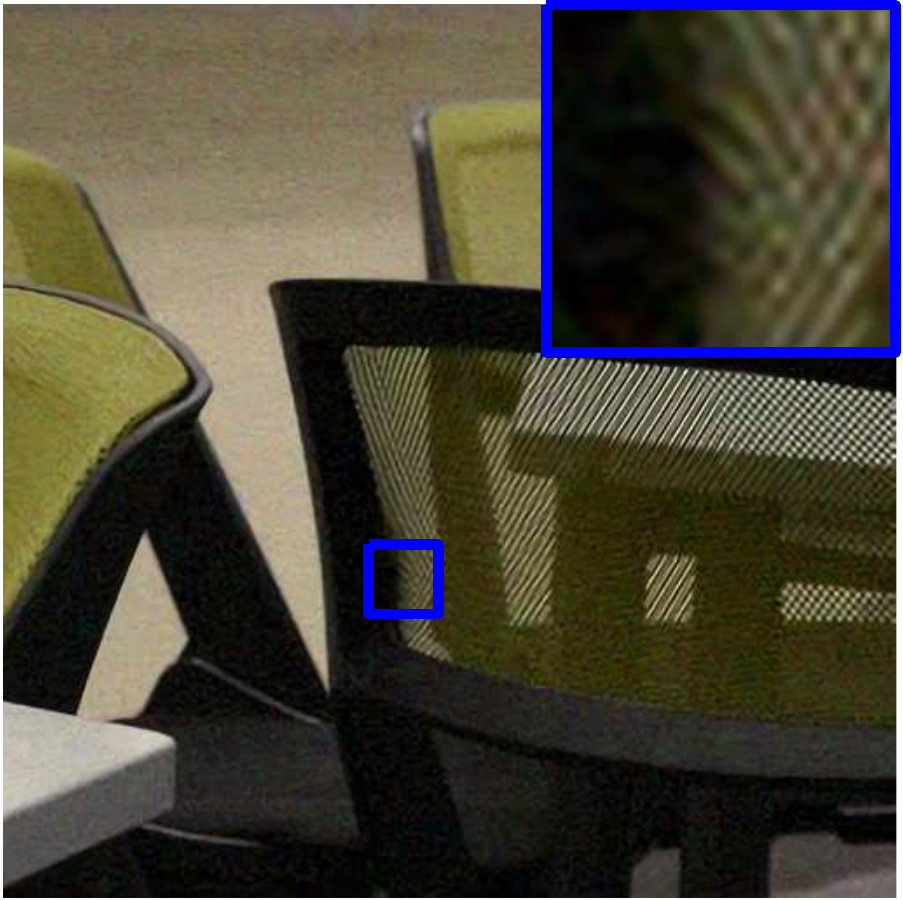}
{ \footnotesize{IRCNN (27.88dB)} }
\end{minipage}
\begin{minipage}[htbp]{0.24\linewidth}
\centering
\includegraphics[width=3.9cm]{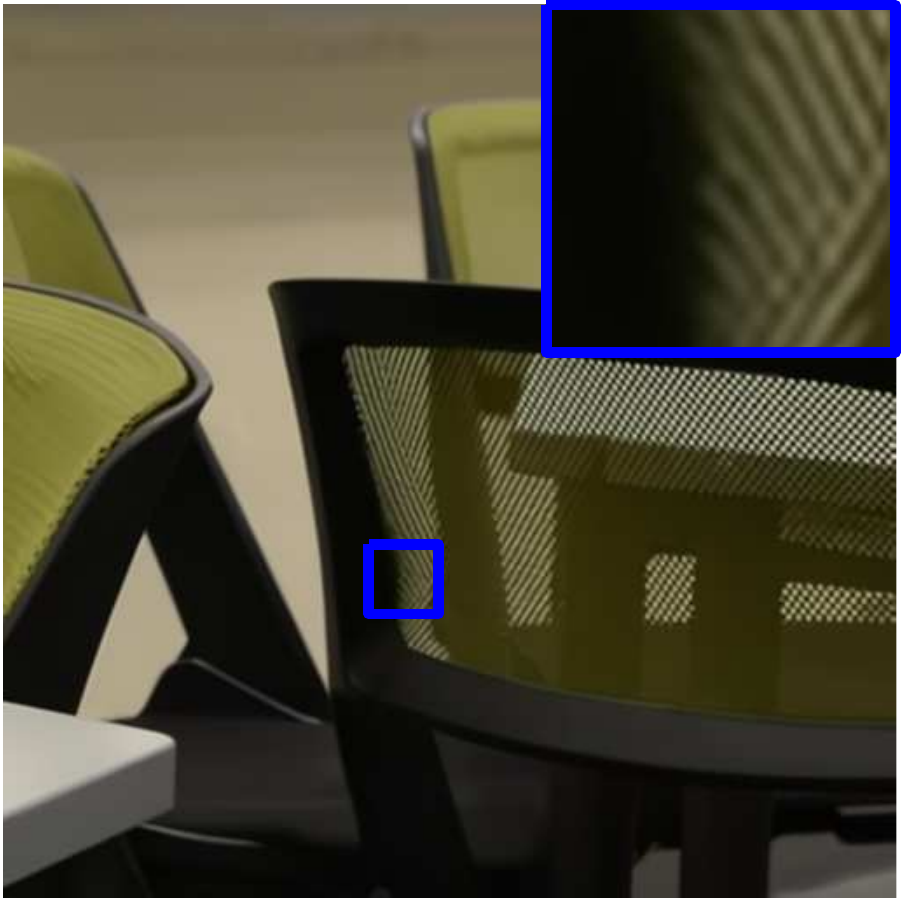}
{ \footnotesize{DPIR (32.31dB)} }
\end{minipage}
\begin{minipage}[htbp]{0.24\linewidth}
\centering
\includegraphics[width=3.9cm]{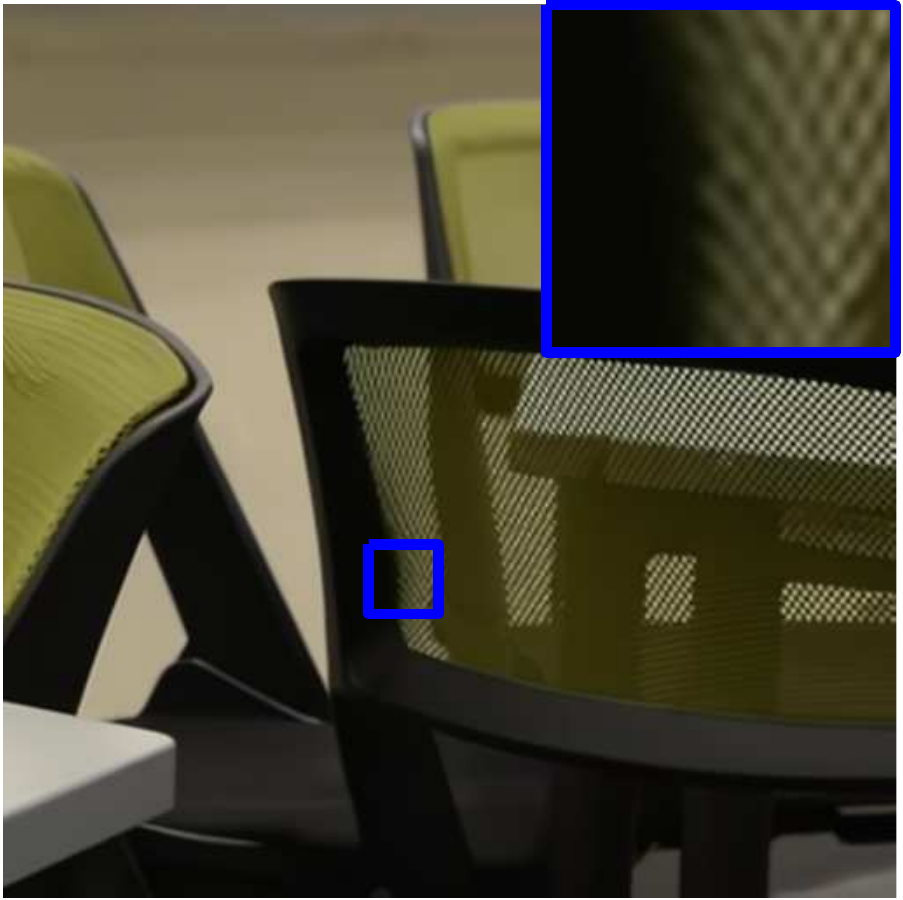}
{ \footnotesize{DPIR+ (32.51dB)} }
\end{minipage}
\begin{minipage}[htbp]{0.24\linewidth}
\centering
\includegraphics[width=3.9cm]{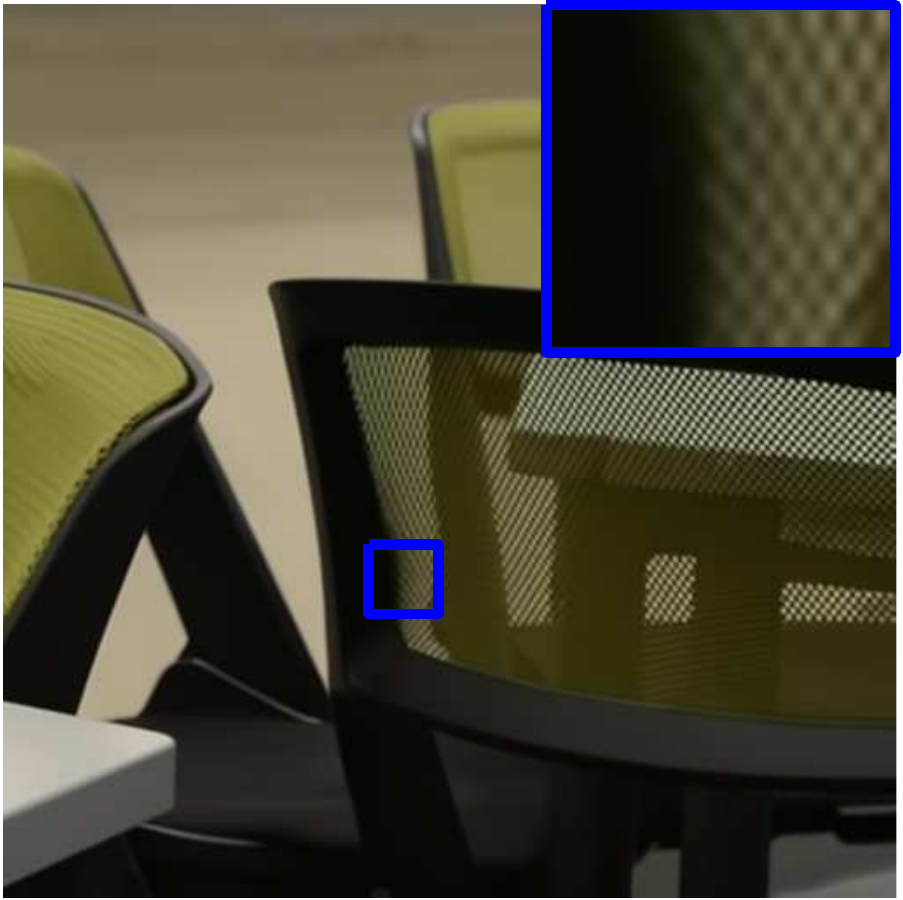}
{ \footnotesize{CurvPnP (32.82dB)} }
\end{minipage}\\
\begin{minipage}[htbp]{0.24\linewidth}
\centering
\includegraphics[width=3.9cm]{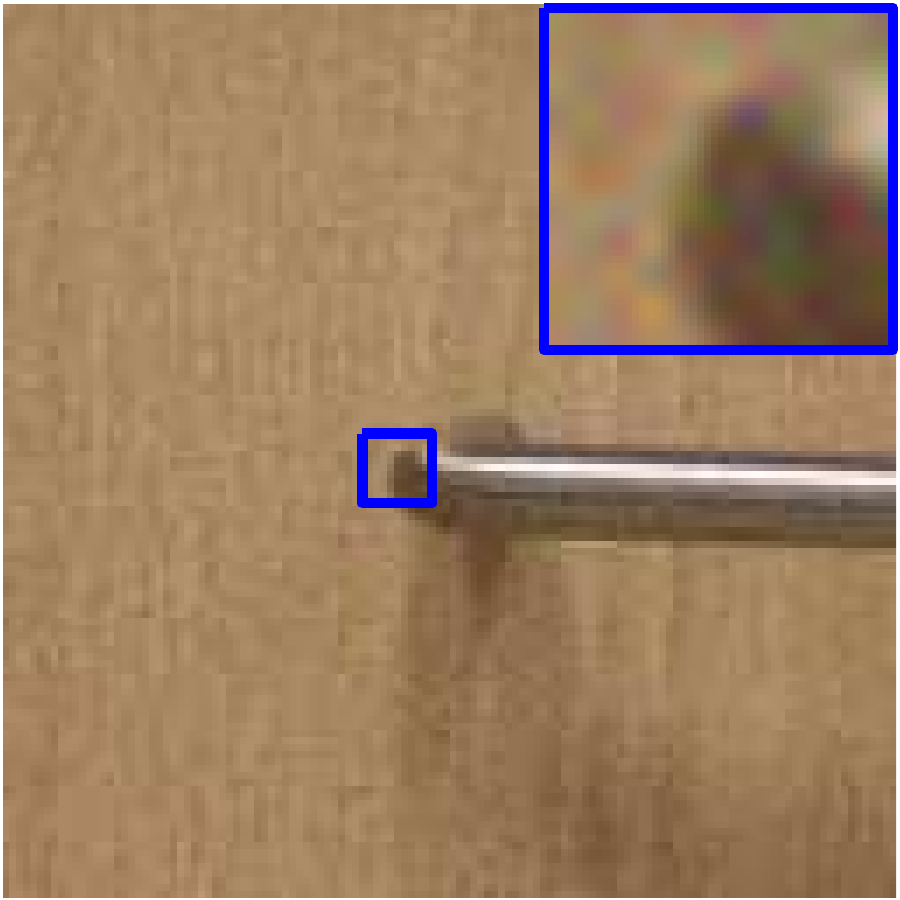}
{ \footnotesize{Low-resolution Image} }
\end{minipage}
\begin{minipage}[htbp]{0.24\linewidth}
\centering
\includegraphics[width=3.9cm]{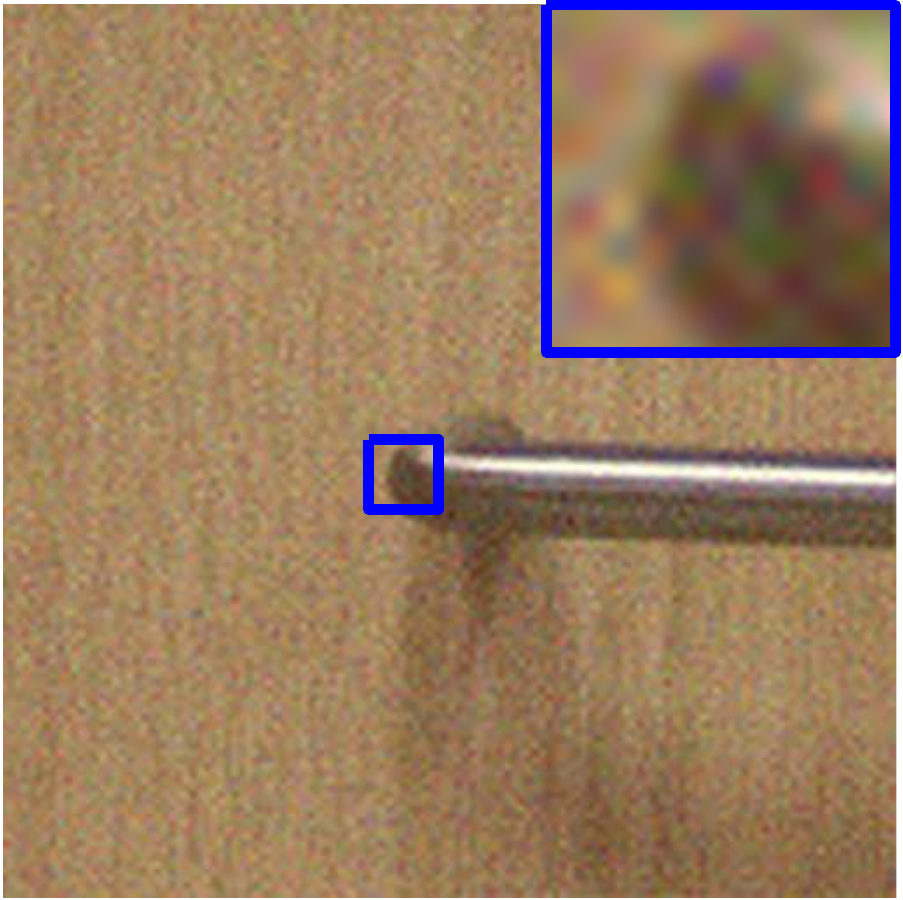}
{ \footnotesize{TFOV (28.94dB)} }
\end{minipage}
\begin{minipage}[htbp]{0.24\linewidth}
\centering
\includegraphics[width=3.9cm]{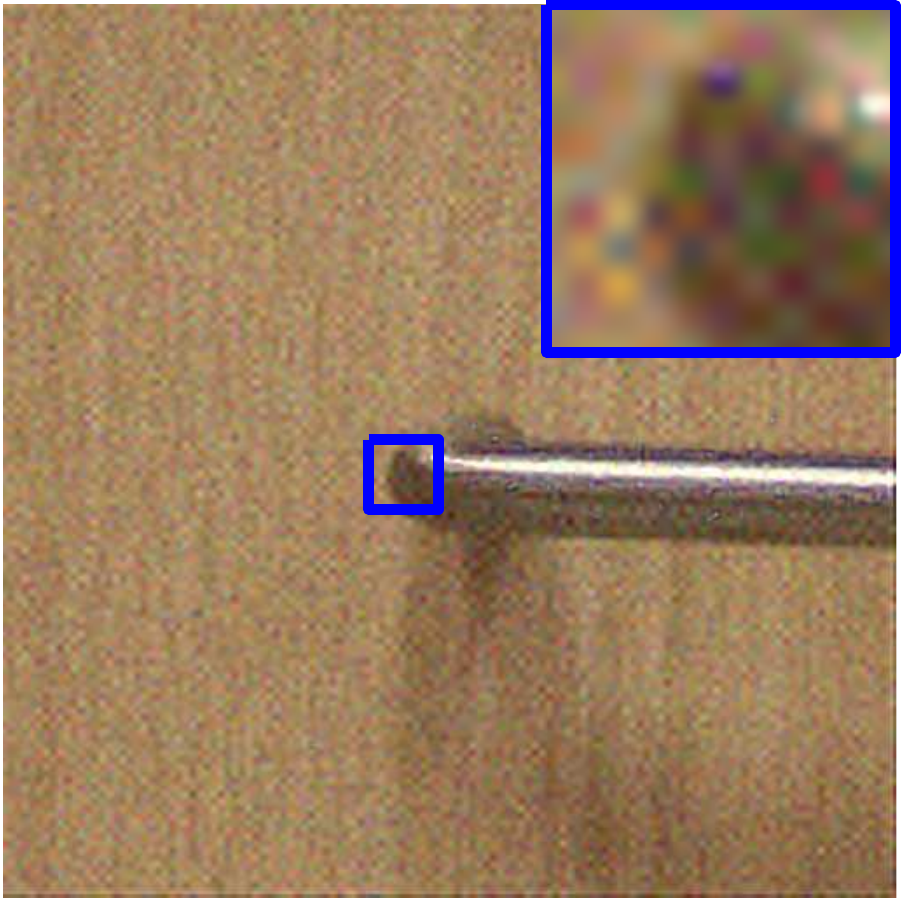}
{ \footnotesize{ZSSR (27.28dB)} }
\end{minipage}
\begin{minipage}[htbp]{0.24\linewidth}
\centering
\includegraphics[width=3.9cm]{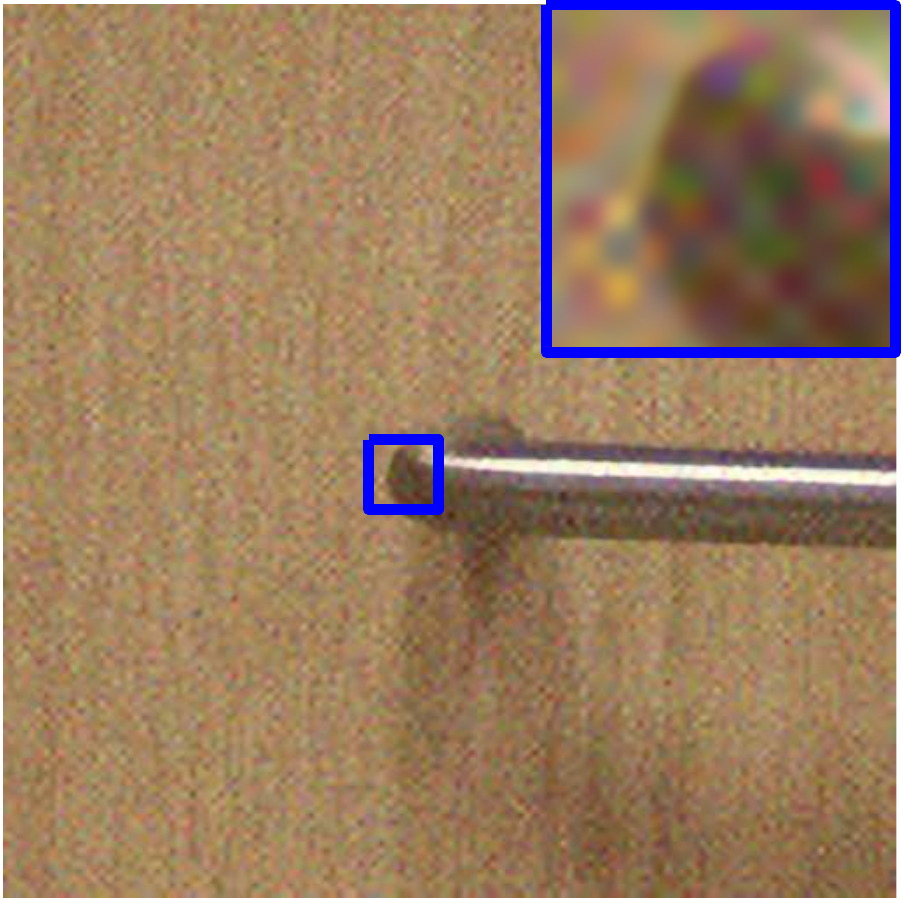}
{ \footnotesize{SRFBN (27.72dB)} }
\end{minipage}\\
\begin{minipage}[htbp]{0.24\linewidth}
\centering
\includegraphics[width=3.9cm]{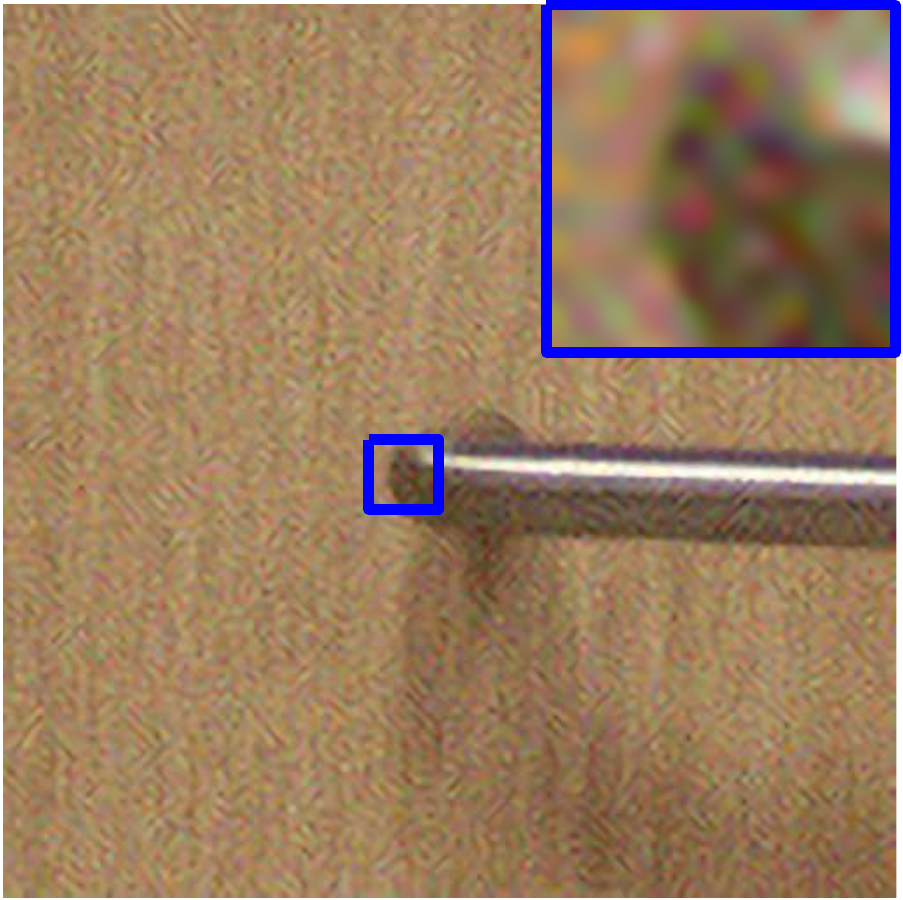}
{ \footnotesize{IRCNN (26.98dB)} }
\end{minipage}
\begin{minipage}[htbp]{0.24\linewidth}
\centering
\includegraphics[width=3.9cm]{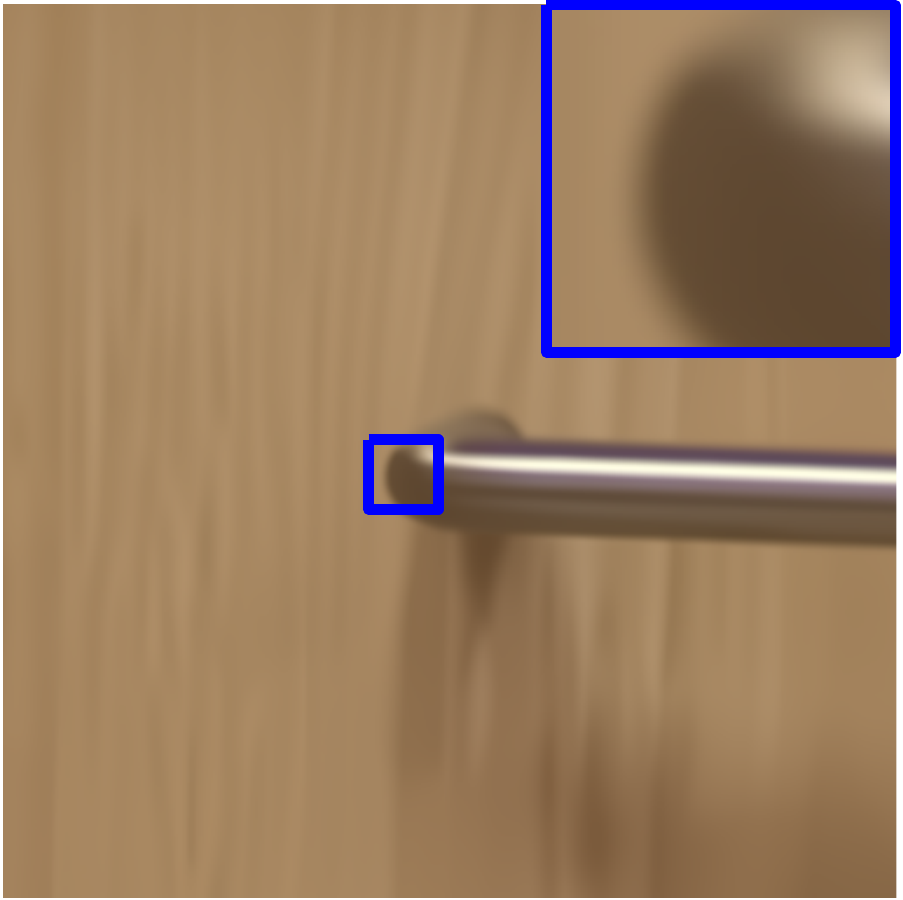}
{ \footnotesize{DPIR (35.39dB)} }
\end{minipage}
\begin{minipage}[htbp]{0.24\linewidth}
\centering
\includegraphics[width=3.9cm]{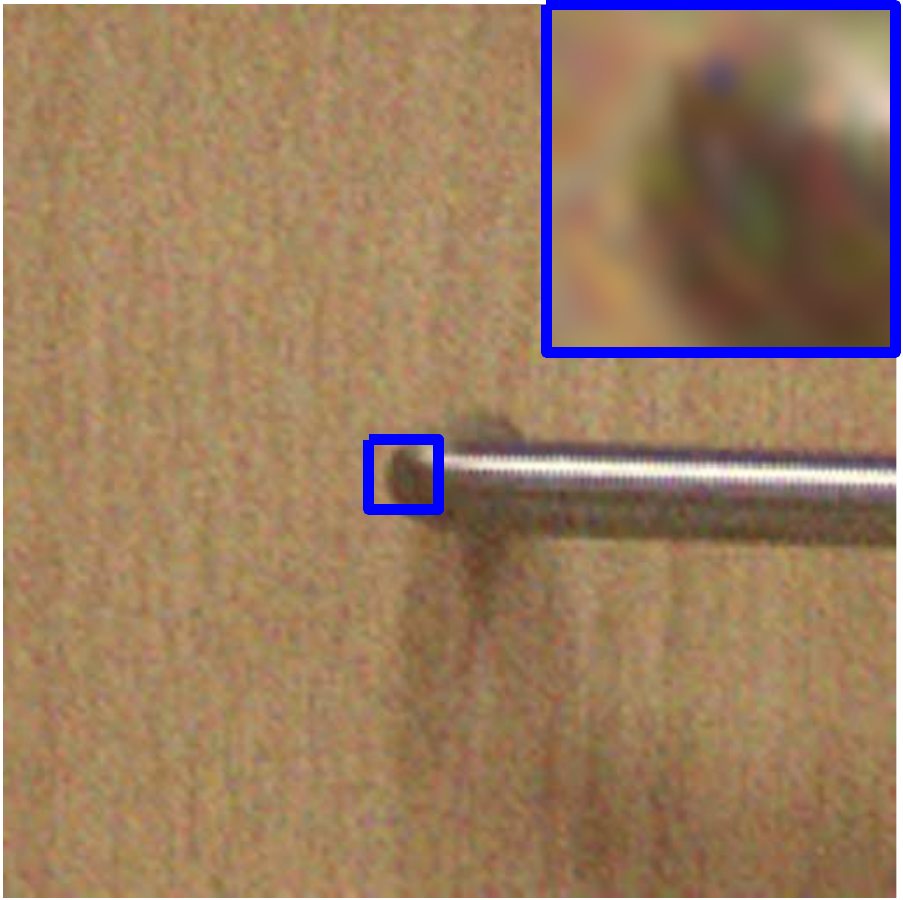}
{ \footnotesize{DPIR+ (31.52dB)} }
\end{minipage}
\begin{minipage}[htbp]{0.24\linewidth}
\centering
\includegraphics[width=3.9cm]{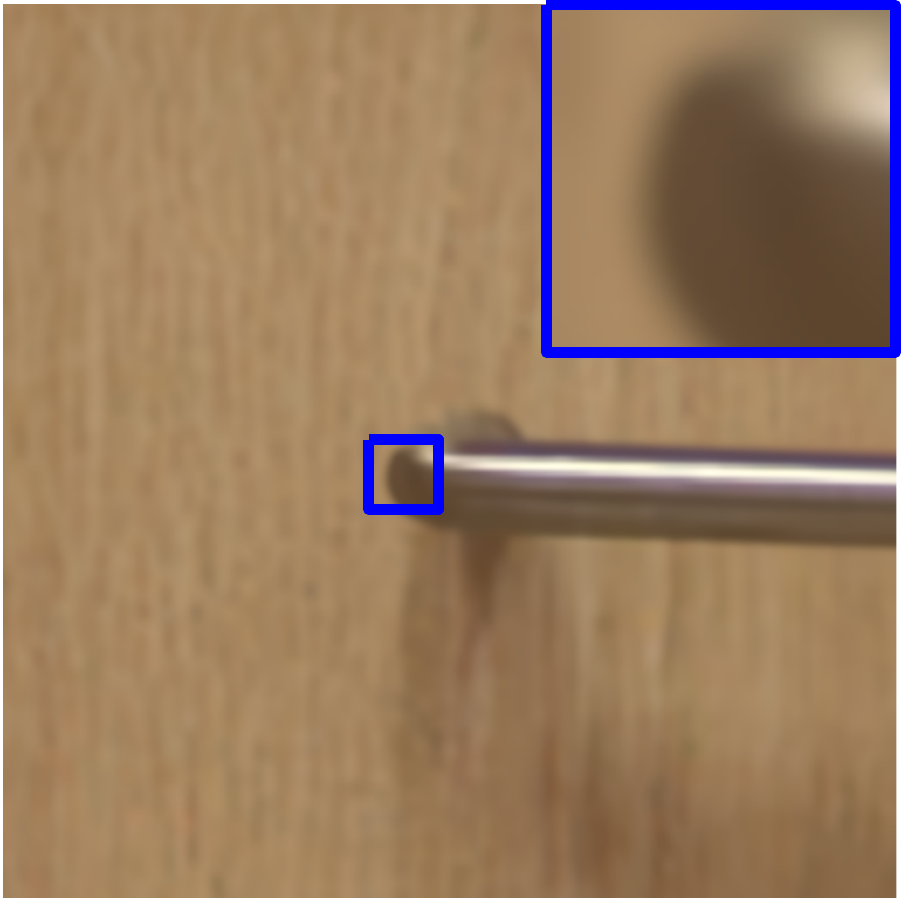}
{ \footnotesize{CurvPnP (37.30dB)} }
\end{minipage}
\caption{Single image super-resolution results of different methods on two representative images selected from PolyU dataset.}
\label{figsisr}
\end{figure*}

Single image super-resolution (SISR) aims to improve the resolution and quality of the low-resolution images, which can be described as follows
\begin{equation}
y=(Ax)\downarrow_s+n,
\label{eqsisr}
\end{equation}
where $A$ denotes the blur kernel and $\downarrow_s$ denotes downsampling by a factor of $s$, i.e., selecting the upper-left pixel for each distinct $s\times s$ patch.

According to \citet{Zhao2016fast}, we solve the $x$ subproblem w.r.t. the single image super-resolution problem by the FFT as follows
\[
\scriptsize
x_{k+1}\!\!=\!\!\mathcal{F}^{-1}\Big(\frac{1}{\alpha_k}\big(d-\overline{\mathcal{F}(A)}\odot_s\frac{(\mathcal{F}(A)d)\Downarrow_s}{(\overline{\mathcal{F}(A)}\mathcal{F}(A))\Downarrow_s+\alpha_k\mathcal{I}}\big)\Big),
\]
where
\begin{equation*}
d=\overline{\mathcal{F}(A)}\mathcal{F}(y\uparrow_s)+\alpha_k\mathcal{F}(v_k),
\end{equation*}
and $\alpha_k=\frac{\lambda\sigma_s^2}{\sigma_k^2}$,  $\odot_s$ denotes element-wise multiplication to the $s\times s$ distinct blocks of $\overline{\mathcal{F}(A)}$, and $\Downarrow_s$ denotes averaging the $s\times s$ distinct blocks. Note that the periodic boundary condition is used to guarantee the implementation of FFT.

\subsubsection{Comparison methods}
Except for aforementioned PnP methods, our CurvPnP is also compared with the representative SISR methods including both model-based method and learning-based method, the details of which are described as follows
\begin{itemize}
    \item TFOV \citep{yao2020total}: The Total Fractional-Order Variation (TFOV) is the fractional-order TV regularization SISR method, which is solved by the scalar auxiliary variable algorithm.
    \item ZSSR \citep{shocher2018zero}: The Zero-Shot Super-Resolution (ZSSR) is an unsupervised CNN-based zero-shot SR method, which does not require the training samples and prior training.
    \item SRFBN \citep{li2019feedback}: The Super-Resolution FeedBack Network (SRFBN) is a recurrent neural network with the feedback mechanism, which was trained on DIV2K and Flickr2K datasets degraded by the bicubic downsampling.
\end{itemize}

\begin{table*}[h]
\begin{center}
\begin{minipage}{\textwidth}
\caption{Average runtime comparison (in seconds) among different methods on single image super-resolution datasets.} \label{sisrtime}
\small
\begin{tabular*}{\textwidth}
{@{\extracolsep{\fill}}llllllll@{\extracolsep{\fill}}}
\toprule
Datasets & TFOV & ZSSR  & SRFBN & IRCNN & DPIR & DPIR+ & CurvPnP \\
\midrule
 CC  & 53.6637  & 28.5531 & 0.0914  &  0.2859 & 1.2214  & 0.7311  & 1.0747 \\
 PolyU  & 52.9980  & 26.0500 & 0.0906  &  0.2854 & 1.2114  & 0.6817  & 1.4411 \\
\botrule
\end{tabular*}
\end{minipage}
\end{center}
\end{table*}

\begin{table*}[h]
\begin{center}
\begin{minipage}{\textwidth}
\caption{Ablation study on individual components of the proposed C-UNet on CC dataset and CurvPnP on the image $``Woman''$ with the first kernel. The best scores are \textbf{highlighted}.}\label{ablation}%
\small
\begin{tabular*}{\textwidth}
{@{\extracolsep{\fill}}llllllll@{\extracolsep{\fill}}}
\toprule
 Task & Mehod & Noise Level Map & Curvature Map & $\sigma_s=20$ & $\sigma_s=35$ & $\sigma_s=50$ \\
\midrule
 \multirow{4}{*}{Denoising} & \multirow{4}{*}{UNet} & $\times$  & $\times$ & 36.90 & 36.99  & 35.26 \\
 & & $\checkmark$ & $\times$ & 39.64 & 37.04  & 35.33 \\
 & & $\times$  & $\checkmark$ & 39.65 & 37.05  & 35.34 \\
 & & $\checkmark$  & $\checkmark$ & \textbf{39.70} &  \textbf{37.11} & \textbf{35.40} \\
\midrule
 Task & Mehod & Noise Level Map & Curvature Map  & $\sigma_s=4$ & $\sigma_s=6$ & $\sigma_s=8$  \\
\midrule
 \multirow{4}{*}{Deblurring}  & \multirow{4}{*}{PnP} & $\times$  & $\times$ & 27.53  & 25.82 & 24.57 \\
 & & $\checkmark$ & $\times$ & 28.07 & 26.41 &  25.23  \\
 & & $\times$  & $\checkmark$ & 27.71 & 26.05 & 24.87  \\
 & & $\checkmark$  & $\checkmark$ & \textbf{28.28} & \textbf{26.69} & \textbf{25.53}   \\
\botrule
\end{tabular*}
\end{minipage}
\end{center}
\end{table*}

\subsubsection{Quantitative and qualitative comparison}
To test and verify the performance of our CurvPnP, we apply both the degradation of the Gaussian blur kernel with standard deviation 0.7 and the bicubic degradation to the testing images. Since the bicubic downsampling can be approximated by setting a proper blur kernel \citep{zhang2020deep}, the solution of the $x$-sub minimization problem can also be used to solve the bicubic degradation problem. Simultaneously, we consider different combinations of the downsampling factors and noise levels. For the Gaussian blur degradation, we choose the scale factor to be 2 and the noise level of 8 and 16. For the bicubic degradation, both the scale factor 3 with noise level of 4 and the scale factor 4 with noise level of 8 are used for evaluation.

The PSNR of the comparison methods on the RGB channels and Y channel (in YCbCr color space) for CC and PolyU datasets are listed in Table \ref{tabsisr}. As can be seen, our proposal outperforms the state-of-the-art SISR methods w.r.t. the scale factors and noise levels. In particular, when the scale factor is set as 4 for bicubic degradation, our CurvPnP gains the 0.93dB$\sim$1.67dB higher PSNR on RGB channels and 1.3dB$\sim$1.68dB PSNR on Y channel over DPIR. Simultaneously, our CurvPnP achieves much higher PSNR than DPIR+, which demonstrates the effectiveness of the proposed denoiser. Our CurvPnP can automatically estimate the noise level and remove different noises without knowing the exact noise level and using other noise estimation methods.

Fig. \ref{figsisr} shows the visual comparison results of different methods on the images $``NikonD800\_4-5\_160\_1800\_classroom\_9''$ from PolyU dataset degraded by the Gaussian blur kernel with the scale factor 2 and $\sigma_s=8$, and the image $``Canon600D\_3-5\_125\_1600\_waterhouse\_10''$ from PolyU dataset corrupted by the bicubic degradation with the scale factor 4 and $\sigma_s=8$. In addition to quantitative comparison, our CurvPnP can obtain much vivid images  with fine textures and sharp edges as exhibited. As shown in the magnified portions, both DPIR and DPIR+ cannot restore details, while TFOV, ZSSR, SRFBN and IRCNN cannot eliminate noise, but also produce color artifacts. Obviously, our CurvPnP gives the fine structures of the stripes on the chair and the door railing.

\subsubsection{Computational efficiency}
Table \ref{sisrtime} reports the average running time of different methods on the SISR problem with the scale factor 4 and $\sigma_s=8$. Since our CurvPnP is terminated by the PSNR values rather than the iteration number, it consumes similar computational time as the DPIR, which is terminated by the fixed iteration numbers of 24 for achieving satisfied restoration results. Although SRFBN, IRCNN and DPIR+ are faster than CurvPnP, their PSNR values are much lower than CurvPnP. The TFOV and ZSSR perform worse than our CurvPnP in terms of restoration performance and running time. Obviously, our CurvPnP achieves the best balance in reconstruction effect and computational efficiency.

\subsection{Ablation Studies}

The ablation studies are conducted on both image denoising and deblurring problems, where the CC dataset and image $``Woman''$ are used as examples. Our network models are trained on image patches of size 128$\times$128 using a batch size of 16. The numbers of channels are set as 32, 64, 128 and 256 for the first scale to the fourth scale, respectively. Table \ref{ablation} demonstrates the effectiveness of the noise level map and curvature map by removing them from our model. As shown, taking out the noise level map or curvature map results in the  degradation of the restoration. To be specific, PSNR decreases by 0.56dB$\sim$0.66dB and 0.21dB$\sim$0.30dB when noise level map or curvature map are removed for image deblurring. If both noise level map and curvature map modules are removed from the C-UNet, the performance significant drops by 0.75dB$\sim$0.96dB, which demonstrates the advantages of the two modules in investigating and integrating the features.

\section{Conclusion}
\label{conlusion}
In this paper, we have proposed a novel PnP blind image restoration method, which can be used to effectively deal with various image restoration problems such as image deblurring and super-resolution. More specifically, a new image restoration model was built up by regarding the noise level as a variable. The resulting model was then solved by the penalty method and alternating direction method, where the subproblem w.r.t. the data fidelity was solved by the closed-form solution and both noise level and denoised image were handled by the CNN models. Furthermore, we introduced the ConvNeXt block and curvature supervised attention module into the UNet architecture to enhance the useful features. Numerical results have demonstrated that our CurvPnP outperforms the state-of-the-art PnP methods on different image restoration problems. Compared to the end-to-end learning-based methods, our CurvPnP is more flexible to adapt with different image restoration tasks. Our future work includes to investigate the performance of our CurvPnP on other image restoration problems and to explore effective unsupervised image restoration methods.

\section*{Acknowledgement}
The work was partially supported by the National Natural Science Foundation of China (NSFC 12071345, 11701418).



\end{document}